\documentclass[12pt]{article}

\usepackage{amsfonts,amsmath}
\usepackage{scalerel}
\usepackage{amssymb}
\usepackage{MnSymbol}
\usepackage{ytableau}
\usepackage{booktabs}
\usepackage{mathtools} % smashoperator for integrals
\usepackage{tikz-cd} % это тоже для диаграмм

\usepackage{pgfplots}
\pgfplotsset{compat=1.5}
\usepgfplotslibrary{fillbetween} 
\usepackage{tikz}
\usetikzlibrary{arrows.meta,bending,patterns,angles,calc}
\usetikzlibrary{patterns.meta}
\usepackage{siunitx}
\usetikzlibrary{decorations.shapes}
\tikzset{decorate sep/.style 2 args=
{decorate,decoration={shape backgrounds,shape=circle,shape size=#1,shape sep=#2}}}

\pgfplotsset{ignore zero/.style={%
  #1ticklabel={\ifdim\tick pt=0pt \else\pgfmathprintnumber{\tick}\fi}
}} %this is to have only one 0 on plot 

\newcommand{\moast}{\scaleobj{0.8}{\circledast}}
\newcommand{\Oast}{\circledast}
\newcommand{\dr}{{{\rm d}}}

\renewcommand{\theequation}{\thesection.\arabic{equation}}

\makeatletter \@addtoreset{equation}{section} \makeatother

{\vspace{3mm} }

\def\al{\alpha}

\def\*{\star}

\def\E2{\mathbf{E}}

\def\y{\mathbf{y}}

\def\pp{\mathrm{p}}
\def\qq{\mathrm{q}}

\newcommand{\be}{\begin{equation}}
\newcommand{\ee}{\end{equation}}
\newcommand{\bee}{\begin{eqnarray}}
\newcommand{\beee}{\begin{array}}
\newcommand{\eee}{\end{eqnarray}}
\newcommand{\eeee}{\end{array}}
\newcommand{\gb}{\beta}
\newcommand{\gga}{\gamma}

\newcommand{\gd}{\delta}

\newcommand{\gep}{\epsilon}

\newcommand{\gs}{\sigma}
\newcommand{\go}{\omega}

\newcommand{\dal}{\dot \alpha}
\newcommand{\dgb}{\dot \beta}
\newcommand{\dgga}{\dot \gamma}

\newcommand{\nn}{\nonumber}

\newcommand{\p}{\partial}

\newcommand{\ff}{\frac}
\newcommand{\rom}[1]{\uppercase

\expandafter{\romannumeral #1\relax}}

\begin{document}
    
\begin{flushright}
FIAN/TD/02-2026\\
\end{flushright}

\vspace{0.5cm}
\begin{center}
{\large\bf Irregular higher-spin generating equations and chiral perturbation theory}

\vspace{1 cm}

\textbf{V.E.~Didenko}\\

\vspace{1 cm}

\textbf{}\textbf{}\\
 \vspace{0.5cm}
 \textit{I.E. Tamm Department of Theoretical Physics,
Lebedev Physical Institute,}\\
 \textit{ Leninsky prospect 53, 119991, Moscow, Russia }\\
 \par\end{center}

\begin{center}
\vspace{0.6cm}
e-mail: didenko@lpi.ru\\
\par\end{center}

\vspace{0.4cm}

\begin{abstract}
\noindent 
We present a complementary approach to the standard Vasiliev framework for nonlinear higher-spin interactions in four dimensions, aimed at identifying their minimally nonlocal form. Our proposal introduces a generating system for higher-spin vertices at the level of classical equations, which we refer to as irregular, in contrast to the regular case described by Vasiliev. This system extends the recently proposed equations for (anti)holomorphic interactions by incorporating the mixed sector. Its perturbative series encompasses the entire (anti)holomorphic sector in the leading order, with vertices related to powers of the complex parity-breaking parameter $\eta$ or $\bar\eta$. The subsequent corrections facilitate the mixing of the two sectors, with vertices carrying mixed powers of $\eta$ and $\bar\eta$. Consistency relies on the nonlinear algebraic constraint, which is shown to be satisfied, at least, in the quadratic and cubic approximations. As a result, the previously discussed (anti)holomorphic interactions in the literature can be systematically extended to generate vertices of the form $\eta^N \bar\eta^k$ and their conjugate, at least, for $k \leq 2$ and any $N$. As a byproduct of our analysis, we also identify the new higher-spin structure dualities.

\end{abstract}
\newpage
\tableofcontents 
\newpage

\section{Introduction}
Vasiliev's higher-spin theory\footnote{Reviews are available in \cite{Vasiliev:1999ba, Bekaert:2004qos, Didenko:2014dwa}.} (HS) in four dimensions \cite{Vasiliev:1990en, Vasiliev:1992av}, which includes gauge fields of all spins, is considered a promising toy model for quantum-consistent gravity. Although the conventional action for this theory has been unknown for some time (see \cite{Boulanger:2011dd} for a nonstandard version), there are compelling reasons to believe it might exhibit good quantum behavior. 
Firstly, apart from a single parameter related to spatial parity, along with a few reductions and supersymmetry generalizations, HS theory lacks ambiguity, even at the classical level. This makes it a unique theory defined by the symmetry principles. Additionally, the holography conjecture \cite{Flato:1978qz, Sezgin:2002rt,  Klebanov:2002ja, Leigh:2003gk, Sezgin:2003pt} suggests a duality between interacting HS fields and well-defined quantum theories at the three-dimensional AdS boundary, such as $O(N)$ models\footnote{The duality has been tested at the tree level for 3-point correlation functions by Giombi and Yin \cite{Giombi:2009wh, Giombi:2010vg} using the Vasiliev equations and also in \cite{Didenko:2017lsn}, where the problematic parity broken structures were identified. A general argument supporting the duality at the 3-point level is provided by Maldacena and Zhiboedov \cite{Maldacena:2011jn, Maldacena:2012sf}. For calculations of all free  $O(N)$  correlators derived from the undeformed higher-spin symmetry, refer to \cite{Colombo:2012jx, Didenko:2012tv} and \cite{Gelfond:2013xt}; see also \cite{Engquist:2005yt, Sezgin:2005pv, Colombo:2010fu, Sezgin:2011hq} for earlier related ideas.  More recent studies on this topic can be found in \cite{Scalea:2023dpw}. The reconstruction of the cubic HS action from boundary 3-point correlators is detailed in \cite{Sleight:2016dba}, where the matching of conformal structures and cubic vertices was confirmed.}. This further supports the idea that Vasiliev's HS theory may lead to consistent quantum behavior.

However, even at the classical level, there are unresolved problems. In recent years, the longstanding problem of the locality of HS interactions \cite{Prokushkin:1998bq} has become a pressing issue \cite{Giombi:2009wh, Boulanger:2015ova} for the concrete predictions that HS theory should offer. This is particularly relevant for testing holography beyond cubic order, where nonlocalities are likely to arise. In \cite{Bekaert:2015tva}, where the validity of the Klebanov-Polyakov conjecture \cite{Klebanov:2002ja} was assumed, the reconstruction of the HS quartic vertex from the 3D boundary 4-point correlator has been attempted by inverting the Witten diagrams. However, this process encounters significant nonlocality, which leads to a breakdown of the standard field-theory methods \cite{Sleight:2017pcz}, leaving the quartic HS vertex somewhat beyond reach from the boundary. 

The challenges associated with quartic holographic reconstruction could lead to different interpretations. One possibility is that there are loopholes within the reconstruction process itself; for more on this, see Refs. \cite{Ponomarev:2017qab} and \cite{Neiman:2023orj} that discuss related subtleties. Another consideration is that the Klebanov-Polyakov conjecture might not apply beyond cubic order, meaning that reconstruction from a free boundary scalar may not be a viable option; for a detailed discussion of this point, refer to \cite{Vasiliev:2012vf}, \cite{Diaz:2024kpr}, and \cite{Diaz:2024iuz}. Lastly, the significant nonlocality present in HS theory might suggest that it behaves more like a theory of extended objects, such as strings, rather than a conventional field theory (see \cite{Lysov:2022zlw} and \cite{Neiman:2022enh} for related ideas).

Understanding the (non)locality of HS theory is crucial, particularly in relation to its quartic vertex. This vertex is key to clarifying the duality proposed by Klebanov and Polyakov. However, elucidating the local structure of Vasiliev's equations has proven to be notoriously complex. For an overview of this issue, we refer to a substantial list of papers that address it \cite{Vasiliev:2016xui, Vasiliev:2017cae, Gelfond:2018vmi, Didenko:2018fgx, Didenko:2019xzz, Gelfond:2019tac, Didenko:2020bxd, Gelfond:2021two, Vasiliev:2022med, Didenko:2022eso, Vasiliev:2023yzx, Kirakosiants:2025gpd}.

Globally, this paper aims to establish a local or minimally nonlocal framework for HS vertices in four dimensions, including the complete quartic vertex. To address this problem, we begin with the two sectors of HS theory where locality is well established at all orders: the holomorphic and antiholomorphic, also referred to as (anti)self-dual or (anti)chiral. Nonlinear interaction within these two sectors was developed using the so-called generating equations in \cite{Didenko:2022qga}. This approach, which is similar in spirit to Vasiliev's standard framework, implies that the interaction vertices arise from solving auxiliary partial differential equations with respect to two-component spinorial variables $z$ or $\bar z$. The proposed solution to this system provides vertices with the minimal number of derivatives; see \cite{Vasiliev:2022med}, where the criterion for the minimal vertex is put forward. The remaining vertices, referred to as {\it mixed}, emerge from the interaction between the two sectors. The mixed interactions were not captured by the generating equations from \cite{Didenko:2022qga}. In this paper, we fill in this gap by presenting a generating system that allows us to extract mixed vertices from any given order of (anti)holomorphic vertices. Specifically, our findings can be summarized as follows:
\begin{itemize}
    \item The HS generating system proposed in \cite{Didenko:2022qga} is designed to accommodate only (anti)holomor-phic dynamics. However, we note that its central component -- the projector identity, which ensures consistency -- can be generalized to include mixed interactions (see also the discussion in \cite{Korybut:2025vdn}). This is reflected in the generating equations, which take the form of a Vasiliev-type system that we refer to as {\it irregular}. The term originates from a specific large algebra utilized in our approach, which cannot be routinely applied within the standard Vasiliev's framework.

    \item The consistency of the proposed equations is tightly constrained by a nonlinear algebraic condition. This condition stems from the functional classes of the holomorphic and antiholomorphic sectors, which determine the locality of the (anti)holomorphic interactions. Importantly, in the mixed sector, this condition also enforces the master field (0-form) controlling the HS vertices to have a purely chiral and antichiral form with respect to the auxiliary spinor variables  $z$ and $\bar z$.

    \item We introduce a perturbation theory that we refer to as {\it chiral} for the underlying generating connection. It develops in terms of {\it degree} $k\geq 0$ that measures entanglement between the holomorphic and antiholomorphic sectors. Specifically, for the leading order (LO) case, where $k=0$, we generate the complete (anti)holomorphic dynamics. In contrast, subleading mixed vertices appear for $k\geq 1$ as next-to-leading corrections $\text{N}^{k}\text{LO}$. We have demonstrated that our equations have solutions for, at least,  $k=0,1,2$.

    \item The constraints imposed by chiral perturbation theory lead to a series of dual representations for the HS vertices. The lower-order relations also apply to the Vasiliev equations, while the higher-order relations are expected to align our approach with Vasiliev's framework and facilitate the evaluation of the HS vertices.
    
\end{itemize}
The main focus of our investigation is to develop a comprehensive approach to tackling the HS locality problem. Our method generates HS vertices that are maximally local in the (anti)holomorphic sector, which imposes a strong constraint in the mixed sector, where locality issues arise. Specifically, this constraint (as will be detailed in \eqref{cons:BB}) is not supported by arbitrary field redefinitions; instead, it allows for only limited flexibility in how the interaction vertices are evaluated. We will address the explicit form of these vertices and analyze their locality properties in future work.

Our choice to use the holomorphic system from \cite{Didenko:2022qga} as a starting point, rather than the full Vasiliev equations, is based on the fact that the locality of (anti)holomorphic sectors has not yet been established for all orders in the latter. Additionally, the concise representation of all-order vertices in \cite{Didenko:2024zpd} shows great promise for completing the mixed sector. A subtle point here is the following. Initially, there is no guaranty that the holomorphic system presented in \cite{Didenko:2022qga} allows for a mixed-vertex completion. Indeed, unlike the Vasiliev case, these equations do not arise as a reduction of some  full-fledged interactions, which means they may differ\footnote{At the lowest order of interaction, the equivalence between the two systems was established (see \cite{Didenko:2024zpd}). Additionally, arguments supporting their equivalence at the next order can be found in \cite{Didenko:2022eso, Korybut:2025vdn}.} from Vasiliev's holomorphic limit. If this is indeed the case, it would suggest that mixed vertices based on \cite{Didenko:2022qga} do not exist, as it would otherwise imply the existence of two distinct theories based on the same HS algebra. The fact that chiral perturbation theory generates, at least, the first few mixed subleading terms at any holomorphic order provides additional evidence that the HS generating equations of \cite{Didenko:2022qga} should follow from the Vasiliev equations.

The paper is organized as follows: Sec. \ref{sec:HS-ver} revisits the standard unfolded approach to the problem of HS vertices. Section \ref{sec:eqs} addresses and classifies the generating equations. Section \ref{sec:irreg} examines the irregular form of these generating equations.
Section \ref{sec:mixed} applies the generating equations to describe mixed interactions. Section \ref{sec:chiral} introduces and develops chiral perturbation theory in several orders. Section \ref{sec:dual} proposes structure dualities.
Finally, we conclude in Sec. \ref{sec:conc}. Additionally, the paper includes four technical appendices for further reference.

\subsection{Conventions} 
\begin{itemize}
    \item {\bf Two-component spinors} 
    
    Our approach extensively exploits the two-component $sl(2, \mathbb{C})$ formalism. Any spinor $\Xi_{A}=(\xi_{\al}, \bar\xi_{\dal})$ splits into the two-component holomorphic $\xi_{\al}$ and antiholomorphic $\bar\xi_{\dal}$ parts, where $\al, \dal=1,2$ with the following sp(2) index convention:
\begin{subequations}\label{conv:sp(2)}
    \begin{align}
        &A^{\al}=\gep^{\al\gb}A_{\gb}\,,\qquad A_{\al}=A^{\gb}\gep_{\gb\al}\,,\qquad \gep_{\al\gb}=-\gep_{\gb\al}\,,\quad \gep^{\al\gb}=-\gep^{\gb\al}\,,\quad \gep_{\al\gb}\gep^{\gga\gb}=\gd_{\al}{}^{\gga}\,,\\
        &\bar A^{\dal}=\gep^{\dal\dgb}\bar A_{\dgb}\,,\qquad \bar A_{\dal}=\bar A^{\dgb}\gep_{\dgb\dal}\,,\qquad \gep_{\dal\dgb}=-\gep_{\dgb\dal}\,,\quad \gep^{\dal\dgb}=-\gep^{\dgb\dal}\,,\quad \gep_{\dal\dgb}\gep^{\dgga\dgb}=\gd_{\dal}{}^{\dgga}\,,
    \end{align}
\end{subequations}
where $\gep_{\al\gb}$ and $\gep_{\dal\dgb}$ are the canonical sp(2) forms
\begin{equation}
    \gep_{\al\gb}=\begin{pmatrix}
          0 & 1\\
          -1 & 0 
    \end{pmatrix}\,,\qquad \gep_{\dal\dgb}=\begin{pmatrix}
          0 & 1\\
          -1 & 0 
    \end{pmatrix}\,.
\end{equation}
To simplify the exposition, we will frequently use the following short-hand notation for two-component spinor contractions:
\begin{equation}
    \xi_{\al}\eta^{\al}:=\xi\eta=-\eta\xi\,,\qquad \bar\xi_{\dal}\bar\eta^{\dal}:=\bar\xi\bar\eta=-\bar\eta\bar\xi\,.
\end{equation}
\item {\bf Integrals}

We have two types of integrations in the paper. There are integrals over compact domains, which we manifestly indicate, e.g., 
\begin{equation}
    \int_{\Delta_\rho}d^3\rho\,f(\rho_1, \rho_2, \rho_3)\,,\qquad \Delta_{\rho}=\{\sum_{i=1}^3\rho_i=1\,,\quad 0\leq\rho_i\leq 1\}\,,
\end{equation}
and noncompact integrals over spinor variables. The uppercase integration variables, such as $U_A$, introduce integrals over $\mathbb{R}^4$, whereas the lowercase variables $u_{\al}$ are integrated over $\mathbb{R}^2$.  Generally, these integration variables will come in pairs, each associated with a specific measure that depends on the number of integrals involved:
\begin{subequations}
\begin{align}
    &\int_{U,V}f(U,V):=\frac{1}{(2\pi)^4}\int_{\mathbb{R}^8} d^4Ud^4V f(U,V)\,,\\ 
    &\int_{u,v} f(u,v):=\frac{1}{(2\pi)^2}\int_{\mathbb{R}^4}d^2 ud^2 v\,f(u,v)\,.
\end{align}    
\end{subequations}

\end{itemize}

\section{Higher-spin vertices}\label{sec:HS-ver}
The unfolded approach \cite{Vasiliev:1988sa} (see also \cite{Misuna:2022cma, Misuna:2024ccj, Misuna:2024dlx, Iazeolla:2025btr} for further development and applications) is known to be efficient for the HS nonlinear dynamics. To describe field evolution in four dimensions, one introduces the 1-form $\go=\go_{\mu}(y, \bar y|x)\dr x^\mu$ and the 0-form $C=C(y, \bar y|x)$. Apart from space-time dependence, the introduced fields depend on the auxiliary spinor generating variables 
\begin{equation}
    Y_{A}=(y_{\al}, \bar y_{\dal})\,,\qquad A=1,\dots, 4\,,\qquad \al=1,2\,,\qquad \dal=1,2\,.
\end{equation}
These encode an infinite set of fields contained in $\go$ and $C$ via the Taylor expansion
\begin{subequations}
    \begin{align}
    &\go:=\sum_{m,n=0}^{\infty}\frac{1}{m!n!}\go_{\al_1\dots\al_{m},\dal_{1}\dots\dal_{n}}(x)y^{\al_1}\dots y^{\al_m}\bar y^{\dal_1}\dots\bar y^{\dal_n}\,,\\ 
    &C:=\sum_{m,n=0}^{\infty}\frac{1}{m!n!}C_{\al_1\dots\al_{m},\dal_{1}\dots\dal_{n}}(x)y^{\al_1}\dots y^{\al_m}\bar y^{\dal_1}\dots\bar y^{\dal_n}\,.
  \end{align}  
\end{subequations}
Additionally, the commuting variables $Y$ serve to generate the associative Moyal algebra with respect to the star product  identified with the HS algebra
\begin{equation}\label{Moyal}
        f(y, \bar y)*g(y, \bar y)=\int\limits_{U, V}  f(y+u, \bar y+\bar u)g(y+v, \bar y+\bar v)e^{iu_{\al}v^{\al}+i\bar u_{\dal}\bar v^{\dal}}\,,
    \end{equation}
where the integration is performed along $U_A=(u_{\al}, \bar u_{\dal})\in\mathbb{R}^4$ and $V_A=(v_{\al}, \bar v_{\dal})\in\mathbb{R}^4$ with the $sp(2)$ index convention from \eqref{conv:sp(2)}. The measure in \eqref{Moyal} is defined in such a way that $1*1=1$. These give us the standard oscillator commutators 
\begin{equation}
    [y_{\al}, y_{\gb}]_*=2i\gep_{\al\gb}\,,\qquad [\bar y_{\dal}, \bar y_{\dgb}]_*=2i\gep_{\dal\dgb}\,,\qquad [y_{\al}, \bar y_{\dgb}]_*=0\,.
\end{equation}
One frequently uses the sp(4) form of the Moyal product \eqref{Moyal}
\begin{equation}
    f(Y)*g(Y)=\int_{U, V}f(Y+U)g(Y+V)e^{iU_AV^A}
\end{equation}
with a similar to \eqref{conv:sp(2)} index convention for the sp(4) canonical form 
\begin{equation}
    \gep_{AB}=\begin{pmatrix}
          \gep_{\al\gb} & 0\\
          0 & \gep_{\dal\dgb} 
    \end{pmatrix}\,.
\end{equation}
The unfolded equations describing HS dynamics acquire the following form: 
\begin{subequations}\label{unfld}
     \begin{align}
        &\dr_x\go=\sum_{n=0}^{\infty}\sum_{i+j=n}\eta^i\bar\eta^j\mathcal{V}_{i,j}(\go, \go, C^n)\,,\label{unfld:w}\\
        &\dr_x C=\sum_{n=0}^{\infty}\sum_{i+j=n}\eta^i\bar\eta^j\Upsilon_{i,j}(\go, C^{n+1})\,,\label{unfld:C}
    \end{align}
    \end{subequations}
where $\mathcal{V}_{i,j}(\go, \go, C^n)=\mathcal{V}_{i,j}(\go, \go, \underbrace{C,\dots, C}_n)$ are understood as a polylinear 2-form functionals of $n+2$ arguments. Similarly, $\Upsilon_{i,j}(\go, C^{n+1})=\Upsilon_{i,j}(\go, \underbrace{C,\dots,C}_{n+1})$ are polylinear 1-forms. The right-hand sides of \eqref{unfld} are referred to as HS vertices. There is an arbitrary  complex parameter $\eta=e^{i\theta}$, $\bar\eta=e^{-i\theta}$ that breaks parity unless $\eta=1$ or $\eta=i$. The form of the HS vertices is ambiguously fixed by the integrability condition $\dr_x^2=0$. The ambiguity is fully described by an arbitrary field redefinition
\begin{equation}
    \go\to\go+F_{\go}(\go, C\dots C)\,,\qquad C\to C+F_C(C\dots C)\,.
\end{equation}
This makes the problem of identifying vertices $\mathcal{V}$ and $\Upsilon$ cohomological and particularly complex when the locality requirement is additionally imposed. To set up the cohomological reconstruction, one has to specify the initial solution corresponding to the lowest vertices with $n=0$. These vertices should align with the linearized HS dynamics. Their form appears to be defined by the global HS algebra \eqref{Moyal} and its automorphism\footnote{Let us note that the choice of the initial vertex, for example, in the form $\Upsilon_{0,0}=[C,\go]_*$ is perfectly consistent with $\dr_x^2=0$. However, this choice describes a collection of Killing tensors packed in $C$ rather than a dynamical system.} $\pi$ or $\bar\pi$ (see, e.g.,  \cite{Vasiliev:1999ba})
\begin{subequations}\label{initio}
    \begin{align}
        &\mathcal{V}_{0,0}=-\go*\go\,,\label{V0}\\
        &\Upsilon_{0,0}=-\go*C+C*\pi(\go)\,,\label{ups0}
    \end{align}
\end{subequations}
where 
\begin{equation}\label{pi}
    \pi f(y, \bar y)=f(-y, \bar y)\,,\qquad \bar\pi f(y, \bar y)=f(y, -\bar y)\,.
\end{equation}
The system \eqref{unfld} is supplemented by the reality conditions 
\begin{equation}\label{real:fields}
    \go^{\dagger}=-\go\,,\qquad C^\dagger=\pi (C)\,,
\end{equation}
where the involution is defined as follows:
\begin{equation}
    y^{\dagger}_{\al}=\bar{y}_{\dal} \,,\qquad \eta^{\dagger}=\bar\eta\,,\qquad (f*g)^{\dagger}=g^{\dagger}*f^{\dagger}\,.
\end{equation}
In order to be consistent with the reality conditions  \eqref{real:fields} beyond the free level, the system \eqref{unfld} has to be bosonic\footnote{To include interaction with fermions, supersymmetry is needed, which leads to the doubling of all fields; see \cite{Vasiliev:1999ba}.} which implies
\begin{equation}\label{bosonic}
    \go(-y, \bar y)=\go(y, -\bar y)\,,\qquad C(-y, \bar y)=C(y, -\bar y)\,.
\end{equation}
Since supersymmetry does not play a significant role in our analysis, we will focus on the purely bosonic version of HS interactions in what follows.

Reconstructing the vertices $\mathcal{V}$ and $\Upsilon$ in Eqs.  \eqref{unfld} is technically challenging. This task essentially boils down to analyzing the integrability conditions derived from $\dr_x^2=0$ with the initial data \eqref{initio}. The resulting consistency conditions take the form of integral equations, which complicate the problem significantly. In \cite{Vasiliev:1992av}, Vasiliev introduced a generating system that reformulates the problem using partial differential equations in an auxiliary space. Solving this system provides the desired vertices, while the inherent freedom in the solution space reflects gauge ambiguity and field redefinitions. In the following, we will provide the generating approaches inspired by the Vasiliev system, as discussed in \cite{Didenko:2022qga}, and propose further development, particularly concerning the issue of locality. For the original form of the Vasiliev system, we refer to \cite{Vasiliev:1992av}.

\section{Generating equations}\label{sec:eqs}
Our starting point is to rewrite Eq. \eqref{unfld:w} as a zero-curvature condition in a larger space $(Z, Y)$ extended by variables $Z_A=(z_{\al}, \bar z_{\dal})$ 
\begin{equation}\label{W:zero}
    \dr_x W+W\filledstar W=0\,.
\end{equation}
Here, 
\begin{equation}\label{W:embd}
    W=\go(Y|x)+W_1[\go, C](Z;Y)+W_2[\go, C,C](Z; Y)+\dots\,,
\end{equation}
where $W_i$ are yet unknown $Z$ and $Y$ dependent functionals of $\go$ and $C$, and the star product $\filledstar$ is an extension of the Moyal product  \eqref{Moyal} on the space of $Z$ and $Y$ variables. At this stage, we need the extended star product $\filledstar$ to be associative, and it should reduce to the Moyal product \eqref{Moyal} for $Z$-independent functions. 
Substituting $W$ into \eqref{W:zero}, we arrive at
\begin{equation}\label{w:from W}
    \dr_x\go+\go*\go=-\sum_{i>0}\dr_xW_i-\sum_{i+j>0}W_i\filledstar W_j\,,\qquad W_0:=\go\,.
\end{equation}
Equation \eqref{w:from W} is clearly consistent because Eq. \eqref{W:zero} is, and it properly reproduces the initial vertex \eqref{V0}. However, while its left-hand side is manifestly $Z$ independent, the right side is generally not. This means that $Z$ dependence on \eqref{W:zero} should be of a specific form to provide $Z$ independent Eq. \eqref{w:from W}.\footnote{The simplest option is to take $W$ independent of $Z$. This option, however, corresponds to a field redefinition of $\go$, thereby resulting in the absence of dynamics.} It is not difficult to figure out the proper evolution of the field $W$ to guaranty that the right-hand side of \eqref{w:from W} depends on $Y$ only. To this end, let us introduce the covariant differential 
\begin{equation}
    D=\dr_x+[W, \bullet]_{\filledstar}\,,\qquad D^2=0\,.
\end{equation}
Imposing the following evolution: 
\begin{equation}\label{eq: dzW}
    \dr_Z W=-D\Lambda\quad\Rightarrow\quad \dr_Z W+\{W, \Lambda\}_{\filledstar}+\dr_x\Lambda=0\,,
\end{equation}
where $\dr_Z=\dr Z^{A}\frac{\partial}{\partial Z^A}$, while the 1-form 
\begin{equation}\label{intro:lambda}
\Lambda=\dr Z^A\Lambda_A=\dr Z^A(\Lambda_{1\,A}[C](Z;Y)+\Lambda_{2\,A}[C,C](Z;Y)+\dots)
\end{equation}
is not yet specified. It is not hard to verify that the right-hand side of \eqref{w:from W} is now $Z$ independent. This can be seen by hitting $\dr_Z$ on the left-hand side of Eq. \eqref{W:zero}, and using \eqref{eq: dzW}, the result boils down to
\begin{equation}\label{dzVer}
    \dr_Z(\dr_x W+W\filledstar W)=-[\Lambda, \dr_x W+W\filledstar W]_{\filledstar}\,.
\end{equation}
Notice that $\Lambda$ is, at least, linear in $C$ by its definition \eqref{intro:lambda}. This implies that at a particular order $C^n$, the vertex $\mathcal{V}(\go, \go, C^n)$, which is expressed from $(\dr_x W+W*W)|_{C^n}$, depends on $Z$, at least, as $C^{n+1}$, as follows from \eqref{dzVer}. This means it is $Z$ independent. For example, consider a linear in $C$ vertex. Starting with some HS vacuum $\go_0(Y|x)$,
\begin{equation}
    \dr_x\go_0+\go_0*\go_0=0
\end{equation}
and discarding $O(C^2)$ terms, we have
\begin{equation}
    \dr_Z\mathcal{V}(\go_0, \go_0, C):=\dr_Z(-\dr_x W_1-\{\go_0, W_1\})=-[\Lambda, \dr_x\go_0+\go_0*\go_0]_*\equiv 0\,.
\end{equation} 
Let us note that if $\Lambda$ satisfying \eqref{eq: dzW} exists, then no further constraints follow from inspecting $\dr_x^2=0$, as Eqs.   \eqref{W:zero} and \eqref{eq: dzW} become consistent. The role of the auxiliary connection $\Lambda$ is twofold. First, it guaranties $Z$ independence of vertices $\mathcal{V}$ via \eqref{eq: dzW}. Second, it implicitly contains information on vertices $\Upsilon$ from \eqref{unfld:C} because 
\begin{equation}
    \dr_x\Lambda[C]=-\frac{\delta\Lambda[C]}{\delta C}\dr_x C\,,
\end{equation}
meaning that $\Lambda$ depends on $x$ only through the field $C$, the differential of which is expressed as \eqref{unfld:C}. For example, in order to reproduce the initial vertex \eqref{ups0}, the field $\Lambda$ should satisfy the following condition: 
\begin{equation}\label{low order:lambda}
    \dr_z\Lambda=i\eta C\filledstar\gga+i\bar\eta C\filledstar\bar\gga+O(C^2)\,,
\end{equation}
where $\gga$ and $\bar\gga$ are the $\dr Z$ two-forms
\begin{equation}\label{2f:klein}
    \gga=\frac{1}{2}\dr z_{\al}\wedge\dr z^{\al}\kappa(z, y)\,,\qquad \bar\gga=\frac{1}{2}\dr \bar z_{\dal}\wedge\dr \bar z^{\dal}\bar{\kappa}(\bar z, \bar y)
\end{equation}
with the functions $\kappa$ and $\bar\kappa$ (if they exist) that should, in particular, realize the automorphisms $\pi$ and $\bar\pi$ from \eqref{pi} as\footnote{The action of the automorphism $\pi$ on the variable $Z$ is not known {\it a priori}. The definition provided here rests on the particular realization of the large $(Z,Y)$ star-product algebra that will be discussed below.}
\begin{equation}\label{def:klein}
    \kappa\filledstar f(z, y)=f(-z, -y)\filledstar\kappa\,,\qquad \bar\kappa\filledstar f(\bar z, \bar y)=f(-\bar z, -\bar y)\filledstar\bar\kappa\,.
\end{equation}
Now, to ensure that \eqref{low order:lambda} does lead to the vertex specified in \eqref{ups0}, we first need to check that this choice is consistent with $\dr_Z^2=0$ to the order $O(C)$. To that end, we apply $\dr_Z$ to \eqref{eq: dzW} with $W=\go+O(C)$ to obtain
\begin{equation}\label{low order:dZ}
    [\go, \eta\, C\filledstar\gga]_{\filledstar}+\eta\,\dr_x C\filledstar\gga=0\,,\qquad [\go, \bar\eta\, C\filledstar\bar\gga]_{\filledstar}+\bar\eta\,\dr_x C\filledstar\bar\gga=0\,. 
\end{equation}
Taking into account the bosonic condition \eqref{bosonic}, along with the assumption that the kernel of $\kappa$ and $\bar\kappa$ is empty ($f\filledstar\kappa=0\,\Rightarrow\,f=0$), we arrive at \eqref{ups0}:
\begin{equation}\label{low order:C}
    \dr_x C+\go*C-C*\pi (\go)+O(C^2)=0\,,
\end{equation}
where we also used that star product $\filledstar$ acts as the Moyal \eqref{Moyal} in the last equation. 
Notice that the $Z$ dependence in \eqref{low order:dZ} is resolved and, therefore, the choice of $\Lambda$ in \eqref{low order:lambda} is consistent, provided Eq. \eqref{low order:C} is consistent, which is automatically the case because the generating  Eq. \eqref{eq: dzW} is consistent with respect to $\dr_x^2=0$. Further details depend on the particular realization of the large star product and the corresponding properties of elements $\kappa$, $\bar\kappa$ defined in \eqref{def:klein}. 

\subsection{Star products}
As we previously stressed, the extended star product must be associative and reduce to \eqref{Moyal} for products of $Z$-independent functions. A one-parameter class of such star products was proposed in  \cite{Didenko:2019xzz, Didenko:2022qga} (see also \cite{DeFilippi:2021xon} for the earlier inspiring discussion of the related problem) in the analysis of the HS locality issue. This family looks as follows:
\begin{align}\label{beta: star}
    &f(Z, Y)\star_{\gb}g(Z,Y):=\\
    &:=\int f(Z+U', Y+U)g(Z-(1-\gb)V-V', Y+V+(1-\gb)V')e^{iU_{A}V^A+iU'_{A}V'^{A}}\,,\nn
\end{align}
where $\gb$ is an arbitrary real number. So-defined product is (i) associative and (ii) acts exactly as in \eqref{Moyal} on $Z$-independent functions. It provides us with the following defining relations for the generating variables $Y$ and $Z$:  
\begin{subequations}\label{def:beta}
    \begin{align}
        &Y_A\star_{\gb}=Y_A+i\frac{\p}{\p Y^A}-i(1-\gb)\frac{\p}{\p Z^A}\,,\qquad  \star_{\gb}Y_A=Y_A-i\frac{\p}{\p Y^A}-i(1-\gb)\frac{\p}{\p Z^A}\,,\\
        &Z_A\star_{\gb}=Z_A-i\frac{\p}{\p Z^A}+i(1-\gb)\frac{\p}{\p Y^A}\,,\qquad  \star_{\gb}Z_A=Z_A+i\frac{\p}{\p Z^A}+i(1-\gb)\frac{\p}{\p Y^A}\,.
    \end{align}
\end{subequations}
Notice that $Y$ and $Z$ star-commute with each other for any $\gb$. The elements $\kappa$ and $\bar\kappa$ that extend the Moyal algebra automorphisms $\pi$ and $\bar\pi$ to the large algebra of $Z$ and $Y$ make these automorphisms inner. They can be straightforwardly found using the definition \eqref{def:klein}. The final result is (see \cite{Didenko:2022qga})
\begin{equation}\label{beta:klein}
    \kappa=\frac{1}{(1-\gb)^2}\exp\left(\frac{i}{1-\gb}z_{\al}y^{\al}\right)\,,\qquad  \bar\kappa=\frac{1}{(1-\gb)^2}\exp\left(\frac{i}{1-\gb}\bar z_{\dal}\bar y^{\dal}\right)\,.
\end{equation}
These, also known as Klein operators \cite{Vasiliev:1992av}, satisfy 
\begin{equation}
    \kappa\star_\gb\kappa=\bar\kappa\star_\gb\bar\kappa=1\,,\qquad \gb\neq 1\,.
\end{equation}
The parameter $\gb$ can be thought of as a parameter that interpolates between various orderings of the operators associated with the symbols $Y$ and $Z$. In particular, it distinguishes several important cases:
\begin{itemize}
    \item {\bf Vasiliev ordering: $\gb=0$}
    
    This case corresponds to the original large star product introduced by Vasiliev in \cite{Vasiliev:1992av}, which we denote here as
    \begin{equation}
        \star:=\star_{\gb=0}\,.
    \end{equation}
    For $\gb=0$, a part of the integration in Eq. \eqref{beta: star} can be completed, as the final result reads
    \begin{equation}\label{star:Vasilev}
        f(Z,Y)\star g(Z,Y)=\int_{U,V} f(Z+U, Y+U)g(Z-V, Y+V)e^{iU_{A}V^{A}}\,.
    \end{equation}
    The Vasiliev case introduces the normal ordering of $Y\pm Z$ operators. It also admits the supertrace operation 
    \begin{equation}\label{str}
        \mathrm{str}\,f(Z,Y):=\int_{U,V} f(U, V)e^{-iU_A V^A}\,,
    \end{equation}
    which satisfies
    \begin{equation}\label{str:prop}
        \mathrm{str}\,(f\star g)=\mathrm{str}\,(g\star\boldsymbol{\pi}\bar{\boldsymbol{\pi}}\,f)=\mathrm{str}\,(\boldsymbol{\pi}\bar{\boldsymbol{\pi}}\,g\star f)\,,\qquad \boldsymbol{\pi}\bar{\boldsymbol{\pi}} f(Z, Y):=f(-Z, -Y)\,,
    \end{equation}
    where 
    \begin{equation}\label{pi:zy}
        \boldsymbol{\pi} f(z,\bar z; y, \bar y)=f(-z, \bar z; -y, \bar y)\,,\qquad \bar{\boldsymbol{\pi}} f(z,\bar z; y, \bar y)=f(z, -\bar z; y, -\bar y)\,.
    \end{equation}
    \item {\bf Weyl ordering: $\gb=1$}
    
    This case disentangles the operators $Y$ and $Z$ and provides symmetric orderings for $Y$'s and $Z$'s separately. A peculiar feature of this ordering is that Klein operators specified in Eq. \eqref{beta:klein} no longer exist as analytic functions; rather, they appear as distributions,
    \begin{equation}
        \kappa\sim\gd(y)\gd(z)\,,\qquad \bar\kappa\sim\gd(\bar y)\gd(\bar z)\,.
    \end{equation}
    \item {\bf anti-Vasiliev ordering: $\gb=2$}
    
    This case is similar to the standard Vasiliev star product \eqref{star:Vasilev}, the difference being the interchange of the normal-ordered operators $Y\pm Z$.  
\end{itemize}
The map between various orderings is realized in terms of the reordering operator. For example, the $\beta$-ordered product \eqref{beta: star} can be obtained from the Vasiliev one \eqref{star:Vasilev} as
\begin{equation}
    f\star_{\gb} g=O_{\gb}\left(O_{\gb}^{-1}f\star O_{\gb}^{-1}g\right)\,,
\end{equation}
where $O_{\gb}$ is the reordering operator that maps symbols from the Vasiliev ordering \eqref{star:Vasilev} into \eqref{beta: star}. Its manifest form found in \cite{Didenko:2019xzz} reads
\begin{equation}
    O_{\gb}f(Z,Y)=\int_{U,V} f(Z+V, Y+\gb U)e^{iU_A V^A}\,,\qquad O^{-1}_{\gb}:=O_{-\gb}\,.
\end{equation}
The choice of ordering is important in the context of the relevant functional class, one operates with, in a given star-product algebra. For example, the original choice of $\gb=0$,  \cite{Vasiliev:1992av} deals with regular Klein operators \eqref{beta:klein}, unlike the case with $\gb=1$; see also \cite{DeFilippi:2021xon}.

An important remark is the Klein two-forms \eqref{2f:klein} defined using Eq. \eqref{beta:klein} stay invariant under reordering, supplemented by a rescaling $Z\to (1-\gb)Z$, 
\begin{equation}\label{inv}
    \mathcal{O}_{\gb}\gga=\gga\,,\qquad \mathcal{O}_{\gb}\bar\gga=\bar\gga\,,
\end{equation}
where 
\begin{equation}\label{reorder}
    \mathcal{O}_{\gb}:=O_{\gb}\Big|_{Z\to(1-\gb)Z}\,.
\end{equation}
The invariance \eqref{inv} holds, including the limit $\gb\to\infty$, which plays a significant role in the analysis of HS locality; see \cite{Didenko:2019xzz}. The reordering with a rescaling allows us to define the contraction $\gb\to\infty$ at the level of the star product \eqref{beta: star}.

\subsubsection{Limiting star product: $\gb\to\infty$} 
Star products in \eqref{beta: star} are shown to be equivalent up to a reordering for finite $\beta$. Interestingly, there is an infinite point at $\gb=\infty$, where the associative product still exists and results in the local holomorphic HS interaction \cite{Didenko:2019xzz, Didenko:2022qga}. Specifically, the rescaling $Z\to (1-\gb)Z$ followed by the limit $\gb\to\infty$ in \eqref{def:beta} results in the following star action:
\begin{subequations}\label{def:infty}
    \begin{align}
        &Y_A*=Y_A+i\frac{\p}{\p Y^A}-i\frac{\p}{\p Z^A}\,,\qquad  *Y_A=Y_A-i\frac{\p}{\p Y^A}-i\frac{\p}{\p Z^A}\,,\\
        &Z_A*=Z_A+i\frac{\p}{\p Y^A}\,,\qquad  *Z_A=Z_A+i\frac{\p}{\p Y^A}\,.
    \end{align}
\end{subequations}
At the level of the definition \eqref{beta: star}, this limit can be formally defined as
\begin{equation}
    f*g:=\lim_{\gb\to\infty}(f\star_{\gb}g)\Big|_{Z\to(1-\gb)Z}\,.
\end{equation}
The result respecting associativity takes the form
\begin{equation}\label{limst}
(f*g)(Z, Y)= \int f(Z+U', Y+U)
g(Z-V,Y+V+V')
\exp({iU_{A}V^{A}+iU'_{A}V'^{A}})\,.
\end{equation}
Notice that the variables $Z$ commute with each other in the limiting case, unlike \eqref{beta: star}. Similar to the Vasiliev case, the limiting star product admits the supertrace operation \eqref{str} satisfying \eqref{str:prop} with $*$ in place of $\star$. Specifically, we will need
\begin{equation}\label{lim:str}
    \text{str} f(Z; Y)=\int_{U,V} f(U;V)e^{-iUV}\,,
\end{equation}
and we also extend \eqref{pi} for $z$-dependent functions
\begin{equation}\label{pi:def}
    \pi f(z,\bar z; y, \bar y)=f(z, \bar z; -y, \bar y)\,,\qquad \bar\pi f(z,\bar z; y, \bar y)=f(z, \bar z; y, -\bar y)\,.
\end{equation}

As we already mentioned, the 2-forms $\gga$ and $\bar\gga$ exist in the limiting case and have the standard form
\begin{equation}\label{gamma}
    \gga=\frac{1}{2}\dr z_{\al}\wedge\dr z^{\al}\exp{izy}\,,\qquad \bar\gga=\frac{1}{2}\dr \bar z_{\dal}\wedge\dr \bar z^{\dal}\exp{i\bar z\bar y}\,.
\end{equation}
However, unlike the Vasiliev case,
\begin{equation}
    e^{izy}*e^{izy}=e^{i\bar z\bar y}*e^{i\bar z\bar y}=\infty\,.
\end{equation}
The important property of \eqref{gamma} is 
\begin{equation}\label{Klein:prop}
    \gga*\phi(y, \bar y)=\pi\phi(y, \bar y)*\gga\,,\qquad \bar\gamma*\phi(y, \bar y)=\bar\pi\phi(y, \bar y)*\bar\gga\,.
\end{equation}
Just as in the case of finite $\beta$, some products with Eq. \eqref{limst} cannot be distinguished from those evaluated with Vasiliev's star product \eqref{star:Vasilev}. Specifically, 
\begin{equation}\label{prod:equiv}
    f(Y)*g(Z,Y)=f(Y)\star g(Z,Y)\,,\qquad g(Z,Y)*f(Y)=g(Z,Y)\star f(Y)\,.
\end{equation}

The further construction of the HS generating systems significantly depends on the type of large algebra and its product. We distinguish between the two cases: {\it regular} or Vasiliev-like, and {\it irregular}. An example of the latter is the generating system for the holomorphic HS sector from \cite{Didenko:2022qga} and the off-shell HS equations in arbitrary dimensions \cite{Didenko:2023vna}.   

\subsection{Regular case} The generating system that should produce the HS vertices \eqref{unfld} includes the zero-curvature equation \eqref{W:zero} and its $Z$-independence condition \eqref{eq: dzW} written using one or another large star products. The form of these equations assumes that the following star products generally exist:
\begin{equation}\label{noinfty}
    (W*W\,,\quad W*\Lambda\,,\quad \Lambda*W)\neq\infty\,.
\end{equation}
If, in addition to \eqref{noinfty}, 
\begin{equation}\label{case:reg}
\Lambda*\Lambda\neq\infty\,,
\end{equation}
in general, then we call the case {\it regular}. This is the situation with the Vasiliev generating equations. Assuming regularity, we can derive Vasiliev's system in a few steps following \cite{Didenko:2022qga}. We begin with Eqs. \eqref{W:zero} and \eqref{eq: dzW}, where the star product is chosen to be from \eqref{star:Vasilev}
\begin{subequations}\label{2eqs:vas},
    \begin{align}
        &\dr_x W+W\star W=0\,,\\
        &\dr_Z W+\{W, \Lambda\}_\star+\dr_x\Lambda=0\,.\label{dzW:vas}
    \end{align}
\end{subequations}
These are the $\dr_x$-consistent equations, provided $\Lambda$, which resolves the $Z$ dependence, exists. Field $\Lambda$ is constrained by $\dr_Z^2=0$. Applying $\dr_Z$ to \eqref{dzW:vas} and using it again, we find
\begin{equation}\label{L:cons}
    D(\dr_Z\Lambda+\Lambda\star\Lambda)=0\,,\qquad D=\dr_x+[W,\bullet]_\star\,.
\end{equation}
Under a few mild assumptions, including Lorentz symmetry, reality conditions, and the lower order constraint \eqref{low order:lambda}, Eq. \eqref{L:cons} is resolved by 
\begin{equation}\label{L:eq}
    \dr_Z\Lambda+\Lambda\star\Lambda=i\eta B\star\gga+i\bar\eta B\star\bar\gga\,,
\end{equation}
where 
\begin{equation}
B(Z,Y|x)=C+B_2[C,C](Z,Y)+\dots    
\end{equation}
is a new field for which we have, from Eq. \eqref{L:cons},
\begin{equation}\label{dB:vas}
    \dr_x B+W\star B-B\star\boldsymbol{\pi} (W)=0\,,
\end{equation}
where we recall $\boldsymbol{\pi}$ is defined in \eqref{pi:zy}. To the lowest order, one recovers \eqref{low order:C}. Equation \eqref{dB:vas} is supposed to recover the vertices \eqref{unfld:C} provided it is $Z$ independent. To ensure its $Z$ independence, one uses the same strategy as with Eq. \eqref{W:zero}.  Applying  $\dr_Z$ to the left-hand side of \eqref{dB:vas} and using  \eqref{dzW:vas}, one arrives at 
\begin{equation}
    \dr_Z(\dr_x B+W\star B-B\star\boldsymbol{\pi}(W))=-D(\dr_Z B+\Lambda\star B-B\star\boldsymbol{\pi}(\Lambda))\,,
\end{equation}
which should be equal to zero. This requirement can be resolved by imposing 
\begin{equation}\label{dZ:vas}
    \dr_Z B+\Lambda\star B-B\star\boldsymbol{\pi}(\Lambda)=0\,.
\end{equation}
A set of equations \eqref{2eqs:vas}, \eqref{L:eq}, \eqref{dB:vas}, and \eqref{dZ:vas} is fully consistent, generating the HS vertices \eqref{unfld} via
\begin{subequations}
    \begin{align}
         &\dr_x\go+\go\star\go=-\sum_{i>0}\dr_xW_i-\sum_{i+j>0}W_i\star W_j\,,\\
         &\dr_x C+\go\star C-C\star \pi(\go)=-\sum_{i>1}\dr_x B_i-\sum_{i+j>0}(W_i\star B_{j+1}-B_{j+1}\star \boldsymbol{\pi}(W_i))\,,
    \end{align}
\end{subequations}
where the $Z$-evolution of the fields $W_i$ and $B_i$ is determined from Eqs. \eqref{dzW:vas} and \eqref{dZ:vas}, also using \eqref{L:eq} with the initial data
\begin{equation}
    W_0:=\go\,,\qquad B_1:=C\,.
\end{equation}
A set of five equations \eqref{2eqs:vas}, \eqref{L:eq}, \eqref{dB:vas}, and \eqref{dZ:vas} is called the Vasiliev system\footnote{The original form of Vasiliev's equations differs from the one presented here by the following field redefinition $\Lambda=S+\frac{1}{2i}\dr Z^AZ_{A}$.} for bosonic HS interactions:
\begin{subequations}\label{vas}
    \begin{align}
       &\dr_x W+W\star W=0\,,\label{vas1}\\
        &\dr_x B+W\star B-B\star\boldsymbol{\pi}(W)=0\,,\label{vas2}\\
        &\dr_Z\Lambda+\Lambda\star\Lambda=i\eta B\star\gga+i\bar\eta B\star\bar\gga\,,\label{vas3}\\
        &\dr_Z B+\Lambda\star B-B\star\boldsymbol{\pi}(\Lambda)=0\,,\label{vas4}\\
        &\dr_Z W+\{W, \Lambda\}_\star+\dr_x\Lambda=0\,.\label{vas5}  
    \end{align}
\end{subequations}
The great advantage of this system is that it has no obstruction and allows for a reconstruction of the HS vertices order-by-order modulo field redefinition. The latter freedom emerges as an ambiguity in solving for the fields $W$ and $B$ of \eqref{dzW:vas} and \eqref{dZ:vas}. However, it comes with certain issues in practical use, which include: 
\begin{itemize}
    \item {\bf Locality issue} 
    
    While system \eqref{vas} can reproduce HS vertices to any order, the locality constraint complicates the process of defining a clear method for fixing field freedom when solving equations \eqref{vas2}, \eqref{vas3}, and \eqref{vas4}. This makes it challenging to derive either local or minimally nonlocal HS vertices. At few orders, the local HS vertices have been obtained in papers \cite{Vasiliev:2016xui, Didenko:2018fgx, Didenko:2019xzz, Gelfond:2021two}, but the full solution is not yet available; see \cite{Vasiliev:2023yzx} for recent progress.

    \item {\bf Regularity issue} 
    
    Although we refer to the Vasiliev approach as regular, there are instances where this regularity breaks down. In particular, while the product $\Lambda\star\Lambda$ is expected to exist, it may not hold true for certain HS data represented by the function $C(Y)$. For example, if
    \begin{equation}\label{div:proj}
    C(Y)=\exp{q_{\al\dal}y^{\al}\bar y^{\dal}}\,,\quad\text{where}\quad q_{\al\dgb}q^{\gb\dgb}=\gd_{\al}{}^{\gb}\,,\quad q_{\al\dgb}q^{\al\dal}=\gd_{\dgb}{}^{\dal}\,,
\end{equation}
then the product $\Lambda\star\Lambda$ becomes ill-defined\footnote{It is important to note that the issue of regularity may be addressed within the differential homotopy approach of \cite{Vasiliev:2023yzx} for Vasiliev's system, where the generating equations are treated modulo the so-called weak terms.} at second order. This, in turn, implies that the field $B$ is also ill-defined to this order, as follows from \eqref{vas3}. The choice of \eqref{div:proj} is special in that $C$ forms a projector $C*C\sim C$. The respective first order correction in $\Lambda$,
\begin{equation}
    \Lambda^{(1)}=i\eta\,\dr z^{\al}z_{\al}\int_{0}^{1}d\tau\tau C(-\tau z,\bar y)e^{i\tau\,zy}=i\eta\,\dr z^{\al}z_{\al}\int_{0}^{1}d\tau\tau e^{-i\tau\,q_{\al\dal}z^{\al}\bar y^{\dal}+i\tau\,zy}\,,
\end{equation}
develops divergencies in interactions via $\Lambda^{(1)}\star\Lambda^{(1)}$. However, the divergences that arise are artificial and do not manifest in the final vertices. For instance, in \cite{Didenko:2021vdb}, a second-order analysis using data of the form \eqref{div:proj} has been carried out at the level of HS vertices. Nevertheless, the issue of regularity may complicate the analysis of the exact solutions obtained from the specified initial data. For example, it remains unclear how to verify that the recently found exact solution of the self-dual theory, presented in \cite{Didenko:2025xca}, satisfies Vasiliev's equations.  
\end{itemize}

Whether the generating equations admit the regular form depends on the large star product. In addition to the original Vasiliev case, which uses the star product \eqref{star:Vasilev}, regularity is also possible for the one-parameter star products \eqref{beta: star} when $\gb\neq 1$ and $\gb\neq\infty$. This means that we can replace the Vasiliev product $\star$ with $\star_{\gb}$ in all equations \eqref{vas}, and substitute the exponentials in \eqref{gamma} with those in \eqref{beta:klein}. This leads to a consistent system that generates \eqref{unfld}. 

The case where $\gb=\infty$, corresponding to the large algebra \eqref{limst}, is especially interesting.  As shown in \cite{Didenko:2019xzz}, the locality of HS vertices in the Vasiliev approach can be effectively reached by contracting the large algebra \eqref{beta: star} down to \eqref{limst}, at least, for the first few orders. However, the equations in \eqref{vas} are not applicable to the star product defined in \eqref{limst}; see \cite{Didenko:2022qga} for more details. In particular, neither $B*\gga$,  $B*\bar\gga$, nor $\Lambda*\Lambda$ is defined in this case in perturbations for {\it any} initial data in $C$. Despite this limitation, there is a way to define the generating equations for large star products that do not meet the regularity condition \eqref{case:reg}. In \cite{Didenko:2022qga}, such a system was proposed to describe the holomorphic sector of HS theory. Its advantage lies in the clear selection of field variables that accurately reproduce all-order local HS vertices.

\section{Irregular generating systems}\label{sec:irreg}
We refer to the large star product as {\it irregular} if \eqref{noinfty} holds, but
\begin{equation}\label{case:irreg}
    \Lambda*\Lambda\quad\text{-- ill-defined}
\end{equation}
for general  $C(Y)$. The corresponding {\it irregular} generating system is built upon this irregular star product. This situation contrasts with the regular case, where $\Lambda\star\Lambda$ generally exists, provided that the initial data, represented by $C(Y)$, are consistent with regularity (see the regularity issue above). 

Let us illustrate how the irregularity \eqref{case:irreg} appears in practice. Choosing the large star product in the form \eqref{limst}, to the lowest order in $C$, we have \eqref{low order:lambda}, where the Klein 2-forms are defined in \eqref{gamma}. The solution for $\Lambda$ can be chosen in the standard form 
\begin{equation}
    \Lambda=\dr z^{\al}\Lambda_{\al}+\dr\bar z^{\dal}\bar\Lambda_{\dal}\,,
\end{equation}
where
\begin{equation}
    \Lambda_{\al}=i\eta\,z_{\al}\int_{0}^{1}d\tau\tau C(-\tau z, \bar y)e^{i\tau zy}\,,\qquad \bar\Lambda_{\dal}=i\bar\eta\, \bar z_{\dal}\int_{0}^{1}d\tau\tau C(y, -\tau \bar z)e^{i\tau \bar z\bar y}\,.
\end{equation}
A straightforward calculation using \eqref{limst} results in, e.g., 
\begin{align}\label{int:div}
    &\Lambda_{\al}*\Lambda_{\gb}=-\eta^2\int_{[0,1]^2}d\tau_1 d\tau_2\frac{\gs(\tau_1, \tau_2)\gs(\tau_2, \tau_1)}{(1-\tau_1\tau_2)^2}z_{\al}z_{\gb}\exp i\left([\gs(\tau_1, \tau_2)+\gs(\tau_2, \tau_1)]zy\right)\times\\
    &\times C(-\gs(\tau_1, \tau_2)z, \bar y)\bar{*} C(-\gs(\tau_2, \tau_1)z, \bar y)\,,\nn
\end{align}
where 
\begin{equation}
    \gs(\tau_1, \tau_2)=\frac{\tau_1(1-\tau_2)}{1-\tau_1\tau_2}\in [0,1]\,,\qquad 0\leq\tau_{1,2}\leq 1\,,
\end{equation}
while the product $\bar *$ stands for the  Moyal product of the variables $\bar y$ in \eqref{Moyal}. The integral \eqref{int:div} diverges at $\tau_1=\tau_2=1$. In particular, the way this product arises in expressions like \eqref{vas3} is via 
\begin{equation}
    \Lambda_{\al}*\Lambda^{\al}=0\cdot\infty\,,
\end{equation}
where $0$ appears as $z_{\al}z^{\al}\equiv 0$. Notice that the divergence in \eqref{int:div} occurs for any analytic function $C(y, \bar y)$, leaving no option to consider the irregular case as regular for the specified data $C$. Another comment is, while $\Lambda_{\al}*\Lambda_{\gb}$ is ill-defined and similarly $\bar\Lambda_{\dal}*\bar\Lambda_{\dgb}$, the cross products, e.g., $\Lambda_{\al}*\bar\Lambda_{\dal}$, are perfectly fine. This will have important implications for the analysis of the mixed sector of HS interactions.

With the introduction of the irregular star product, we face the challenge of generating \eqref{unfld} using it. To address this, we embed the HS interaction within the large 0-curvature condition and require it to be independent of $Z$. This brings us back to the system defined by \eqref{W:zero} and \eqref{eq: dzW}. Both equations are consistent with the conditions $\dr_x^2=0$ and $\{\dr_x, \dr_Z\}=0$. Consequently, we only need to specify the field $\Lambda[C]$, which must also satisfy the condition $\dr_Z^2=0$ derived from \eqref{eq: dzW}:
\begin{equation}\label{L:main}
    \dr_Z\{W,\Lambda\}_*-\dr_x\dr_Z\Lambda=0\,.
\end{equation}

In the regular case, applying the Leibniz rule along with \eqref{eq: dzW} leads to the condition \eqref{L:cons}. However, in the irregular case, we cannot verify Eq. \eqref{L:cons} because $\Lambda*\Lambda$ is not well defined. This limitation does not imply that \eqref{eq: dzW} becomes inconsistent; rather, we simply lack access to its consistency consequence in the form of \eqref{L:cons}. This situation prevents us from following the regular path and excludes the crucial equation \eqref{vas3}, which defines how the field $\Lambda$ depends on $Z$.

To move forward, we need a $\Lambda$ that solves \eqref{L:main}. For instance, if there exists an explicit $\Lambda[C]$ that resolves the $Z$ dependence in \eqref{L:main} for {\it any} $W$ in its evolution class, then there would be no additional constraints on $W$ beyond the original condition \eqref{eq: dzW}. This possibility is indeed feasible for the star product defined in \eqref{limst}; see \cite{Didenko:2022qga}. However, demonstrating this requires a careful definition of the appropriate functional classes to which the fields $W$ and $\Lambda$ belong.

\subsection{Classes of functions}

Equations \eqref{W:zero} and \eqref{eq: dzW} establish the evolution of the fields $W$ and $\Lambda$ within a specific functional class. In the context of four-dimensional HS theory, which can be viewed as an interacting theory comprising two sectors --- the holomorphic and antiholomorphic --- it is natural to introduce a space of functions represented as a tensor product:
\begin{equation}
    \mathbf{C}\otimes\bar{\mathbf{C}}\,,
\end{equation}
where  $\mathbf{C}$  describes functions of $y$ and $z$ associated with the holomorphic sector, while $\bar{\mathbf{C}}$  represents the conjugate part, consisting of functions of  $\bar y$ and $\bar z$. We will specify these two sectors separately, starting with the class $\mathbf{C}$.

The Eqs. \eqref{W:zero}, \eqref{eq: dzW}, and \eqref{L:main} introduce a grading for the  $\dr z$-differential form with ranks  $r=0,1,2$. Consequently, we denote the class as  $\mathbf{C}^r$ to indicate the  $\dr z$  rank. To remain closed under the evolution equations, functions from  $\mathbf{C}^r$  must satisfy the following natural properties:
\begin{subequations}\label{def:class}
\begin{align}
    &\mathbf{C}^r+\mathbf{C}^r\in \mathbf{C}^r\,,\\
    &\dr_z \mathbf{C}^r\in \mathbf{C}^{r+1}\,,\\
    &\mathbf{C}^{r_1}*\mathbf{C}^{r_2}\in \mathbf{C}^{r_1+r_2}\,.\label{respect:star}
\end{align}
\end{subequations}
Additionally, we do not impose constraints on the space of solutions for $W$ in Eq. \eqref{eq: dzW}. At each order of perturbation, there remains freedom in choosing $z$-independent functions. Therefore, we further assume that all analytic functions in  $Y$ belong\footnote{Strictly speaking, not all analytic functions in $Y$  have well-defined star products. At this point, we do not need to rigorously define the dependence on $Y$. For the related discussion on this topic and a proposal for the $Y$-classes, refer to \cite{DeFilippi:2021xon}.} to $\mathbf{C}^{0}$, implying, in particular,
\begin{equation}\label{coh:C0}
    C(Y)\in \mathbf{C}^{0}\,,\qquad \go(Y)\in \mathbf{C}^{0}\,.
\end{equation}
The class $\mathbf{C}^{r}$, satisfying \eqref{def:class} and \eqref{coh:C0}, has been identified in \cite{Didenko:2022qga}. It is represented by the following functions:\footnote{The representation of $\mathbf{C}^{r}$ slightly differs from that in \cite{Didenko:2022qga}, but it is equivalent.}
\begin{equation}\label{class}
    \mathbf{C}^r[\phi_r]:=\int_{0}^{1}d\tau\frac{1-\tau}{\tau}\int_{u,v}\phi_r\left(\tau(z-v), (1-\tau)y+u; \frac{\tau}{1-\tau}\dr z\right)e^{i\tau zy+iuv}\,,
\end{equation}
where 
\begin{equation}
    \phi_r(z,y; \dr z)\sim (\dr z)^r\phi(z,y)\,,\qquad r=0, 1, 2\,.
\end{equation}
Functions $\phi(z,y)$ are analytic in $z$ and $y$. Additionally, the integration in \eqref{class} should be well defined, implying, in particular, $\phi_0(0, y; \dr z)=0$. Let us summarize the main features of the space $\mathbf{C}^{r}$.
\begin{itemize}
    \item The conditions specified in \eqref{class} are satisfied, with an important clarification. The well-defined products are
    \begin{equation}
        \mathbf{C}^{r_1}*\mathbf{C}^{r_2}\neq\infty\,,\qquad r_1+r_2<2\,,
    \end{equation}
    while the products  $\mathbf{C}^{1}*\mathbf{C}^{1}$, $\mathbf{C}^{0}*\mathbf{C}^{2}$, and $\mathbf{C}^{2}*\mathbf{C}^{0}$ are ill-defined, except in the case where $\dr_z \mathbf{C}^{0}=0$. Consequently, this has implications for the generating system based on class \eqref{class}; it cannot be regular because $\Lambda*\Lambda$ is not well defined when $\Lambda\in \mathbf{C}^{1}$.  Additionally, certain manipulations involving the Leibniz rule become ill-defined for $\dr z$ 1-forms. For instance, the left-hand side of the expression  $\dr_z(\mathbf{C}^{0}*\mathbf{C}^{1})=\dr_z\mathbf{C}^{0}*\mathbf{C}^{1}+\mathbf{C}^{0}*\dr_z \mathbf{C}^{1}$ is well defined, but the two terms on the right do not exist separately,\footnote{To avoid confusion, it is important to emphasize that the failure of the Leibniz rule is not an issue with the differential $\dr_z$, which is a standard de Rham differential. Instead, it is a characteristic of the specific star product \eqref{limst} and the chosen functional class \eqref{class}.} unless $\dr_z \mathbf{C}^{0}=0$.

    \item Although it may not be immediately apparent, the $z$-independent analytic functions satisfy the requirement \eqref{coh:C0}; see \cite{Didenko:2022qga} for a proof. Furthermore, the 2-form $\gga$ from \eqref{gamma} belongs to $\mathbf{C}^{2}$.  

    \item A notable characteristic of the functions in \eqref{class} is their invariance under reordering, as described in \eqref{reorder}. Specifically, for any $f\in \mathbf{C}^{r}$, we have
    \begin{equation}\label{inv:func}
        \mathcal{O}_{\gb} f=f\,,\qquad\forall\beta\,.
    \end{equation}
    This property has implications for lower-order HS vertices. In particular, the vertices $\mathcal{V}(\go, \go, C)$ and $\Upsilon(\go, C,C)$ in \eqref{unfld} appear to be unaffected by the choice of star product  \eqref{beta: star} for any $\gb$, unlike higher-order vertices; see also \cite{Didenko:2019xzz}. This happens because the action of all star-products \eqref{beta: star} in lower orders, where only the products \eqref{prod:equiv} emerge, is the same for any $\gb$, including $\gb=\infty$.  
    
    \item The function space described in \eqref{class} emerged from analyzing the locality of the holomorphic HS sector. It is smaller than what was previously proposed in \cite{Gelfond:2019tac} (refer also to \cite{Vasiliev:2015wma}, where related studies began). However, this new class accurately captures the maximally local holomorphic vertices to all orders, as discussed in \cite{Didenko:2022qga}, and the unconstrained vertices in D dimensions in \cite{Didenko:2023vna}. Moreover, this class coincides with the evolution space of master fields from the generating system in \cite{Didenko:2022qga}. In other words, $\mathbf{C}^{0}$ from Eq. \eqref{class} is constituted by the field $W$ evolution itself. The star product \eqref{limst} plays a crucial role in its definition. Despite Eq. \eqref{class} being invariant under reordering as shown in \eqref{inv}, only the star product \eqref{limst} respects \eqref{respect:star}, unlike any other $\gb$-products in \eqref{beta: star}.
    \end{itemize}
Analogously, one defines the conjugate class, 
\begin{equation}\label{class:conj}
\bar{\mathbf{C}}^r[\phi_r]:=\int_{0}^{1}d\tau\frac{1-\tau}{\tau}\int_{\bar u, \bar v}\phi_r\left(\tau(\bar z-\bar v), (1-\tau)\bar y+\bar u; \frac{\tau}{1-\tau}\dr \bar z\right)e^{i\tau \bar z\bar y+i\bar u\bar v}\,.
\end{equation}
It has the same properties as the class ${\mathbf{C}}^r$ for the conjugate variables $\bar y$ and $\bar z$. 

\subsection{Further properties and identities}
Here, we collect additional properties of the functional class \eqref{class} specific to the star product \eqref{limst}. These include a crucial identity that allows us to move forward with defining the field $\Lambda$ for the irregular generating equations. To have a better grasp of functions from \eqref{class}, it is convenient to introduce sources; see \cite{Didenko:2018fgx}.

\subsubsection{Source realization}
An analytic function $\phi(z,y)$ can be defined using its Taylor series:
\begin{equation}\label{Taylor}    \phi(z,y)=e^{izB+iyA}\phi(z',y')\Big|_{z'=y'=0}\,,\qquad
B_{\al}=i\p^{z'}_{\al}\,,\quad A_{\al}=i\p^{y'}_{\al}\,.
\end{equation}
We will frequently ignore the source part $\phi(z',y')$ in Eq. \eqref{Taylor}, aiming at the exponential operator. With this prescription, evaluating the $u,v$ integrals, we have for $\mathbf{C}^0$ and $\mathbf{C}^1$:  
\begin{subequations}\label{class:source}
\begin{align}
&\mathbf{C}^0[e^{iyA+izB}]=\int_{0}^{1}d\tau \frac{1-\tau}{\tau}
e^{i\tau z(y+B)+i(1-\tau)yA+i\tau AB}\,,\label{C0:source}\\
&\mathbf{C}^1[e^{iyA+izB}]=\int_{0}^{1}d\tau
e^{i\tau z(y+B)+i(1-\tau)yA+i\tau AB}\,.
\end{align}
\end{subequations}
Recall that the pole in \eqref{C0:source} is absent when functions $\phi$ are, at least, linear in $z$. The source prescription assumes that the integral along $\tau$ is performed after the generating exponential acts on the source. 

\subsubsection{Convolution $\circledast$} The integrand of \eqref{class:source} admits a nice representation in terms of the following convolution \cite{Didenko:2022eso}:
\begin{equation}\label{convl}
    f(y)\moast g(z,y)=\int_{u,v} f(y+u)g(z-v, y)e^{iuv}
\end{equation}
that factorizes exponentials of \eqref{class:source} as
\begin{equation}
    e^{i\tau z(y+B)+i(1-\tau)yA+i\tau AB}=e^{iyA}\moast\, e^{i\tau z(y+B)}\,.
\end{equation}
Importantly, at $\tau=1$, the convolution $\moast$ reveals a projective behavior. Indeed, from \eqref{convl}, we find 
\begin{equation}\label{convl:proj}
    f(y)\moast e^{iz(y+B)}=f(-B)e^{iz(y+B)}\,.
\end{equation}
The introduced representation is convenient for practical computations in terms of the following generating functions:
\begin{subequations}\label{C:circle}
    \begin{align} &\mathbf{C}^0[e^{iyA+izB}]=e^{iyA}\moast\int_{0}^{1}d\tau \frac{1-\tau}{\tau}
    e^{i\tau z(y+B)}\,,\label{C0:circle}\\
    &\mathbf{C}^1[z_{\al}\,e^{iyA+izB}]=e^{iyA}\moast\int_{0}^{1}d\tau \tau z_{\al}\,e^{i\tau z(y+B)}\,.
    \end{align}
\end{subequations}
In these terms, various products can be parameterized as follows:
\begin{subequations}\label{prod}
    \begin{align}
&\mathbf{C}^0[e^{iyA_1+izB_1}]*\mathbf{C}^0[e^{iy    A_2+izB_2}]=\int_{[0,1]^2}d\tau d\gs\frac{1}{\gs(1-             \gs)}\left(e^{iyA_1}*e^{iyA_2}\right)\moast     \frac{1-\tau}{\tau}e^{i\tau         z(y+B_{1,2})}\,,\label{C0*C0}\\
&\mathbf{C}^0[e^{iyA_1+izB_1}]*\mathbf{C}^1[z_{\al}\,e^{iy    A_2+izB_2}]=\int_{[0,1]^2}d\tau d\gs\frac{1-\gs}{\gs}\left(e^{iyA_1}*e^{iyA_2}\right)\moast     \tau z_{\al}\,e^{i\tau         z(y+B_{1,2})}\,,\label{C0*C1}\\
&\mathbf{C}^1[z_{\al}\,e^{iyA_1+izB_1}]*\mathbf{C}^0[e^{iy    A_2+izB_2}]=\int_{[0,1]^2}d\tau d\gs\frac{\gs}{1-             \gs}\left(e^{iyA_1}*e^{iyA_2}\right)\moast     \tau z_{\al}\,e^{i\tau         z(y+B_{1,2})}\,,\label{C1*C0}
\end{align}
\end{subequations}
where
\begin{equation}
    B_{1,2}=\gs(B_1+A_2)+(1-\gs)(B_2-A_1)\,.
\end{equation}
Recall that if functions entering \eqref{class} are, at least, linear in $z$, i.e., $\phi_{0}(0,y)=0$, then the product $\mathbf{C}^0*\mathbf{C}^0$ is well defined, as the poles in $\gs$ and $\tau$ in \eqref{C0*C0} cancel out, and similarly with $\mathbf{C}^0*\mathbf{C}^1$ and $\mathbf{C}^1*\mathbf{C}^0$. Analogous expressions can be written down in the antiholomorphic sector, where one defines
\begin{equation}\label{convl:bar}
    f(\bar y)\bar\moast g(\bar z,\bar y)=\int_{\bar u, \bar v} f(\bar y+\bar u)g(\bar z-\bar v, \bar y)e^{i\bar u\bar v}\,,
\end{equation}
which yields, for example, 
\begin{align}
    &\bar{\mathbf{C}}^0[e^{i\bar y\bar A_1+i\bar z\bar B_1}]*\bar{\mathbf{C}}^0[e^{i\bar y    \bar A_2+i\bar z\bar B_2}]=\int_{[0,1]^2}d\tau d\gs\frac{1}{\gs(1-             \gs)}\left(e^{i\bar y\bar A_1}*e^{i\bar y\bar A_2}\right)\bar\moast     \frac{1-\tau}{\tau}e^{i\tau \bar z(\bar y+\bar B_{1,2})}\,,\\
    &\bar B_{1,2}=\gs(\bar B_1+\bar A_2)+(1-\gs)(\bar B_2-\bar A_1)\,.
\end{align}
The two conjugate convolutions naturally form the tensor product $\Oast=\moast\otimes\bar\moast$, relevant in the mixed sector:
\begin{equation}\label{convl:mixed}
    f(Y)\Oast g(Z,Y)=\int_{U,V} f(Y+U)g(Z-V,Y)e^{iUV}\,.
\end{equation}

As a side remark, an observation from \cite{Didenko:2022eso} that highlights the relevance of the operation \eqref{convl} is that, as soon as contributions to HS vertices from the generating system are considered, it takes the form of a sum of \eqref{convl}, while the locality of the final result appears to depend on the locality of the left multipliers on the left-hand side of \eqref{convl}.   

\subsection{Projector identities}
The key observation that leads to the consistent evolution of the irregular holomorphic equations discussed in \cite{Didenko:2022qga} is the unique identities associated with the irregular star product \eqref{limst}. These identities help establish that the integrability condition outlined in Eq. \eqref{L:main} has a solution within a specific class of functions. Previously, these identities were only found for self-dual interactions. In this work, we present a generalization that enables us to consider mixed interactions within the framework of the irregular system.

Consider functions from $\mathbf{C}^1$,
\begin{equation}
    \mathbf{C}^1[\phi]=\dr z^{\al}\int_{0}^{1}d\tau\int_{u,v}\phi_\al\left(\tau(z-v), (1-\tau)y+u\right)e^{i\tau zy+iuv}\,.
\end{equation}
Among these, there are functions generated by $\phi_{\al}=z_{\al}\phi(z,y)$. The class $\mathbf{C}^1$ generated by these functions will play an important role in the following analysis. We denote this class as $\mathcal{C}^1\in \mathbf{C}^1$:
\begin{equation}\label{C1:Lambda}
   \mathcal{C}^1[\phi]= \dr  z^{\al}\int_{0}^{1}d\tau\tau\int_{u,v} (z-v)_{\al}\phi\left(\tau(z-v), (1-\tau)y+u\right)e^{i\tau zy+iuv}\,.
\end{equation}
Its Taylor form \eqref{Taylor} provides the following simple representation in terms of \eqref{convl}: 
\begin{equation}\label{C1:L}
    \mathcal{C}^1=\dr z^{\al}\int_{0}^{1}d\tau\tau e^{iyA}\circledast z_{\al}e^{i\tau z(y+B)}\,.
\end{equation}
Let us now enlist the important properties of functions $\Lambda[\phi]\in\mathcal{C}^1$. First of all, $\mathcal{C}^1$ is closed under product with $\mathbf{C}^0$
\begin{equation}
  \mathcal{C}^1*\mathbf{C}^0\in \mathcal{C}^1\,,\qquad  \mathbf{C}^0*\mathcal{C}^1\in \mathcal{C}^1\,.
\end{equation}
Now, recall that the limiting star product \eqref{limst} possesses the supertrace operation \eqref{lim:str}. Using it, define the following projector on $z$-independent functions:
    \begin{equation}\label{proj}
        \mathbf{h}_af(z,y)=\mathrm{str} f(z+a, y):=\int_{u,v} f(a+v, u)e^{iuv}\,,
    \end{equation}
    where supertrace satisfies \eqref{str:prop} with the star product \eqref{limst}, while $a$ could be any spinor parameter, including $y$. It is straightforward to show using \eqref{App:pre-proj} that 
    \begin{equation}\label{dzLambda}
        \dr_z\Lambda[\phi]=\mathbf{h}_y\phi(-z, y)*\gga\,,
    \end{equation}
    where $\gga$ is a two-form defined in \eqref{gamma}. In proving \eqref{dzLambda}, it is crucially important that the functions $\phi_{\al}$ defining the class $\mathcal{C}^1$ contain the factor of $z_{\al}$. Notice that, while $\phi(z,y)$ manifestly depends on $z$ in an arbitrary analytic fashion, the dependence on $z$ in \eqref{dzLambda} emerges via $\gamma$ only.  

There is a nontrivial generalization of the identity \eqref{dzLambda} that involves the class $\mathbf{C}^0$. Specifically, it can be shown that the following identities hold for an arbitrary function $f(z,y)\in\mathbf{C}^0$:
    \begin{subequations}\label{idnt:proj}
        \begin{align}
            &\dr_z(f(z,y)*\Lambda[\phi])=\mathbf{h}_{y}'[f(-z', y)*\mathbf{h}_y\phi(-z, y+y')]*\gga\,,\label{proj:fL}\\
            &\dr_z(\Lambda[\phi]*f(z,y))=\mathbf{h}_{y}'[\mathbf{h}_y\phi(-z,y+y')*f(-z', -y)]*\gga\,,\label{proj:Lf}
        \end{align}
    \end{subequations}
    where $\mathbf{h}'$ acts on the variables $z'$ and $y'$, as previously defined, i.e.,
    \begin{equation}\label{proj:prime}
        \mathbf{h}_
        {a}'f(z',y')=\int_{u,v}f(a+v, u)e^{iuv}\,,
    \end{equation}
while $\mathbf{h}$ acts on $z$ and $y$.     
The relations \eqref{idnt:proj} are crucial for formulating the irregular generating system for HS fields. We refer to these as {\it projector identities}, with the proof provided in Appendix B. Regardless of how  $f(z,y)\in \mathbf{C}^0$ and $\phi(z,y)$  depend on  $z$, these identities project it  into the Klein exponential $e^{izy}$, which explains their name. Notice that when $f = 1$, Eqs. \eqref{idnt:proj} simplify to \eqref{dzLambda}. Also note that the product $*$ on the right-hand sides of \eqref{idnt:proj} acts as the Moyal product \eqref{Moyal} with respect to the variable $y$, as opposed to its action on the left-hand sides, where it acts as prescribed by \eqref{limst}. Equations \eqref{idnt:proj} generalize\footnote{See also \cite{Korybut:2025vdn}, where an extension of the projector identities \eqref{proj:old} specialized to the so-called shifted homotopies \cite{Gelfond:2018vmi} was proposed.} the projector identity previously found in \cite{Didenko:2022qga} for $\phi(z,y)=C(-z)$, where $C$ is a function of a single spinor variable. In this case, \eqref{dzLambda} simplifies to 
     \begin{equation}
         \dr_z\Lambda=C(y)*\gga\,,
     \end{equation}
     while Eqs. \eqref{idnt:proj} take the form \cite{Didenko:2022qga}
     \begin{subequations}\label{proj:old}
         \begin{align}
             &\dr_z(f(z,y)*\Lambda[C])=[f(-z', y)*C(y)]\Big|_{z'=y}*\gga\,,\\
            &\dr_z(\Lambda[C]*f(z,y))=[C(y)*f(-z', -y)]\Big|_{z'=y}*\gga\,,
         \end{align}
     \end{subequations}
     Explicitly, the identities \eqref{idnt:proj} can be written down as
     \begin{subequations}
         \begin{align}
             &\dr_z(\mathbf{C}^0(z,y)*\Lambda[\phi])=\left(\int_{u,v}\mathbf{C}^0(-z'-v, y)*\phi(-y-v, u)e^{iuv}\right)\Big|_{z'=y}*\gga\,,\\
             &\dr_z(\Lambda[\phi]*\mathbf{C}^0(z,y))=\left(\int_{u,v}\phi(-y-v, u)*\mathbf{C}^0(-z'-v, -y)e^{iuv}\right)\Big|_{z'=y}*\gga\,.
         \end{align}
     \end{subequations}

\subsection{(Anti)holomorphic HS sector}
The irregular generating approach has been successfully implemented for HS dynamics of the (anti)holomorphic sector\footnote{Studies of holomorphic (chiral) or self-dual interactions date back to the works of Metsaev \cite{Metsaev:1991mt}, \cite{Metsaev:PhD}, in which it was demonstrated, using the light-cone approach, that these interactions form a closed sector that terminates at the cubic order in Minkowski space. These ideas were later extended to AdS in \cite{Metsaev:2018xip} and revisited in flat space in, e.g., \cite{Ponomarev:2017nrr, Serrani:2025owx}; see also a peculiar generalization to higher dimensions \cite{Basile:2024raj}. In the covariant framework, these interactions form a closed sector of the Vasiliev theory \cite{Vasiliev:1992av}. Their local behavior on any HS background has been perturbatively established order-by-order in \cite{ Didenko:2018fgx, Didenko:2019xzz, Didenko:2020bxd, Gelfond:2021two, Didenko:2022qga}; see also \cite{Sharapov:2022nps}, where the so-called chiral vertices were proposed, which were  claimed to be consistent in \cite{Sharapov:2023erv}.} \cite{Didenko:2022qga}. The corresponding generating systems require either $z_{\al}$ or $\bar{z}_{\dal}$, but not both. The two sectors are essentially complex, being conjugates of one another. The complete unitary HS interaction requires both, while its consistency leads to a nonlinear mixing of the two sectors, forming the so-called {\it mixed} sector. To generate the latter, one needs the full set $Z_A=(z_\al, \bar z_{\dal})$ of auxiliary variables. Here, we briefly reproduce the result of \cite{Didenko:2022qga}, where the holomorphic sector was described. To this end, we need the $(Y,z)$-part of the star product \eqref{limst}, which has the following form:
\begin{equation}\label{limst:hol}
(f*g)(z, Y)= \int f(z+u', y+u)\bar{*}
g(z-v,y+v+v')
e^{iuv+iu'v'}\,,
\end{equation}
where we left the variable $\bar y$ implicit, and we denoted by $\bar{*}$ the Moyal part \eqref{Moyal} acting on $\bar y$. We now introduce the master fields $W=W(z, Y)$ and $\mathbf{\Lambda}[C]$ and impose the generating equations \eqref{W:zero} and \eqref{eq: dzW}:
\begin{subequations}
\begin{align}\label{hol:gen}
&\dr_xW+W*W=0\,,\qquad\\    
&\dr_zW+\{W,\mathbf{\Lambda}\}_*+\dr_x\mathbf{\Lambda}=0\,.\label{hol:dzW}
\end{align}
\end{subequations}
In order to be well defined and closed under the generating system operation, we require $W\in\mathbf{C}^0$ and $\mathbf{\Lambda}\in\mathbf{C}^1$. The system has been established to be consistent with the conditions $\dr_x^2 = 0$ and $\{\dr_x, \dr_z\} = 0$. Now, we need to define the field $\mathbf{\Lambda}[C]$ in Eq. \eqref{hol:dzW}. An arbitrary choice for $\mathbf{\Lambda}$ imposes additional constraints on the field $W$, expressed by the equation
\begin{equation}\label{hol:cons}
    \dr_z\{W, \mathbf{\Lambda}\}_*-\dr_x\dr_z\mathbf{\Lambda}=0\,.
\end{equation}
This equation must hold for any value of $z$, which could limit the dependence of $W$ on $z$ and potentially lead to inconsistencies. Therefore, it is essential to choose a $\mathbf{\Lambda}$ that resolves the $z$-dependence of \eqref{hol:cons} without imposing constraints on the evolution of $W$ with respect to the variable $z$.  This option is viable for $\mathbf{\Lambda}\in \mathcal{C}^1$. Indeed, taking 
\begin{equation}\label{L:gen}
    \mathbf{\Lambda}=\dr  z^{\al}\int_{0}^{1}d\tau\tau\int_{u,v}(z-v)_{\al}\phi\left(\tau(z-v); (1-\tau)y+u\right)e^{i\tau zy+iuv}
\end{equation}
with an arbitrary $\phi(z;y, \bar y)$, and using Eqs. \eqref{idnt:proj} and \eqref{dzLambda}, we see that the dependence on $z$ factors out from \eqref{hol:cons} in the form of the Klein operator $e^{izy}$ for {\it any} $W\in \mathbf{C}^0$. Consequently, we conclude that any choice of $\phi(z;Y)$ in \eqref{L:gen} leads to a consistent system \eqref{hol:gen}. The simplest option used in \cite{Didenko:2022qga} is to take $\phi$ independent of $y$; see \eqref{proj} and \eqref{dzLambda},
\begin{equation}
    \phi(z; y, \bar y)=i\eta\,C(-z, \bar y)\,,
\end{equation}
where we recall that the fields also depend on $\bar y$. The choice of the minus sign is conventional. This gives us
\begin{equation}\label{hol:L}
    \mathbf{\Lambda}:=i\eta\,\dr z^{\al}z_{\al}\int_{0}^1d\tau\tau C(-\tau z, \bar y)e^{i\tau zy}\,,\qquad \dr_z\mathbf{\Lambda}=i\eta\,C(y, \bar y)*\gga\,.
\end{equation}
Substituting into \eqref{hol:cons}, one arrives at
\begin{equation}\label{hol:dC}
    \dr_x C+\left(W(-z', y, \bar y)*C-C*W(-z', -y, \bar y)\right)\Big|_{z'=y}=0\,,
\end{equation}
which, together with \eqref{hol:gen}, generates holomorphic vertices. The fact that $\mathbf{\Lambda}[C]$ is linear in $C$ drastically simplifies the analysis of the holomorphic vertices entering \eqref{unfld}. The explicit all-order local result was obtained in \cite{Didenko:2024zpd} using the above irregular holomorphic generating equations.   

The antiholomorphic sector is generated in the same fashion. It can be reproduced from the holomorphic sector by conjugation. Specifically, one introduces 
\begin{equation}\label{ahol:L}
    \bar{\mathbf{\Lambda}}:=i\bar\eta\,\dr \bar z^{\dal}\bar z_{\dal}\int_{0}^1d\tau\tau C(y, -\tau \bar z)e^{i\tau \bar z\bar y}\,,\qquad \dr_{\bar z}\bar{\mathbf{\Lambda}}=i\bar\eta\,C(y, \bar y)*\bar\gga
\end{equation}
and replaces $\dr_z$ with $\dr_{\bar z}$ in \eqref{hol:dzW}, as well as the Moyal product $\bar *$ with * in \eqref{limst:hol}.

\section{Mixed sector}\label{sec:mixed} 
It has already been noted that the complete HS interaction in four dimensions emerges from the consistent combination of two closed conjugate sectors: the holomorphic sector and the antiholomorphic sector. The holomorphic sector contributes HS vertices that involve powers of the phase parameter $\eta$, while the antiholomorphic sector introduces powers of $\bar\eta$. The consistency implies that when both sectors are present, there should be cross terms that result from their mixing. These cross terms typically contain arbitrary powers of the form $\eta^m\bar\eta^n$, parameterized by the {\it degree} 
\be
k=\mathrm{min}(m,n)\,.
\ee
The generating equations require two types of auxiliary variables, denoted as $Z_A=(z_{\al}, \bar z_{\dal})$. The corresponding large star-product algebra represents the tensor product of the large holomorphic algebra (as defined in Eq. \eqref{limst:hol}) and its conjugate, resulting in \eqref{limst}. 

It is important to note that the projector identities \eqref{idnt:proj} fundamentally rely on the fact that $z_{\al}$ is a two-component variable. Therefore, simply replacing $z_{\al}$ with $Z_A$ in these identities will not work. This implies that a straightforward generalization of the holomorphic sector generating equation \eqref{hol:dC} by extending the variable $z$ will lead to inconsistencies.
Nonetheless, our initial approach remains valid, as we must still impose the conditions \eqref{W:zero} and \eqref{eq: dzW}, which can be expressed as 
\begin{subequations}\label{mixed}
    \begin{align}
        &\dr_x W+W*W=0\,,\\
        &\dr_zW+\{W,\Lambda\}_*+\dr_x\Lambda=0\,,\label{mixed:dzW}\\
        &\dr_{\bar z}W+\{W, \bar\Lambda\}_*+\dr_x\bar\Lambda=0\,,\label{mixed:bardzW}
    \end{align}
\end{subequations}
where 
\begin{equation}\label{L:real}
    \Lambda[C]=\dr z^{\al}\Lambda_{\al}[C](Z,Y)\,,\qquad \bar\Lambda[C]=\dr\bar z^{\dal}\bar\Lambda_{\dal}[C](Z, Y)\,.
\end{equation}
To identify the integrability conditions of Eqs. \eqref{mixed:dzW} and \eqref{mixed:bardzW}, one must specify the space of functions for the fields $W$, $\Lambda$, and $\bar\Lambda$.

\subsection{Functional class} The space of functions describing the mixed sector should contain functions from the holomorphic sector $\mathbf{C}^r$ and the antiholomorphic sector $\bar{\mathbf{C}}^r$. Naturally, the class of functions for the mixed sector takes the form of the following tensor product:
\begin{equation}
    \mathbf{C}^{p,q}:=\mathbf{C}^p\otimes\bar{\mathbf{C}}^{q}\,,\qquad p,q=0,1,2\,,
\end{equation}
where $\mathbf{C}^p$ is given in \eqref{class}, while its conjugate $\bar{\mathbf{C}}^{q}$ is defined in \eqref{class:conj}. Explicitly, the functions from $\mathbf{C}^{p,q}$ are of the following form:
\begin{align}\label{mixed:class}
  &\mathbf{C}^{p,q}[\phi]:=  \int_{[0,1]^2}d\tau d\tau'\frac{1-\tau} {\tau} \frac{1-\tau'} {\tau'}\int_{U,V}e^{i\tau zy+i\tau'\bar z\bar y+iUV}\times\\
  &\times\phi_{p,q}\left(\tau(z-v), \tau'(\bar z-\bar v); (1-\tau)y+u,  (1-\tau')\bar y+\bar u; \frac{\tau}{1-\tau}\dr z, \frac{\tau'}{1-\tau'}\dr \bar z\right)\,,\nonumber
\end{align}
where 
\begin{equation}
    \phi_{p,q}(Z,Y; \dr Z)\sim (\dr z)^p(\dr\bar z)^q\phi(Z,Y)\,,\qquad 0\leq p, q\leq 2\,.
\end{equation}
The HS master fields belong to the class \eqref{mixed:class} in accordance with their $\dr Z$ rank. Specifically, 
\begin{equation}
    W\in\mathbf{C}^{0,0}\,,\qquad\Lambda\in\mathbf{C}^{1,0}\,,\qquad \bar\Lambda\in\mathbf{C}^{0,1}\,.
\end{equation}
A consequence of the above splitting is that some products are well defined, while others are not. So, the well defined are
\begin{subequations}
\begin{align}
        &W*W\in\mathbf{C}^{0,0}\,,\quad (W*\Lambda\,,\quad\Lambda*W)\in\mathbf{C}^{1,0}\,,\quad (W*\bar\Lambda\,,\quad\bar\Lambda*W)\in\mathbf{C}^{0,1}\,,\\  
        &(\Lambda*\bar\Lambda\,,\quad \bar\Lambda*\Lambda)\in\mathbf{C}^{1,1}\,,
\end{align}        
\end{subequations}
while the ill-defined include
\begin{equation}
\Lambda*\Lambda\qquad \text{and}\qquad \bar\Lambda*\bar\Lambda\,.
\end{equation}
Consequently, the regular Vasiliev-type generating equations \eqref{vas} make no sense with the star product \eqref{limst} because Eq. \eqref{vas3} is ill-defined. The irregular generating systems rely on the projector identities \eqref{idnt:proj}. To make use of them, we restrict 
\begin{equation}
    \Lambda\in\mathcal{C}^{1,0}:=\mathcal{C}^1\otimes\bar{\mathbf{C}}^0\,,\qquad\bar\Lambda\in\mathcal{C}^{0,1}:=\mathbf{C}^{0}\otimes\bar{\mathcal{C}}^{1}\,,
\end{equation}
where $\mathcal{C}^1$ is defined in \eqref{C1:Lambda}. The manifest form of such $\Lambda$' reads
\begin{equation}\label{mixed: L}
    \Lambda=\dr  z^{\al}\int_{0}^{1}d\tau\tau\int_{u,v}(z-v)_{\al}\phi\left(\tau(z-v), \bar z;\, (1-\tau)y+u, \bar y\right)e^{i\tau zy+iuv}\,,
\end{equation}
where $\phi$ is an arbitrary analytic function of $z$ and $y$, and a function from $\bar{\mathbf{C}}^0$ of $\bar z$ and $\bar y$, i.e., 
\begin{equation}\label{mixed:phiC0}
\phi(z, \bar z;\,y,\bar y)=\int_{0}^{1}d\tau'\frac{1-\tau'}{\tau'}\int_{\bar u, \bar v}\phi_0\left(z, \tau'(\bar z-\bar v);\,y,  (1-\tau')\bar y+\bar u\right)e^{i\tau' \bar z\bar y+i\bar u\bar v}\,,
\end{equation}
where $\phi_0(z,0; y, \bar y)=0$ and otherwise is an arbitrary analytic function. 
Similarly, $\bar\Lambda$ is of the following form 
\begin{equation}\label{mixed: barL}
    \bar{\Lambda}=\dr  \bar z^{\dal}\int_{0}^{1}d\tau\tau\int_{\bar u,\bar v}(\bar z-\bar v)_{\dal}\bar\phi\left(z, \tau(\bar z-\bar v);\, y, (1-\tau)\bar y+\bar u\right)e^{i\tau \bar z\bar y+i\bar u\bar v}
\end{equation}
with $\bar\phi$ from $\mathbf{C}^0$ in the variables $z$ and $y$. Equations \eqref{mixed: L} and \eqref{mixed: barL} imply (see \eqref{idnt:proj}) that 
\begin{equation}\label{chrl:L}
    \dr_z\Lambda=i\eta\,\bar{\mathcal{B}}(\bar z; Y)*\gga\,,\qquad \dr_{\bar z}\bar\Lambda=i\bar\eta\,\mathcal{B}(z; Y)*\bar\gga\,,
\end{equation}
where $\mathcal{B}, \bar{\mathcal{B}}\in \mathbf{C}^{0,0}$ are the 0-forms expressed as follows; see \eqref{dzLambda}:
\begin{equation}
    i\bar\eta\,\mathcal{B}(z; Y):=\bar{\mathbf{h}}_{\bar y}\bar\phi(z, -\bar z;\,y, \bar y)\,,\qquad i\eta\,\bar{\mathcal{B}}(\bar z; Y):=\mathbf{h}_y\phi(-z, \bar z;\, y, \bar y)\,.
\end{equation}
Notice that the right-hand sides of \eqref{chrl:L} have a {\it chiral} form with respect to the dependence on $z$ and $\bar z$, with $\mathcal{B}$ being independent of $\bar z$ and $\bar{\mathcal{B}}$ -- independent of $z$:
\begin{equation}
    \dr_z\bar{\mathcal{B}}=0\,,\qquad \dr_{\bar z}\mathcal{B}=0\,.
\end{equation}
The chiral form guaranties that the right-hand sides of \eqref{chrl:L} are well defined (we recall that $\gga\in\mathbf{C}^2$, while $\mathbf{C}^0*\mathbf{C}^2$ is generally sick unless $\dr_z\mathbf{C}^0=0$). 

\subsection{Generating equations and consistency} Equations \eqref{mixed} are consistent with respect to $\dr_x^2=\{\dr_x, \dr_{Z}\}=0$. Now we need to collect conditions that guaranty consistency with $\dr_z^2=\dr_{\bar z}^2=\{\dr_z, \dr_{\bar z}\}=0$. The last condition $\{\dr_z, \dr_{\bar z}\}=0$ is satisfied if we impose
\begin{equation}\label{L:cross}
    \dr_{\bar z}\Lambda+\dr_z\bar\Lambda+\{\Lambda, \bar\Lambda\}_*=0\,.
\end{equation}
Indeed, applying $\dr_{\bar z}$ to \eqref{mixed:dzW} and $\dr_z$ to \eqref{mixed:bardzW}, and then substituting the equations using the Leibniz rule, one arrives at the identity, provided Eq. \eqref{L:cross} is satisfied. Notice that each contribution to \eqref{L:cross} is well defined and belongs to $\mathbf{C}^{1,1}$, provided $\Lambda\in\mathcal{C}^{1,0}$ and $\bar\Lambda\in\mathcal{C}^{0,1}$. 

Equation \eqref{L:cross} induces further constraints. In particular, the consistency with $\dr_z^2=0$ and $\dr_{\bar z}^2=0$ amounts to the following equations that specify the $z$ and $\bar z$ dependence of the fields $\mathcal{B}$ and $\bar{\mathcal{B}}$, respectively:
\begin{subequations}\label{mixed: dzBeq}
    \begin{align}
        &\dr_z \mathcal{B}+\bar{\mathbf{h}}_{\bar y}'[\Lambda(z, -\bar z';\,Y)*\mathcal{B}_{\bar y'}-\mathcal{B}_{\bar y'}*\bar\pi\Lambda(z, -\bar z';\, Y)]=0\,,\label{mixed:dzB}\\
        &\dr_{\bar z} \bar{\mathcal{B}}+\mathbf{h}_{y}'[\bar\Lambda(-z', \bar z;\,Y)*\bar{\mathcal{B}}_{ y'}-\bar{\mathcal{B}}_{y'}*\pi\bar\Lambda(-z', \bar z;\, Y)]=0\,,\label{mixed:bardzB}
    \end{align}
\end{subequations}
where, in deriving \eqref{mixed: dzBeq}, we made use of the projector identities \eqref{idnt:proj} and introduced the notation 
\begin{subequations}\label{def:By}
\begin{align}
        &i\bar\eta\,\mathcal{B}_{\bar y'}(z;Y):=\bar{\mathbf{h}}_{\bar y}\bar\phi(z, -\bar z;\,y, \bar y+\bar y')\,,\\ &i\eta\,\bar{\mathcal{B}}_{y'}(\bar z;Y):= {\mathbf{h}}_y\phi(-z, \bar z;\,y+y', \bar y)\,,
\end{align}
and $\pi$, $\bar \pi$ act as defined in \eqref{pi:def},
\begin{equation}
    \pi\bar\Lambda(Z;\, y, \bar y):=\bar\Lambda(Z;\, -y, \bar y)\,,\quad \bar\pi\Lambda(Z;\,y, \bar y):=\Lambda(Z;\,y, -\bar y)\,. 
\end{equation}
\end{subequations}
Typically of the projector identities \eqref{idnt:proj}, the star products \eqref{limst} act on unprimed variables in \eqref{mixed: dzBeq}, while the projectors $\mathbf{h}'$ and $\bar{\mathbf{h}'}$ act on the primed ones in accordance with \eqref{proj:prime}. Additional constraints emerge from the integrability conditions $\dr_z^2=0$ and $\dr_{\bar z}^2=0$ for the Eqs. \eqref{mixed: dzBeq}. By applying $\dr_z$ to \eqref{mixed:dzB} and utilizing $\Lambda\in\mathcal{C}^{1,0}$, $\mathcal{B}_{\bar y'}\in\mathbf{C}^{0,0}$, and the commutation of $[\dr_z, \bar{\mathbf{h}}']=0$, we can leverage the projector identities \eqref{idnt:proj} to arrive at
\begin{equation}\label{mixed:BB}
    \mathbf{h}_{y}'\bar{\mathbf{h}}_{\bar y}'\left[\mathcal{B}_{\bar y'}(-z'; Y)*\bar\pi\bar{\mathcal{B}}_{y'}(-\bar z'; Y)-\bar{\mathcal{B}}_{y'}(-\bar z'; Y)*\pi\mathcal{B}_{\bar y'}(-z'; Y)\right]=0\,.
\end{equation}
Applying $\dr_{\bar z}$ to \eqref{mixed:bardzB} leads to exactly the same condition \eqref{mixed:BB}. Notice that Eq. \eqref{mixed:BB} is by definition of the projectors $\mathbf{h}$ and $\bar{\mathbf{h}}'$ is $Z$ independent. Consequently, it does not introduce additional consistency constraints. 

Let us now investigate the consistency of Eqs. \eqref{mixed:dzW} and \eqref{mixed:bardzW}. We need to check whether $\dr_z^2=0$ and $\dr_{\bar z}^2=0$ introduce any further constraints. Applying $\dr_{\bar z}$ to \eqref{mixed:bardzW} and using \eqref{idnt:proj} along with the definitions \eqref{chrl:L} and \eqref{def:By}, we find 
\begin{equation}\label{eq:dB}
    \dr_x\mathcal{B}+\bar{\mathbf{h}}_{\bar y}'\left[W(z, -\bar z'; Y)*\mathcal{B}_{\bar y'}-\mathcal{B}_{\bar y'}*\bar\pi W(z, -\bar z'; Y)\right]=0\,.
\end{equation}
Analogously, applying $\dr_z$ to \eqref{mixed:dzW}, we arrive at the conjugate equation
\begin{equation}\label{eq:dbarB}
    \dr_x\bar{\mathcal{B}}+\mathbf{h}_{y}'\left[W(-z', \bar z; Y)*\bar{\mathcal{B}}_{y'}-\bar{\mathcal{B}}_{y'}*\pi W(-z', \bar z; Y)\right]=0\,.
\end{equation}
The obtained Eqs. \eqref{eq:dB} and \eqref{eq:dbarB} are $z$ and $\bar z$ independent, respectively. To see this, one can differentiate the left-hand sides of these equations using $\dr_{\bar z}$ and $\dr_z$ to observe that the result is zero on shell, provided Eqs. \eqref{mixed: dzBeq} hold. 

Summarizing the results obtained at this point, we have arrived at the following generating system:
\begin{itemize}
    \item {\bf Generating equations:}
    \begin{subequations}\label{mixed:gen}
    \begin{align}
        &\dr_x W+W*W=0\,,\quad &&W\in\mathbf{C}^{0,0}\label{gen:zero}\\
        &\dr_zW+\{W,\Lambda\}_*+\dr_x\Lambda=0\,,\quad&&\Lambda[\phi]\in\mathcal{C}^{1,0}\label{gen:dzW}\\
        &\dr_{\bar z}W+\{W, \bar\Lambda\}_*+\dr_x\bar\Lambda=0\,,\quad&&\bar\Lambda[\bar\phi]\in\mathcal{C}^{0,1}\,,\label{gen:bardzW}\\
        &\dr_{\bar z}\Lambda+\dr_z\bar\Lambda+\{\Lambda, \bar\Lambda\}_*=0\,.\label{gen:L}
    \end{align}
\end{subequations}
Equations \eqref{gen:dzW}-\eqref{gen:L} restore the dependence of the fields $W$ and $\Lambda$, $\bar\Lambda$ on $Z$ in terms of the unspecified functions $\phi$ and $\bar\phi$.

\item {\bf Consistency constraints:}
\begin{subequations}\label{cons:constr}
    \begin{align}
    &\dr_z\Lambda=i\eta\,\bar{\mathcal{B}}(\bar z; Y)*\gga\,,\label{gen:dL}\\
    &\dr_z \mathcal{B}+\bar{\mathbf{h}}_{\bar y}'[\Lambda(z, -\bar z';\,Y)*\mathcal{B}_{\bar y'}-\mathcal{B}_{\bar y'}*\bar\pi\Lambda(z, -\bar z';\, Y)]=0\,,\label{gen:dzB}\\
        &\dr_x\mathcal{B}+\bar{\mathbf{h}}_{\bar y}'\left[W(z, -\bar z'; Y)*\mathcal{B}_{\bar y'}-\mathcal{B}_{\bar y'}*\bar\pi W(z, -\bar z'; Y)\right]=0\,.\label{gen:dB}
    \end{align}
\end{subequations}
The above three equations are supplemented by the three conjugates 
\begin{subequations}\label{cons:constr-cc}
    \begin{align}
    &\dr_{\bar z}\bar\Lambda=i\bar\eta\,{\mathcal{B}}(z; Y)*\bar\gga\,,\label{gen:dL-cc}\\
    &\dr_{\bar z} \bar{\mathcal{B}}+{\mathbf{h}}_{y}'[\bar\Lambda(-z', \bar z;\,Y)*\bar{\mathcal{B}}_{y'}-\bar{\mathcal{B}}_{y'}*\pi\bar\Lambda(-z', \bar z;\, Y)]=0\,,\label{gen:dzB-cc}\\
        &\dr_x\bar{\mathcal{B}}+{\mathbf{h}}_{y}'\left[W(-z', \bar z; Y)*\bar{\mathcal{B}}_{y'}-\bar{\mathcal{B}}_{y'}*\pi W(-z', \bar z; Y)\right]=0\,,\label{gen:dB-cc}
    \end{align}
\end{subequations}

and the condition \eqref{mixed:BB} which, if not satisfied, leads to an obstruction of the generating system \eqref{mixed:gen}. Equivalently, if \eqref{gen:L} admits a solution, then Eqs. \eqref{gen:dzB} and \eqref{mixed:BB} hold true.  
\item {\bf $b$-constraint:} 

The consistency of Eq. \eqref{gen:L} constrains the fields $\mathcal{B}$ and $\bar{\mathcal{B}}$ with Eq. \eqref{mixed:BB}, which has a form of twisted-projective commutativity, 
\begin{equation}\label{cons:BB}
    \mathbf{H}_{Y}'\left[\mathcal{B}_{\bar y'}(-z'; Y)*\bar\pi\bar{\mathcal{B}}_{y'}(-\bar z'; Y)-\bar{\mathcal{B}}_{y'}(-\bar z'; Y)*\pi\mathcal{B}_{\bar y'}(-z'; Y)\right]=0\,,
\end{equation}
where the projector $\mathbf{H}$ acts in both spinor sectors as follows:
\begin{equation}
    \mathbf{H}_Af(Z;Y):=\mathbf{h}_a\bar{\mathbf{h}}_{\bar a}f(z, \bar z;\, y, \bar y )=\int_{U,V}f(A+U;\, V)e^{iUV}\,,\qquad A=(a_{\al}, \bar a_{\dal})\,.
\end{equation}
Recall that the prime notation in $\mathbf{H}'$ indicates that the projector operates on  $Z'$  and  $Y'$, rather than on $Z$ and $Y$.

The condition \eqref{cons:BB} is automatically satisfied at the free level and in the (anti)self-dual case, where either $\mathcal{B}=0$ or $\bar{\mathcal{B}}=0$. However, it becomes important at quadratic and higher orders in the unitary case. In particular, from $\mathcal{B}=C+\eta\, O(C^2)$ and $\bar{\mathcal{B}}=C+\bar\eta \,O(C^2)$, it follows from \eqref{cons:BB} that
\begin{equation}\label{constr:CC}
    C*\bar \pi (C)-C*\pi(C)=0\,,
\end{equation}
which is satisfied in the bosonic case, where $C(y, -\bar y)=C(-y, \bar y)$. The higher-order extension \eqref{cons:BB} is referred to as {\it b-constraint}. 

Let us point out in this regard that the interacting (anti)self-dual sector, for which, e.g., $\mathcal{B}=0$, is not necessarily bosonic. The constraint \eqref{constr:CC} implies, however, that there is no consistent completion of self-dual interactions in the presence of fermions in our approach, when each field in the HS spectrum comes single. This observation perfectly fits the fact that the fields should appear in, at least, two copies if fermions are included \cite{Vasiliev:1992av}.

\item {\bf Reality conditions:} 

In order to describe the real dynamics, the system is supplemented by the following reality conditions:
\begin{subequations}\label{real}
\begin{align}
    &y_{\al}^{\dagger}=\bar y_{\dal}\,,\qquad z_{\al}^{\dagger}=-\bar z_{\dal}\,,\qquad \eta^{\dagger}=\bar\eta\,,\\
    &W^\dagger=-W\,,\qquad \Lambda^{\dagger}=-\bar\Lambda\,,\qquad \mathcal{B}^\dagger=\pi\bar{\mathcal{B}}\,,\qquad \bar{\mathcal{B}}^\dagger=\bar\pi\mathcal{B}\,,
\end{align}
\end{subequations}

\item {\bf $Z$ independence:} 

Recall that Eq. \eqref{gen:zero} is independent of the variable $Z$, even though $W$  clearly depends on it. This independence arises from the evolution equations \eqref{gen:dzW} and \eqref{gen:bardzW}, which enable us to access the vertices in Eq. \eqref{unfld:w}. Similarly, Eq. \eqref{gen:dB} also appears to be independent of $Z$ and provides access to the vertices from \eqref{unfld:C}. Indeed, it is clearly independent of $\bar z$ since both $\mathcal{B}$ and the component involving the projector $\bar{\mathbf{h}}'$ do not depend on $\bar z$. Demonstrating that Eq. \eqref{gen:dB} is also independent of $z$  is a bit more complex. Instead of directly proving that this equation, which we will refer to as $E$, lacks $z$  dependence, we can show that the product $E*\bar\gga$  does not depend on $z$ . This implies that the equation $E$ itself does not involve $z$. Multiplying Eq. \eqref{gen:dB} by $\bar\gga$ and using the projector identity \eqref{idnt:proj}, along with the definition \eqref{chrl:L}, we have 
\begin{equation}
    E*\bar\gga=\dr_{\bar z}\{W, \bar\Lambda\}_*-\dr_x\dr_{\bar z}\bar\Lambda\,.
\end{equation}
It is not hard to show that 
\begin{equation}\label{z-indep}
    \dr_z(E*\bar\gga)=-\dr_{\bar z}[\Lambda, \dr_{\bar z}W+\{W, \bar\Lambda\}_*+\dr_x\bar\Lambda]\,,
\end{equation}
where, in deriving this expression, we used Eqs. \eqref{gen:L} and \eqref{gen:dzW}, and 
\begin{equation}
    \dr_{\bar z}([W, \dr_{\bar z}\Lambda]-[\dr_{\bar z}W, \Lambda])=-\dr_{\bar z}^2\{W, \Lambda\}=0\,.
\end{equation}
The condition in \eqref{z-indep} relates to the equation of motion \eqref{gen:dzW}, which means that \eqref{gen:dB} is independent of  $z$ . It is important to highlight that throughout this process, we did not encounter any ill-defined star products of the form  $\mathbf{C}^{p,q}*\mathbf{C}^{p', q'}$ with $p+p'>1$ or $q+q'>1$.

\end{itemize}

\section{Chiral perturbation theory}\label{sec:chiral}
Generating equations \eqref{mixed:gen} and \eqref{cons:constr} can be processed perturbatively
\begin{subequations}\label{BL:pert}
   \begin{align}
 &\mathcal{B}=\mathbf{B}[C]+\mathcal{B}_2[C,C]+\dots\,,\qquad \bar{\mathcal{B}}=\bar{\mathbf{B}}[C]+\bar{\mathcal{B}}_2[C,C]+\dots\,,\\
 &\Lambda=\mathbf{\Lambda}[C]+\Lambda_2[C,C]+\dots\,,\qquad \bar\Lambda=\bar{\mathbf{\Lambda}}[C]+\bar\Lambda_2[C,C]+\dots\,.
 \end{align} 
\end{subequations}
This data defines the perturbation of $W$ via the solution of Eqs. \eqref{gen:dzW} and \eqref{gen:bardzW}, which exists provided $\Lambda$ exists. A remarkable feature of the decomposition \eqref{BL:pert} is that its first order in $C$ approximation restores the all-order (anti)holomorphic sector of the theory, rather than just the linearized equations. The higher order (anti)holomorphic corrections arise through the nonlinear corrections in $W$, while the (anti)holomorphic part of $\mathcal{B}$ and $\mathbf{\Lambda}$ is linearly exact:
\begin{equation}\label{B:init}
    \mathbf{B}=C(Y|x)\,,\qquad \bar{\mathbf{B}}=C(Y|x)
\end{equation}
and
\begin{equation}\label{L:init}
    \mathbf{\Lambda}=i\eta\,\dr z^{\al}z_{\al}\int_{0}^{1}d\tau\tau C(-\tau z,\bar y)e^{i\tau zy}\,,\qquad  \bar{\mathbf{\Lambda}}=i\bar\eta\,\dr \bar z^{\dal}\bar z_{\dal}\int_{0}^{1}d\tau\tau C(y, -\tau \bar z)e^{i\tau \bar z\bar y}\,.
\end{equation}
Indeed, to reach out to the holomorphic sector, we set $\bar\eta=0$, implying 
\begin{subequations}\label{self-dual}
\begin{align}
    &\mathcal{B}=0\,,\qquad \bar{\mathcal{B}}=\bar{\mathbf{B}}\,,\\
    &\Lambda=\mathbf{\Lambda}\,,\qquad \bar\Lambda=0\,.
\end{align}    
\end{subequations}
As a result, we obtain that 
\begin{equation}
    \dr_z\mathbf{\Lambda}=i\eta\,\bar{\mathbf{B}}*\gga
\end{equation}
and Eq. \eqref{gen:dL} is satisfied. Equations \eqref{gen:L} and \eqref{gen:bardzW} are also satisfied for $W$ independent of $\bar z$. Importantly, the constraint \eqref{cons:BB} is automatically fulfilled for $\mathcal{B}=0$.  Eventually, to arrive at the holomorphic vertices at a given order $C^n$, it is sufficient to solve for $W$ from Eq. \eqref{gen:dzW} using the standard contracting homotopy,
\begin{equation}\label{W:hom}
    W_{n}=-\Delta(\{W_{n-1}, \mathbf{\Lambda}\}_*)\,,\qquad W_0:=\go(y, \bar y|x)\,,
\end{equation}
where 
\begin{equation}
    \Delta(\dr z^{\al}f_{\al}(z; Y)):=z^{\al}\int_{0}^{1}d\tau\tau f_{\al}(\tau z; Y)\,.
\end{equation}
In connection with the (anti)holomorphic sector, let us highlight the difference between the regular and irregular approaches. In the former case, neither $\Lambda$ nor $B$ terminates at the linearized order. This happens because the two are related to each other through the nonlinear equations \eqref{vas3} and \eqref{vas4}. In the latter case, the projector identities from \eqref{idnt:proj} allow us to avoid the nonlinear map at the cost of having a different large $(Z,Y)$-algebra. 

We have demonstrated that the $C$-linear terms in \eqref{BL:pert} are responsible for the all-order (anti)self-dual sector of the HS theory. A natural question arises regarding the significance of the higher-order $C$-corrections in $\mathcal{B}_n$ and $\Lambda_n$. These corrections contribute to the nonlinear mixing of the holomorphic and antiholomorphic sectors. 
One might have hoped to extend the self-dual conditions in \eqref{self-dual} by setting $\mathcal{B}=\mathbf{B}$ and $\bar\Lambda=\bar{\mathbf{\Lambda}}$, thereby accounting for the full HS interactions without introducing higher-order corrections. However, this choice is inconsistent with Eq. \eqref{gen:L}. Consequently, we expect that $\mathcal{B}_n$ and $\Lambda_n$ will be nonzero for $n \geq 2$. Their dependence on $\eta$ and $\bar\eta$ can be understood from \eqref{gen:L}; for example, $\Lambda_2\sim \eta\bar\eta$. 

Having $\Lambda_2$ and $\bar\Lambda_2$ one can access $W$ that solves \eqref{gen:dzW} and \eqref{gen:bardzW} up to the following orders:   
\begin{equation}
   W=\sum_{n\geq 0}\eta^n\bar\eta\, W_{n,1}(Z,Y|x)+h.c.+\dots\,, 
\end{equation}
where the missing terms, denoted as $\dots$, require $\Lambda_3$ and higher, while $h.c.$ stands for the Hermitian conjugation. Consequently, having the connection $\Lambda$ only up to order $C^2$ gives us access to all-order vertices, such as $\mathcal{V}_{n,1}$ and $\Upsilon_{n,1}$, from \eqref{unfld}. Similarly, $\Lambda_3$ opens access to $\mathcal{V}_{n,2}$ and $\Upsilon_{n,2}$ for any $n$, and so forth. In general, the decomposition of $\Lambda$' in terms of the phase parameter acquires the following form:
\begin{equation}\label{Lambda:decomp}
    \Lambda=\sum_{i\geq 1,\, j\geq 0}\eta^i\bar\eta^j\Lambda_{i,j}\,,\qquad \bar\Lambda=\sum_{i\geq 0,\, j\geq 1}\eta^i\bar\eta^j\bar \Lambda_{i,j}\,,\quad \mathbf{\Lambda}:=\eta\Lambda_{1,0}\,,\quad \bar{\mathbf{\Lambda}}=\bar\eta\bar\Lambda_{0,1}\,,
\end{equation}
where $\Lambda_{i,j}$ and $\bar\Lambda_{i,j}$ are independent of $\eta$ and $\bar\eta$. From the definition \eqref{gen:dL}, where $\Lambda$ is proportional to $\eta$, and similarly, $\bar\Lambda$ is, at least, linear in $\bar\eta$, it  follows that
\begin{equation}
    \Lambda_{0,i}=\bar\Lambda_{i,0}=0\,.
\end{equation}
Additionally,    
\begin{equation}\label{chrl:decomp}
    \Lambda_{i,0}=\bar\Lambda_{0,i}=0\qquad\text{for}\qquad i>1\,,
\end{equation}
manifesting that the (anti)holomorphic sector is linearly exact for the connection $\Lambda$. Similarly, from \eqref{gen:dL} we have that
\begin{equation}\label{B:decomp}
    \mathcal{B}=\sum_{i,j\geq 0}\eta^i\bar\eta^j\,\mathcal{B}_{i,j}\,,\qquad \bar{\mathcal{B}}=\sum_{i,j\geq 0}\eta^i\bar\eta^j\,\bar{\mathcal{B}}_{i,j}\,,\quad \mathcal{B}_{0,0}=\bar{\mathcal{B}}_{0,0}:=C(Y|x)\,.
\end{equation}
The phase components of the fields $\mathcal{B}$'s are related to the underlying connections $\Lambda$'s through the following identification:
\begin{equation}
    \dr_z\Lambda_{l+1,j}=i\bar{\mathcal{B}}_{l,j}*\gga\,,\qquad \dr_{\bar z}\bar\Lambda_{l,j+1}=i\mathcal{B}_{l,j}*\bar\gga\,.
\end{equation}
Consequently, Eq. \eqref{chrl:decomp} entails that 
\begin{equation}\label{chrl B: decomp}
    \mathcal{B}_{0,i}=\bar{\mathcal{B}}_{i,0}=0\qquad\text{for}\qquad i>0\,.
\end{equation}

The chiral splitting in terms of variables $z$ and $\bar z$ of the fields $\mathcal{B}(z; Y)$, $\bar{\mathcal{B}}(\bar z; Y)$, \eqref{chrl:L}, and the initial data \eqref{B:init} and \eqref{L:init} that establish the perturbation series \eqref{BL:pert} from the self-dual sector singled out by \eqref{chrl:decomp} (equivalently by \eqref{chrl B: decomp}) leads us to refer to our approach as {\it chiral} perturbation theory. Its leading order is the (anti)self-dual vertices, while the subleading perturbations correspond to the degree two sectors penetrating each other. 

\paragraph{Vertices from chiral perturbations} The vertex reconstruction procedure relies on the order-by-order reconstruction of the connections $\Lambda$ and $\bar\Lambda$ in terms of the Weyl 0-form $C(Y|x)$. The order $O(C^{k+1})$ solution generates a jet of HS vertices $\mathcal{V}_{N,k}$, $\mathcal{V}_{k,N}$ in \eqref{unfld:w} and $\Upsilon_{N,k}$, $\Upsilon_{k,N}$ in \eqref{unfld:C} for all $N$ as follows:
\begin{itemize}
    \item {\bf Degree} $k=0$ 

    At this stage, the solution for $\Lambda$'s is given in \eqref{L:init} and, consequently, $\mathcal{B}=C$. It allow us to restore the components $W_{n,0}$ and $W_{0,n}$ via the recurrence equation \eqref{W:hom},
    \begin{equation}\label{W:rec}
        W_{n,0}(z; Y)=-\Delta\left(\{W_{n-1,0}\,, \Lambda_{0,1}\}_*\right)\,,\qquad W_{0,n}(\bar z; Y)=-\Delta\left(\{W_{0,n-1}\,,\Lambda_{0,1}\}_*\right)\,,
    \end{equation}
    using the initial data $W_{0,0}=\go(Y|x)$. Having these components, one recovers the corresponding vertices via\footnote{Setting $Z=0$ is convenient because this choice eliminates the contribution from $\dr_xW_{i,0}$, which is proportional to $z$ due to \eqref{W:hom}.} \eqref{gen:zero} and \eqref{gen:dB},
    \begin{subequations}
    \begin{align}
        &\mathcal{V}_{N,0}=-\eta^N\sum_{i=0}^N(W_{i,0}*W_{N-i,0})\Big|_{Z=0}\,,\qquad \mathcal{V}_{0,N}=-\mathcal{V}_{N,0}^\dagger\,,\\ 
        &\Upsilon_{N,0}=-\eta^N\left(W_{N,0}(z'; Y)*C-C*\pi W_{N,0}(z'; Y)\right)\Big|_{z'=-\y}\,,\quad \Upsilon_{0, N}^\dagger=\bar\pi \Upsilon_{N,0}\,. 
    \end{align} 
    \end{subequations}
    This way, the minimal-derivative (anti)holomorphic vertices were found in \cite{Didenko:2024zpd} for all $N$. 

    \item {\bf Degree} $k=1$ 
    
    The quadratic corrections to the connection $\Lambda$ emerge as solutions of \eqref{gen:L}. These are contained in $\Lambda_{1,1}$ and $\bar\Lambda_{1,1}$. Through \eqref{gen:dzW} and \eqref{gen:bardzW}, one determines the associated 1-forms $W_{n,1}$ and $W_{1,n}$, which, along with the already available $W_{n,0}$ and $W_{0,n}$, evaluate 
    \begin{equation}\label{ver:N1}
        \mathcal{V}_{N,1}=-\eta^N\bar\eta\left(\sum_{i=1}^{N} \dr_xW_{i, 1}+\sum_{i+j=N} \{W_{i,0}, W_{j, 1}\}_*\right)_{Z=0}\,,\qquad \mathcal{V}_{1,N}=-\mathcal{V}_{N,1}^{\dagger}\,.
    \end{equation}
    Note that although $W_{i,0}(0; Y)=0$, as follows from \eqref{W:rec}, this may not hold true for  $W_{i,1}(0; Y)$. The solution for $W_{i,1}$  is not necessarily provided by the standard contracting homotopy. This explains the presence of $\dr_x W_{i,1}$ in \eqref{ver:N1}.  

    The degree $k=1$ vertices $\Upsilon$ for 0-forms from \eqref{unfld:C} can be extracted from \eqref{gen:dB} or its conjugate. Specifically, from $\dr_{z}\Lambda_{1,1}=i\bar{\mathcal{B}}_{0,1}*\gga$, we read off field $\bar{\mathcal{B}}_{0,1}$, which contributes the following vertices:
    \begin{align}
        \Upsilon_{N,1}=-\eta^N\bar\eta\,\Big(\dr_x\bar{\mathcal{B}}_{0,1}+(W_{N,1}(z', \bar z; Y)*C-C*\pi W_{N,1}(z', \bar z; Y))\Big|_{z'=-y}+\\
        +\mathbf{h'}_y[W_{N,0}(-z'; Y)*(\bar{\mathcal{B}}_{y'})_{0,1}-(\bar{\mathcal{B}}_{y'})_{0,1}*\pi W_{N,0}(-z'; Y)]\Big)\,.\nn
    \end{align}
    The higher-order degree $k\geq2$ case applies analogously. 
\end{itemize}

It is important to stress that while the existence of any-order vertices is guaranteed in the regular case, based on Eqs. \eqref{vas}, this is not necessarily true in the irregular case. For the latter, the (in)existence of a vertex depends on whether \eqref{cons:BB} has a solution. In other words, the large star product \eqref{limst} imposes constraints on the space of functions for our master fields, resulting in a chiral form of the connection $\Lambda$, represented by \eqref{chrl:L}. This form may not be supported by Eq. \eqref{gen:L}. If the $b$-constraint \eqref{cons:BB} encounters an obstruction at a certain order, then the holomorphic vertices studied in \cite{Didenko:2022qga} cannot be lifted to a complete 4D theory. Although we do not know if \eqref{cons:BB} leads to an obstruction at any point, we will demonstrate that this is not the case for $\Lambda_2$ and $\Lambda_3$. This means that the self-dual HS theory proposed in \cite{Didenko:2022qga} admits a unitary completion, at least, up to the vertices $\mathcal{V}_{N,k}$, $\Upsilon_{N, k}$ from \eqref{unfld} and their conjugates with $N$ and $k$ interchanged, where $N$ is arbitrary and $k=1,2$. These vertices include the important quartic vertex, represented by $\Upsilon_{1,1}$, which exhibits a locality issue.

\subsection{Quadratic order}
In this section, we show that the expansion \eqref{BL:pert} is well defined to the second order in $C$, corresponding to the degree $k=1$. In addition, we explicitly find $\Lambda_2$ and $\bar\Lambda_2$ that solve \eqref{gen:L}. To this end, we start with the leading order fields \eqref{B:init} and \eqref{L:init}. Equation \eqref{gen:L} admits a solution, provided the $b$-constraint \eqref{cons:BB} is satisfied. In this order, this constraint reduces to \eqref{constr:CC}. Consequently, $\Lambda_2$ and $\bar\Lambda_2$ exist. In the quadratic approximation, Eq. \eqref{gen:L} reduces to 
\begin{equation}\label{L2:eq}
    \dr_{\bar z}\Lambda_2+\dr_z\bar\Lambda_2+\{\mathbf{\Lambda}, \bar{\mathbf{\Lambda}}\}_*=0\,,
\end{equation}
where, in accordance with \eqref{Lambda:decomp}, 
\begin{equation}
    \Lambda_2=\eta\bar\eta\,\Lambda_{1,1}\,,\qquad \bar\Lambda_2=\eta\bar\eta\,\bar\Lambda_{1,1}\,,\qquad \bar\Lambda_{1,1}=-\Lambda_{1,1}^{\dagger}\,.
\end{equation}
Let us highlight once again the difference with Vasiliev's perturbation expansion. In our case, the chiral decomposition implies $\Lambda_{2,0}=\bar\Lambda_{0,2}=0$.

We proceed using a convenient source prescription for the 0-form $C(Y|x)$; see \cite{Didenko:2018fgx}. Specifically, let us introduce the Taylor expansion, 
\begin{equation}\label{C:Taylor}
    C(y, \bar y|x)=e^{-iy\pp-i\bar y\bar \pp}C(y', \bar y'|x)\Big|_{y'=\bar y'=0}\,,\qquad \pp_{\al}=-i\frac{\p}{\p y^{'\al}}\,,\qquad \bar \pp_{\dal}=-i\frac{\p}{\p\bar{y}^{'\dal}}\,.
\end{equation}
We will omit, for brevity, the $C(Y')$ part by writing
\begin{equation}
    C(Y)\to e^{-iy\pp-i\bar y\bar \pp}\,.
\end{equation}
The bosonic condition \eqref{bosonic} implies the following important equivalence relation under simultaneous sign flipping $\pp\to-\pp$ and $\bar \pp\to-\bar \pp$:
\begin{equation}\label{equiv:bos}
   e^{-iy\pp-i\bar y\bar \pp}\sim e^{iy\pp+i\bar y\bar \pp}\,. 
\end{equation}
At higher orders, where several $C$' emerge, we will label $\pp_i$ with the index $i$ that counts each appearance of $C$ from the left. For example,
\begin{equation}
    C(y, \bar y)C(y, \bar y):=e^{-iy\pp_1-iy\pp_2-i\bar y\bar \pp_1-i\bar y\bar \pp_2}\,C(y'_1, \bar y'_1)C(y'_2, \bar y'_2)\Big|_{Y'_{1,2}=0}\to e^{-iy\pp_1-iy\pp_2-i\bar y\bar \pp_1-i\bar y\bar \pp_2}\,.
\end{equation}
With this prescription, Eq. \eqref{L:init} reads
\begin{subequations}\label{L:source}
    \begin{align}
        &\mathbf{\Lambda}=i\eta\,\dr z^{\al} z_{\al}\int_{0}^{1}d\tau\tau e^{i\tau z(y+\pp)-i\bar y \bar \pp}:=\eta\Lambda_{1,0}\,,\\
        &\bar{\mathbf{\Lambda}}=i\bar \eta\,\dr \bar z^{\dal} \bar z_{\dal}\int_{0}^{1}d\tau\tau e^{i\tau \bar z(\bar y+\bar \pp)-i y\pp}:=\bar\eta\bar\Lambda_{0,1}\,.
    \end{align}
\end{subequations}
Recalling now the convolution \eqref{convl}, we rewrite \eqref{L:source} as
\begin{equation}
    \mathbf{\Lambda}=i\eta\,\dr z^{\al} e^{-i\bar y \bar \pp}\Oast \int_{0}^{1}d\tau\tau z_{\al} e^{i\tau z(y+\pp)}\,,\qquad \bar{\mathbf{\Lambda}}=i\bar \eta\,\dr \bar z^{\dal} e^{-i y\pp}\Oast \int_{0}^{1}d\tau\tau \bar z_{\dal} e^{i\tau \bar z(\bar y+\bar \pp)}\,.
\end{equation}
Although the convolution $\Oast$ in the above formulas is merely a point-wise product, we introduce it because it will facilitate the subsequent star-product calculations. So, performing a simple integration with \eqref{limst}, we have
\begin{subequations}\label{L*L}
    \begin{align}
&\mathbf{\Lambda}*\bar{\mathbf{\Lambda}}=-\eta\bar\eta\,\dr z^{\al}\wedge \dr \bar z^{\dal}\,
e^{-iy\pp_2-i\bar y\bar \pp_1}\circledast\int_{[0,1]^2}d\tau
d\tau'\tau\tau' z_{\al}\bar z_{\dal}e^{i\tau z(y+\pp_1-\pp_2)+i\tau'
\bar z(\bar y+\bar \pp_1+\bar
\pp_2)}\,,\label{LbarL}\\
&\bar{\mathbf{\Lambda}}*\mathbf{\Lambda}=-\eta\bar\eta\,\dr\bar z^{\dal}\wedge\dr z^{\al}\,
e^{-iy\pp_1-i\bar y\bar \pp_2}\circledast\int_{[0,1]^2}d\tau
d\tau'\tau\tau' z_{\al}\bar z_{\dal}e^{i\tau z(y+\pp_1+\pp_2)+i\tau'
\bar z(\bar y+\bar \pp_1-\bar \pp_2)}\,.\label{barLL}
\end{align}
\end{subequations}
Notice that the integrands in \eqref{L*L} may appear different at first glance. However, they are actually equivalent when we apply the bosonic equivalence \eqref{equiv:bos}. Specifically, by changing $\pp_2\to-\pp_2$ and $\bar\pp_2\to-\bar\pp_2$ in \eqref{barLL}, we find that the integrands coincide and \begin{align}\label{comm:LbL}
    \{\mathbf{\Lambda}, \bar{\mathbf{\Lambda}}\}_*=-\eta\bar\eta\,\dr z^{\al}\wedge\dr \bar z^{\dal}\,
    \left(e^{-iy\pp_2-i\bar y\bar \pp_1}-e^{-iy\pp_1+i\bar y\bar
    \pp_2}\right)\circledast\\
    \int_{[0,1]^2}d\tau d\tau'\tau\tau'
    z_{\al}\bar z_{\dal}e^{i\tau z(y+\pp_1-\pp_2)+i\tau' \bar z(\bar
    y+\bar \pp_1+\bar \pp_2)}\,.\nonumber
\end{align}      
It is convenient to further reduce \eqref{comm:LbL} using the following identity:
\begin{equation}
    f(x)-f(y)=\int_{0}^{1}d\gs\frac{\p}{\p\gs}f(\gs x+(1-\gs)y)\,,
\end{equation}
which enables us to write 
\begin{equation}\label{LL:com}
    \{\mathbf{\Lambda}, \bar{\mathbf{\Lambda}}\}_*=-i\, \dr z^{\al}\wedge \dr\bar z^{\dal}\,[(y+\qq)^{\gb}\phi_{\gb}-(\bar y+\bar \qq)^{\dgb}\bar\phi_{\dgb}]\Oast\int_{[0,1]^2}d\tau d\tau' \tau\tau' z_{\al}\bar
    z_{\dal}e^{i\tau z(y+\qq)+i\tau' \bar z(\bar y+\bar \qq)}\,,
\end{equation}
where 
\begin{subequations}\label{def:phi}
    \begin{align}
    &\phi_{\al}=\eta\bar\eta\,y_{\al}\int_{0}^{1}d\gs e^{-i\gs (y\pp_2+\bar y\bar\pp_1)+i(1-\gs)(-y\pp_1+\bar y\bar \pp_{2})}+(y+\qq)_{\al}\psi(y, \bar y)\,,\\
    &\bar\phi_{\dal}=\eta\bar\eta\,\bar y_{\dal}\int_{0}^{1}d\gs e^{-i\gs (y\pp_2+\bar y\bar\pp_1)+i(1-\gs)(-y\pp_1+\bar y\bar \pp_{2})}+(\bar y+\bar\qq)_{\dal}\bar \psi(y, \bar y))
\end{align}
\end{subequations}
with $\psi(Y)$ and $\bar\psi(Y)$ being arbitrary functions that do not contribute to the anticommutator \eqref{LL:com} and 
\begin{equation}\label{q:quadratic}
    \qq_{\al}=(\pp_1-\pp_2)_{\al}\,,\qquad \bar \qq_{\dal}=(\bar\pp_1+\bar\pp_2)_{\dal}\,.
\end{equation}
Using the identity \eqref{App:dzC0}, we observe that \eqref{LL:com} represents a sum of $\dr_z$- and $\dr_{\bar z}$-exact pieces. This allows us to write down the solution to Eq. \eqref{L2:eq}:
\begin{subequations}\label{L2}
    \begin{align}
        &\Lambda_2=\dr z^{\al}\bar\phi^{\dal}\Oast\int_{[0,1]^2}d\tau d\tau' \tau(1-\tau')z_{\al}\bar z_{\dal}e^{i\tau z(y+\qq)+i\tau'\bar z(\bar y+\bar \qq)}+\dr_z\gep:=\eta\bar\eta\,\Lambda_{1,1}\,,\\
        &\bar\Lambda_2=\dr \bar z^{\dal}\phi^{\al}\Oast\int_{[0,1]^2}d\tau d\tau' \tau'(1-\tau)z_{\al}\bar z_{\dal}e^{i\tau z(y+\qq)+i\tau'\bar z(\bar y+\bar \qq)}+\dr_{\bar z}\gep:=\eta\bar\eta\,\bar\Lambda_{1,1}\,,
    \end{align}
    \end{subequations}
where $\gep(Z,Y|x)\in\mathbf{C}^{0,0}$ is an arbitrary function from the specified class, representing the gauge ambiguity.      
The obtained solution features freedom in two functions $\psi$ and $\bar\psi$. We fix this freedom by setting 
\begin{equation}\label{psi=0}
    \psi=\bar\psi=0\,.
\end{equation}
We will explain later why the choice \eqref{psi=0} is the proper one from the locality perspective. We also fix the gauge freedom as $\gep=0$. Notice that $\Lambda_2\in\mathcal{C}^{1,0}$ (see \eqref{C1:L}) and $\bar\Lambda_2\in\mathcal{C}^{0,1}$, as this should be. Hence, using \eqref{App:pre-proj}, we easily obtain that
\begin{equation}
    \dr_z\Lambda_2=i\eta\,\bar{\mathcal{B}}_2*\gga\,,\qquad \dr_{\bar z}\bar\Lambda_2=i\bar\eta\,\mathcal{B}_2*\bar\gga\,,
\end{equation}
where
\begin{subequations}\label{B2}
    \begin{align}
        &\mathcal{B}_2=\eta\, y^{\al}\int_{0}^{1}d\gs e^{-i\bar\pp_1\bar\pp_2-i\bar y(\bar \pp_1+\bar\pp_2)-iy(\gs\pp_2+(1-\gs)\pp_1)}\Oast\int_{0}^{1}d\tau (1-\tau)z_{\al}e^{i\tau z( y+\pp_1-\pp_2)}\,,\label{nobar:B2}\\
        &\bar{\mathcal{B}}_2=\bar\eta\,\bar y^{\dal}\int_{0}^{1}d\gs e^{i\pp_1\pp_2+iy(\pp_2-\pp_1)+i\bar y(-\gs\bar\pp_1+(1-\gs)\bar\pp_2)}\Oast\int_{0}^{1}d\tau (1-\tau)\bar z_{\dal}e^{i\tau\bar z(\bar y+\bar\pp_1+\bar\pp_2)}\,.\label{barB2}
    \end{align}
\end{subequations}
While the above result may seem to be in tension with the reality condition \eqref{real}, this is not the case. Indeed, the bosonic condition \eqref{equiv:bos} allows us to, for example, flip $\pp_2\to -\pp_2$ and $\bar \pp_2\to -\bar \pp_2$ in \eqref{barB2}, thereby demonstrating that $\mathcal{B}_2$ and $\bar{\mathcal{B}_2}$ are manifestly in involution. 

Let us now summarize the results of our analysis at the quadratic order.
\begin{itemize}
    \item The existence of solutions to Eq. \eqref{gen:L} at this order was possible for $C$ satisfying \eqref{cons:constr}. This means that the bosonic (anti)holomorphic vertices can be extended to include the mixed HS vertices of the form $\eta^N\bar\eta$ and $\bar\eta^N\eta$ for any integer $N$ that corresponds to the degree $k=1$ case. Indeed, having $\mathbf{\Lambda}$ and $\Lambda_2$ is sufficient to recover $W$ up to the order $\eta^N\bar\eta$ by solving Eq. \eqref{gen:dzW}. Similarly, with $\bar{\mathbf{\Lambda}}$ and $\bar\Lambda_2$, one restores $W$ up to the order $\bar\eta^N\eta$.

    \item The obtained solution $\eqref{L2}$ is not unique. It has a gauge redundancy  parameterized by the parameter $\gep$ and a freedom in two functions, $\psi(Y)$ and $\bar\psi(Y)$. The gauge parameter $\gep$ does not affect $\mathcal{B}_2$ and $\bar{\mathcal{B}_2}$ in \eqref{B2}, while the functions $\psi$ and $\bar\psi$ do have an impact. This freedom may be significant at higher orders. However, we choose to set $\psi =\bar\psi=0$ based on the following observation: the resulting solutions \eqref{B2} corresponding to this choice match, up to normalization, the Vasiliev field $B$ from \cite{Vasiliev:2017cae}, which reproduces the maximally local quadratic holomorphic vertices. This finding may seem surprising,\footnote{Recall that within the irregular approach, the field $\mathcal{B}=C$ does not receive higher-order corrections in the holomorphic sector.} considering that Vasiliev's $B_2$ pertains to the {\it holomorphic} sector, while our solutions in \eqref{B2} pertain to the {\it mixed} sector. This observed matching (checked in Appendix C) suggests intriguing dualities between the various sectors of the HS theory, which we hope to explore further in the future. From a technical standpoint, this matching could only be possible if the chiral factorization \eqref{chrl:L} occurs. Indeed, since $\mathbf{\Lambda}$, \eqref{L:init} is independent of $\bar z$, the action of the projector $\bar{\mathbf{h}}'$ in \eqref{gen:dzB} can be easily found, as the $z$-evolution equation \eqref{gen:dzB} for the field $\mathcal{B}_2$ becomes
    \begin{equation}\label{B2:z-evol}
        \dr_z\mathcal{B}_2+\mathbf{\Lambda}*C-C*\bar\pi_{\bar y}\mathbf{\Lambda}=0\,.
    \end{equation}
This equation resembles the one that arises in the second-order analysis of the holomorphic sector in Vasiliev's theory. It becomes identical (except for a normalization factor) when we consider that the star product $*$  defined in \eqref{limst} acts like the Vasiliev star product $\star$ from \eqref{star:Vasilev}, when one of the multipliers (in our case $C$) is independent of  $Z$.

    \item The result \eqref{B2} manifests a pure star-product form with respect to the variables $\bar y$ in \eqref{nobar:B2} and $y$ in \eqref{barB2}. For example, in \eqref{nobar:B2}, we have   
    \begin{equation}
        e^{-\bar\pp_1\bar\pp_2-i\bar y(\bar\pp_1+\bar\pp_2)}=e^{-i\bar y\bar\pp_1}*e^{-i\bar y\pp_2}\,.
    \end{equation}
    Using the definition \eqref{convl}, we can calculate $\mathcal{B}_2$, with the result being
    \begin{equation}\label{B2:final}
        \mathcal{B}_2=i\eta\,e^{izy}\int_{[0,1]^2}d\tau d\gs (1-\tau)\frac{\p}{\p \mathrm{y}^{\al}_q}\left(\mathrm{y}^{\al}_q\Phi(\mathrm{y}_q)\right)C(y_1', \bar y)\bar{*} C(y_2', \bar y)\Big|_{y_{1,2}'=0}\,,
    \end{equation}
    where
    \begin{equation}
        \Phi(y)=e^{iy(z-\gs\pp_2-(1-\gs)\pp_1)}\,,\qquad \mathrm{y}_q^{\al}=((1-\tau) y-\tau q)^{\al}
    \end{equation}
and we recall that we use the Taylor form defined in \eqref{C:Taylor}, while $\qq$ is given in \eqref{q:quadratic}, and the star product $\bar*$ acts only on $\bar y$ variables. 
\end{itemize}

\subsection{Cubic order} 
In the cubic order $O(C^3)$, related to the degree $k=2$, whether Eqs. \eqref{mixed:gen} produce consistent interactions in the mixed sector depends on whether the $b$-constraint \eqref{cons:BB} is satisfied. In the order $O(C^3)$, it takes the following form:
\begin{equation}\label{consis:CCC}
    \left(\mathcal{B}_2(z; Y)*\bar\pi C-C*\pi\mathcal{B}_2(z; Y)\right)\Big|_{z=-y}=0\,,
\end{equation}
In obtaining \eqref{consis:CCC}, we also used the fact that the contribution proportional to $\eta$ must be equal to zero independently\footnote{As a matter of principle, we should not dismiss the possibility that the constraint \eqref{cons:BB} may not be satisfied for any arbitrary value of $\eta$, but could hold true for a specific value instead. However, we will not explore this option in this discussion.} of the contribution proportional to $\bar\eta$. The two contributions are related by conjugation.

The Eq. \eqref{consis:CCC} is indeed valid, though its verification is not entirely straightforward. The cancellation of the two terms in \eqref{consis:CCC} arises from the integration process in \eqref{B2}. This integration can be conveniently expressed as an integration over a simplex; see \cite{Vasiliev:2017cae}. We will not perform this check here. Instead, we will solve \eqref{gen:L}, implying the validity of Eq. \eqref{consis:CCC}.

Let us look for solutions to Eq. \eqref{gen:L} in cubic order. In accordance with \eqref{Lambda:decomp}, there are two conjugate contributions, $\eta^2\bar\eta$ and $\bar\eta^2\eta$:
\begin{equation}
\Lambda_3=\eta^2\bar\eta\,\Lambda_{2,1}+\eta\bar\eta^2\,\Lambda_{1,2}\,,\qquad \bar\Lambda_3=\eta^2\bar\eta\,\bar\Lambda_{2,1}+\eta\bar\eta^2\,\bar\Lambda_{1,2}\,,
\end{equation}
with the conjugation 
\begin{equation}\label{L3:conj}
    \bar\Lambda_{2,1}=-\Lambda_{1,2}^{\dagger}\,,\qquad \bar\Lambda_{1,2}=-\Lambda_{2,1}^{\dagger}\,.
\end{equation}
Zooming in on, for example, the $\eta^2\bar\eta$ piece, we have
\begin{equation}\label{L21:eq}
    \dr_{\bar z}\Lambda_{2,1}+\dr_{z}\bar\Lambda_{2,1}+\{\Lambda_{1,0}, \bar\Lambda_{1,1}\}_*=0\,,
\end{equation}
where the lower-order contributions to $\Lambda$ are presented in \eqref{L:source} and \eqref{L2}. The analysis of the equation above is similar to the quadratic order discussed earlier. Specifically, we need to explicitly demonstrate how the commutator $\{\Lambda_{1,0}, \bar\Lambda_{1,1}\}_*$ can be expressed as a combination of $\dr_z$- and $\dr_{\bar z}$-exact pieces. These pieces will then be linked to the desired solutions for $\Lambda_{2,1}$ and $\bar\Lambda_{2,1}$ up to gauge ambiguity. The necessary computation is quite straightforward, and we will provide it in Appendix D, while presenting the final result here:
\begin{subequations}\label{L3}
    \begin{align}
        &\bar\Lambda_{2,1}=-i\dr\bar z^{\dal}\int_{[0,1]^2}d\tau d\tau' \int_{\Delta_{\rho,\gs}}\,
y^{\al} y^{\gb}\phi(y, \bar y)\Oast\tau(1-\tau)\tau'z_{\al}z_{\gb}\bar{z}_{\dal} e^{i\tau z(y+\qq)+i\tau'\bar z(\bar y+\bar \qq')}\,,\\
    &\Lambda_{2,1}=-i\dr z^{\al}\int_{[0,1]^2}d\tau d\tau'\int_{\Delta_{\rho, \gs}}\,
(\bar y+\bar\xi)^{\dal} y^{\gb}\phi(y, \bar y)\Oast\tau^2(1-\tau')z_{\al}z_{\gb}\bar{z}_{\dal} e^{i\tau z(y+\qq)+i\tau'\bar z(\bar y+\bar \qq')}\,,
\end{align}
\end{subequations}
where the integration goes over a square formed by the variables $\tau$, $\tau'$, and a compact domain represented by $\rho_i$ and $\gs$ 
\begin{equation}\label{domain:rho-gs}
    \Delta_{\rho, \gs}=\{\rho_1+\rho_2+\rho_3=1\,,\quad 0\leq\gs\leq 1\,,\quad 0\leq\rho_i\leq 1\,,\quad i=1,2,3\}\,.
\end{equation}
The rest of the ingredients of \eqref{L3} are
\begin{equation}\label{def:phi2}
    \phi(y,\bar y)=\frac{\rho_2}{(\rho_1+\rho_2)(\rho_2+\rho_3)} e^{i\gs\big[y\mathcal{P}_y+(\bar y+\bar
\pp_1)\bar{\mathcal{P}}_{\bar
y}\big]+i(1-\gs)\big[y\mathcal{P}'_y+(\bar y+\bar
\pp_3)\bar{\mathcal{P}}'_{\bar y}\big]}\,,
\end{equation}
where
\begin{subequations}
    \begin{align}
      &\mathcal{P}_y=-\frac{\rho_2}{\rho_2+\rho_3}\pp_3-\frac{\rho_3}{\rho_2+\rho_3}\pp_2\,,\qquad
\bar{\mathcal{P}}_{\bar y}=-\bar
\pp_1-\frac{\rho_2}{\rho_2+\rho_3}\bar\pp_2+\frac{\rho_3}{\rho_2+\rho_3}\bar \pp_3\,,\\
&\mathcal{P}'_y=-\frac{\rho_1}{\rho_1+\rho_2}\pp_2-\frac{\rho_2}{\rho_1+\rho_2}\pp_1\,,\qquad
\bar{\mathcal{P}}'_{\bar y}=\bar
\pp_3-\frac{\rho_1}{\rho_1+\rho_2}\bar
\pp_1+\frac{\rho_2}{\rho_1+\rho_2}\bar \pp_2\,,
    \end{align}
\end{subequations}
and
\begin{subequations}\label{def:q-xi}
    \begin{align}
       &\qq=(\rho_2+\rho_3)\pp_1+(\rho_1-\rho_3)\pp_2-(\rho_1+\rho_2)\pp_3\,,\label{q:def}\\
        &\bar \qq'=\bar \pp_1+\bar \pp_2+\bar \pp_3\,,\label{q':def}\\
        &\bar\xi=\rho_1 \bar
\pp_1-\rho_2\bar \pp_2+\rho_3\bar \pp_3\,.\label{xi:def}
    \end{align}
\end{subequations}
The missing contribution from $\Lambda_{1,2}$ and $\bar\Lambda_{1,2}$ that we have not calculated is recovered from \eqref{L3:conj}. 

Field $\mathcal{B}$ in the cubic order arises from \eqref{chrl:L} as
\begin{equation}
    \dr_z\Lambda_3=i\eta\bar{\mathcal{B}_3}(\bar z; Y)*\gga\,,\qquad \dr_{\bar z}\bar\Lambda_3=i\bar\eta{\mathcal{B}_3}(z; Y)*\bar\gga\,,
\end{equation}
where 
\begin{equation}\label{B3:split}
    \mathcal{B}_3=\eta\bar\eta\,\mathcal{B}_{1,1}+\eta^2\mathcal{B}_{2,0}\,,\qquad \bar{\mathcal{B}}_3=\eta\bar\eta\,\bar{\mathcal{B}}_{1,1}+\bar\eta^2\bar{\mathcal{B}}_{0,2}\,.
\end{equation}
This way, we find that 
\begin{equation}
    \dr_{\bar z}\bar\Lambda_{2,1}=i\mathcal{B}_{2,0}(z; Y)*\bar\gga\,,\qquad \dr_{z}\Lambda_{2,1}=i\bar{\mathcal{B}}_{1,1}(\bar z; Y)*\gga\,.
\end{equation}
Using identities \eqref{App:pre-proj-cc} and \eqref{App:zz_ident}, one arrives at 
\begin{subequations}\label{B3:cond}
\begin{align}
   &\mathcal{B}_{2,0}(z; Y)=-\int_{\Delta_{ \rho,\gs}} y^{\al}y^{\gb}\phi(y, -\bar\qq')\,e^{-i\bar y\bar\qq'}\moast\int_{0}^{1}d\tau \tau(1-\tau)\,z_{\al}z_{\gb} e^{i\tau z(y+\qq)}\,,\label{B20}\\
   &\bar{\mathcal{B}}_{1,1}(\bar z; Y)=-i\int_{\Delta_{ \rho,\gs}}(\bar y+\bar\xi)^{\dal}\frac{\p}{\p\qq^{\al}}\left(\qq^{\al}\phi(-\qq, \bar y)e^{-iy\qq}\right)\bar\moast\int_{0}^{1}d\tau'(1-\tau')\bar z_{\dal}e^{i\tau' \bar z(\bar y+\bar\qq')}\,.\label{B11}
\end{align}
\end{subequations}
Substituting \eqref{q:def} and \eqref{q':def} into \eqref{B20}, we find that its $\bar y$ and $\bar\pp_i$ dependence composes itself into
\begin{equation}
    e^{-i\bar y(\bar\pp_1+\bar\pp_2+\bar\pp_3)-i(\bar\pp_1\bar\pp_2+\bar\pp_1\bar\pp_3+\bar\pp_2\bar\pp_3)}\,,
\end{equation}
which represents a pure star product $\bar *$ of the three $C$'s:  
\begin{align}\label{B20:res}
    \mathcal{B}_{2,0}(z; Y)=&e^{izy}\int_{\Delta_{ \rho,\gs}}d^3\rho\,d\gs\int_{0}^{1}d\tau\,\tau(1-\tau)\frac{\p^2}{\p \mathrm{y}^{\al}_q \p\mathrm{y}^{\gb}_q}\left(\mathrm{y}^{\al}_q \mathrm{y}^{\gb}_q\Phi(\mathrm{y}_q|\gs, \rho_i)\right)\times\nn\\
    &\times C(y_1', \bar y)\bar *C(y_2', \bar y)\bar *C(y_3', \bar y)\Big|_{y_i'=0}\,,
\end{align}
where 
\begin{equation}
  \Phi(y|\gs, \rho_i)=\frac{\rho_2}{(\rho_1+\rho_2)(\rho_2+\rho_3)}e^{iy(z+\gs\mathcal{P}_y+(1-\gs)\mathcal{P}_y')}\,,\quad\mathrm{y}^{\al}_q=((1-\tau)y-\tau\qq)^{\al}\,.  
\end{equation}
We leave the detailed form of the cubic order field $\mathcal{B}$ to Appendix D. Let us now draw a line here and provide a short summary: 
\begin{itemize}
    \item The existence of the connection $\Lambda$ to this order \eqref{L3} means that the system \eqref{mixed:gen}, at least, reproduces vertices $O(\eta^N\bar\eta^k)$ and their conjugate for any $N$ and $k=0,1,2$. Consequently, the previously found (anti)holomorphic vertices remain a part of the full theory for the degree $k\leq 2$.  

    \item Similar to the quadratic analysis, when solving for \eqref{L21:eq}, we encountered gauge ambiguity that does not affect the fields in \eqref{B3:split}. Additionally, there is a degree of freedom in choosing a single analytic function for each component of $\mathcal{B}$. We set these functions to zero when we reached \eqref{B3:cond}. Notably, this choice of Eq. \eqref{B20} coincides with Vasiliev's field $B_3$ from the cubic analysis from \cite{Didenko:2020bxd}. While Vasiliev's field $B_3$ was derived during the analysis of the holomorphic sector, our result originates from the mixed sector. This connection suggests certain structural dualities, which will be explored in the next section. Furthermore, it is important to highlight that the cubic $b$-constraint \eqref{consis:CCC} is directly applicable in the context of Vasiliev's theory. This is because the star product used in \eqref{consis:CCC} behaves exactly like the original Vasiliev star product \eqref{star:Vasilev} when applied to products involving $Z$-independent function.
    
    \item Equation \eqref{gen:dzB} for $\mathcal{B}_{2,0}$ simplifies at this order because $\mathcal{B}_{1,0}$ and $\Lambda_{1,0}$ are independent of $\bar z$, leading to\footnote{ 
    It is important to note that the $z$-evolution equation for $\mathcal{B}_{2,0}$ differs from the evolution of Vasiliev's $B_3$ from \cite{Didenko:2020bxd}. However, the final results coincide. This discrepancy can be explained as follows: our result \eqref{B20:res} is exact, while the result in \cite{Didenko:2020bxd} is approximate and misses contributions that vanish at the level of the final vertices. By neglecting these terms in the Vasiliev case, we effectively arrive at Eq. \eqref{B20:eq}, which, among other things, assumes a different star product.} 
    \begin{equation}\label{B20:eq}
        \dr_{z}\mathcal{B}_{2,0}+\Lambda_{1,0}*\mathcal{B}_{1,0}-\mathcal{B}_{1,0}*\bar\pi\Lambda_{1,0}=0\,.
    \end{equation}
    Analogously, the evolution of $\mathcal{B}_{1,1}$ determined from \eqref{gen:dzB} reads
    \begin{equation}
        \dr_z{\mathcal{B}}_{1,1}+\left(\Lambda_{1,1}(z, \bar z'; Y)*C-C*\bar\pi\Lambda_{1,1}(z, \bar z'; Y)\right)\Big|_{\bar z'=-\bar y}=0\,.
    \end{equation}
    \item One can assess the degree of (non)locality introduced by the fields $\mathcal{B}_{2,0}$ and $\mathcal{B}_{1,1}$ in HS vertices by focusing on the exponential contractions $\pp_i\pp_j$ and $\bar\pp_i\bar\pp_j$ contained in $\phi(y, -\bar\qq')$ and $\phi(-q, \bar y)$ as shown in Eqs. \eqref{B3:cond}. This way, one observes that $\mathcal{B}_{2,0}$ contains only $\bar\pp_i\bar\pp_j$ contractions (see \eqref{B20:res}), which means its contribution is spin-local according to the definition provided in \cite{Gelfond:2018vmi}. In contrast, $\mathcal{B}_{1,1}$ does not exhibit spin-locality, as the relevant contractions from \eqref{B11} take the following form:
    \begin{equation}
        \exp{\left(i\sum_{i<j}^3a_{ij}\pp_i\pp_j+i\sum_{i<j}^3a_{ij}'\bar\pp_i\bar\pp_j\right)}\,,
    \end{equation}
    where from \eqref{App:B11-fin} we have 
    \begin{subequations}\label{aij:def}
        \begin{align}
            &a_{12}=\rho_3\,,\qquad a_{13}=\rho_2\,,\qquad a_{23}=\rho_1\,,\\
            &a_{12}'=-\frac{\gs\rho_2}{\rho_2+\rho_3}\,,\quad a_{13}'=\frac{\gs\rho_3}{\rho_2+\rho_3}+\frac{(1-\gs)\rho_1}{\rho_1+\rho_2}\,,\quad a_{23}'=-\frac{(1-\gs)\rho_2}{\rho_1+\rho_2}\,.
        \end{align}
    \end{subequations}
    Although $\mathcal{B}_{1,1}$ appears nonlocal, the extent of its nonlocality is constrained by a condition introduced by Gelfond in \cite{Gelfond:2023fwe}, referred to as {\it moderate} nonlocality:
    \begin{equation}\label{mod-non}
        |a_{ij}|+|a_{ij}'|\leq 1\,,\qquad 1\leq i\leq j\leq 3\,.
    \end{equation}
Since $a_{12}'\leq\rho_2$, $a_{13}'\leq\rho_3+\rho_1$, $a_{23}'\leq\rho_2$ and $\rho_1+\rho_2+\rho_3\leq 1$, the coefficients in \eqref{aij:def} satisfy the inequality \eqref{mod-non}, aligning with the nonlocal behavior observed in \cite{Gelfond:2023fwe}. 
\end{itemize}

\section{Structure dualities}\label{sec:dual}
An interesting finding from the analysis of the {\it mixed} sector at quadratic and cubic orders is the emergence of the fields $\mathcal{B}_{1,0}$ and $\mathcal{B}_{2,0}$, which align with the perturbative $B$-module from the {\it holomorphic} sector of the original Vasiliev equations corresponding to minimal-derivative interactions. This alignment was not expected, except perhaps for $\mathcal{B}_{1,0}$. In this case, the match is due to the $z$-evolution equation \eqref{B2:z-evol}, which is identical to that from Vasiliev's system. It is important to emphasize that our approach does not require any higher-order corrections to $\mathcal{B}$ in the (anti)holomorphic sector, where $\eta\bar\eta=0$. Consequently, we expect some algebraic relations at the level of HS vertices. These relations arise from the following structure of the perturbative expansion of $\mathcal{B}$:
    \begin{align}\label{B0i:eq}
        \mathcal{B}_{0,i}=\bar{\mathcal{B}}_{i,0}=\begin{cases}
        0 \,,\quad i\geq 1\,, \\
        C\,,\quad i=0\,.
    \end{cases}
    \end{align}
Let us illustrate them with an example of $O(C^2)$ vertices. From \eqref{gen:dB}, we have that 
\begin{equation}\label{0form:CC}
    \dr_xC+\go*C-C*\pi (\go)=\eta\Upsilon_{1,0}(\go, C, C)+\bar\eta\Upsilon_{0,1}(\go, C, C)\,,
\end{equation}
where 
\begin{subequations}
\begin{align}
    &\Upsilon_{1,0}=-W_{1,0}*C+C*\bar\pi W_{1,0}-\dr_x\mathcal{B}_{1,0}-\go*\mathcal{B}_{1,0}+\mathcal{B}_{1,0}*\bar\pi(\go)\,,\label{wCC:ver}\\
    &\Upsilon_{0,1}=-\left(W_{0,1}(\bar z'; Y)*C-C*\bar\pi W_{0,1}\right)\Big|_{\bar z'=-\bar y}\,.
\end{align}
\end{subequations}
In deriving this result, we used that $\mathcal{B}_{0,1}=\bar{\mathcal{B}}_{1,0}=0$ and also evaluated the action of projectors $\mathbf{h}'$ in \eqref{idnt:proj} using that $W_{1,0}(z; Y)$ is independent of $\bar z$, while $W_{0,1}(\bar z; Y)$ is independent of $z$, as it follows from \eqref{gen:dzW} and \eqref{gen:bardzW}, specified to the first order in $C$. Applying the involution responsible for the reality conditions to \eqref{wCC:ver} and using that
\begin{equation}
    \Upsilon_{1,0}^{\dagger}=\pi\Upsilon_{0,1}\qquad \mathcal{B}_{1,0}^{\dagger}=\pi\bar{\mathcal{B}}_{0,1}\,,\qquad W_{1,0}^{\dagger}=-W_{0,1}\,,\qquad C^{\dagger}=\pi C\,, 
\end{equation}
we arrive at the following $O(C^2)$ identity:
\begin{equation}\label{ident:C2}
    W_{0,1}*C-C*\pi W_{0,1}+\dr_x\bar{\mathcal{B}}_{0,1}+\go*\bar{\mathcal{B}}_{0,1}-\bar{\mathcal{B}}_{0,1}*\pi(\go)=\left(W_{0,1}(\bar z'; Y)*C-C*\bar\pi W_{0,1}\right)\Big|_{\bar z'=-\bar y}\,.
\end{equation}
A similar analysis can be carried out in the cubic order for the purely (anti)holomorphic vertices. To proceed, we use that 
\begin{equation}
    \mathcal{B}_{0,1}=\mathcal{B}_{0,2}=\bar{\mathcal{B}}_{1,0}=\bar{\mathcal{B}}_{2,0}=0\,,\qquad W_{2,0}=W_{2,0}(z; Y)\,,\qquad W_{0,2}=W_{0,2}(\bar z; Y)\,.
\end{equation}
This leads us to the following final result for the conjugate vertices:
\begin{subequations}
    \begin{align}
        &\Upsilon_{2,0}(\go, C^3)=-\dr_x\mathcal{B}_{1,0}-\dr_x\mathcal{B}_{2,0}-W_{2,0}*C+C*\bar\pi W_{2,0}-\label{ver:C3}\\
        &\qquad\quad -W_{1,0}*\mathcal{B}_{1,0}+\mathcal{B}_{1,0}*\bar\pi W_{1,0}-\go*\mathcal{B}_{2,0}+\mathcal{B}_{2,0}*\bar\pi(\go)\nonumber\,,\\
        &\Upsilon_{0,2}(\go, C^3)=-\left(W_{0,2}(\bar z'; Y)*C-C*\bar\pi W_{0,2}\right)\Big|_{\bar z'=-\bar y}\,.
    \end{align}
\end{subequations}
Notice that while $\mathcal{B}_{1,0}$ is quadratic in $C$, its contribution from $\dr_x \mathcal{B}_{1,0}$ to the $O(C^3)$ vertex is nonzero and comes from \eqref{0form:CC}. Now, from the conjugation $\Upsilon_{2,0}^{\dagger}=\pi\Upsilon_{0,2}$, we obtain \begin{align}
    &\dr_x\mathcal{B}_{1,0}+\dr_x\mathcal{B}_{2,0}+W_{2,0}*C-C*\bar\pi W_{2,0}+\label{ident:C3}\\
        &+W_{1,0}*\mathcal{B}_{1,0}-\mathcal{B}_{1,0}*\bar\pi W_{1,0}+\go*\mathcal{B}_{2,0}-\mathcal{B}_{2,0}*\bar\pi(\go)=
        \left(W_{2,0}(z'; Y)*C-C*\pi W_{2,0}\right)\Big|_{z'=-y}\,,\nn
\end{align} 
where in $\dr_x\mathcal{B}_{1,0}$ and $\dr_x\mathcal{B}_{2,0}$ only the contribution to $\eta^3$-vertices is taken into account. 

The purely holomorphic {\it structure relations} \eqref{ident:C2} and \eqref{ident:C3} generalize to any order, provided the $b$-constraint \eqref{cons:BB} holds. In this case, it is straightforward to arrive at the following order-$n$ condition: 
\begin{align}\label{structure:hol}
    \sum_{i=1}^n\dr_x\mathcal{B}_{i,0}+\sum_{i=0}^n(W_{i,0}*\mathcal{B}_{n-i,0}-\mathcal{B}_{n-i,0}*\bar\pi W_{i,0})=\left(W_{n,0}(z'; Y)*C-C*\pi W_{n,0}\right)\Big|_{z'=-y}\,,
\end{align}
where, again, in $\dr_x\mathcal{B}_{i,0}$ we only keep $\eta^n$ contributions. Analogous relations emerge in the mixed sector, relating two representations for the 0-form vertices $O(\eta^m\bar\eta^n)$ and $O(\eta^n\bar\eta^m)$ via involution $\dagger$.

The identities in Eq. \eqref{structure:hol} significantly simplify the evaluation of (anti)holomorphic vertices. The right-hand side of this equation illustrates how all the 0-form holomorphic vertices $\Upsilon_{n,0}(\go, C^n)$ were explicitly derived in \cite{Didenko:2024zpd}. In contrast, the left-hand side reflects the typical form of a vertex used in Vasiliev's framework; see, e.g., \cite{Gelfond:2021two}.
We anticipate that Eq. \eqref{ident:C3} will help establish the equivalence\footnote{At the quadratic level $O(C^2)$, Eq. \eqref{wCC:ver} concisely reproduces the corresponding vertex in the Vasiliev approach. Moreover, Eq. \eqref{ident:C2} can be proven within Vasiliev's framework using a specific quadratic structure relation identified in \cite{Didenko:2019xzz} (see Eq. (6.28) in this reference).} between the holomorphic $C^3$ vertex identified in \cite{Gelfond:2021two} and the one presented in \cite{Didenko:2024zpd}. More broadly, Eq. \eqref{structure:hol} is expected to connect the traditional Vasiliev approach with the recent developments from \cite{Didenko:2022qga}. This is particularly important in the context of the locality issue. Indeed, the all-order locality of the holomorphic sector observed in \cite{Didenko:2022qga} may suggest that the locality is reached through the particular large algebra \eqref{limst:hol} with the commuting $[z_{\al}, z_{\gb}]_*=0$. In contrast, the noncommutativity, as in the regular Vasiliev's framework \eqref{star:Vasilev}, could be a contributing factor to nonlocality. However, this does not hold true for a few orders in the holomorphic sector within the Vasiliev approach, even though establishing a local setup for this framework remains quite challenging.

\section{Discussion and conclusions}\label{sec:conc}
The main goal of this paper was to investigate the potential interaction between the holomorphic and antiholomorphic higher spin sectors recently proposed in \cite{Didenko:2022qga}. The respective vertices that exhibit the minimal number of derivatives were identified in \cite{Didenko:2024zpd}, in line with the requirement from \cite{Vasiliev:2022med} for projectively-compact locality. If such a cross interaction exists, then it would unify the two nonunitary sectors into the full 4D HS theory.
However, there is a challenge: while the (anti)self-dual interaction described in \cite{Didenko:2022qga} is expected to form a closed sector within Vasiliev's theory, this has not yet been proven for higher orders. As a result, it is possible that the two sectors may not have a unitary completion and could instead demonstrate some freedom within self-dual theories. It is important to remind that no such freedom occurs at the quadratic order, as the corresponding vertices at this level are consistent with those of Vasiliev \cite{Vasiliev:2016xui}. Furthermore, there is evidence (see \cite{Didenko:2022eso}) suggesting that the cubic vertex evaluated in \cite{Gelfond:2021two} should also align with the findings of \cite{Didenko:2024zpd}. Consequently, our goal was to check whether the holomorphic system of \cite{Didenko:2022qga} admits well-defined mixed interactions. 

To incorporate the mixed interacting sector, we introduced the so-called irregular generating equations, which generalize the purely (anti)holomorphic system. This has become possible due to a key generalization \eqref{idnt:proj} of the projector identities from \cite{Didenko:2022qga}, which opens the way to the mixed interactions. The term {\it irregular} means that this system does not provide access to some of the standard consistency constraints due to the specific (irregular) star product \eqref{limst} used in the equations. This contrasts with the standard (regular) star product \eqref{star:Vasilev} employed in the Vasiliev framework. Somewhat surprisingly, the irregular star product has effectively emerged as a contraction of the original Vasiliev large algebra during the evaluation of lower-order interaction vertices, which are constrained by locality \cite{Didenko:2019xzz}. As a result, it was natural to use it in the context of HS locality.  

While our equations resemble the standard Vasiliev form, they differ in several important ways. Below, we present a central part of both systems that defines the evolution of the auxiliary connections $\Lambda$ and $\bar\Lambda$, which are responsible for HS vertices:

\begin{subequations}\label{conc:eq}
    \begin{align}
        &\text{\bf Regular (Vasiliev):}\,&&\text{\bf Irregular:}\nn \\
        \nn\\
        &\dr_z\Lambda+\Lambda\star\Lambda=i\eta\,B\star\gamma && \dr_z\Lambda=i\eta\,\bar{\mathcal{B}}(\bar z; Y)*\gga\label{conc:eq1}\\
        &\dr_{\bar z}\bar\Lambda+\bar\Lambda\star\bar\Lambda=i\bar\eta\,B\star\bar\gga && \dr_{\bar z}\bar\Lambda=i\bar\eta\,\mathcal{B}(z; Y)*\bar\gga\label{conc:eq2}\\
        &\dr_{z}\bar\Lambda+\dr_{\bar z}\Lambda+\{\Lambda, \bar\Lambda\}_{\star}=0 &&\dr_{z}\bar\Lambda+\dr_{\bar z}\Lambda+\{\Lambda, \bar\Lambda\}_{*}=0\,.\label{conc:eq3} 
    \end{align}
\end{subequations}
The only common equation between the two systems is Eq. \eqref{conc:eq3}, while the others differ. The system on the left is clearly consistent, provided that the proper $Z$-evolution of the field $B$ is considered. In contrast, the irregular system is consistent under the following conditions: (i) the star product is defined as in \eqref{limst}, (ii) the connections $\Lambda$ and $\bar\Lambda$ belong to the appropriate functional class, which, along with the star product $*$, generates the projector identity \eqref{idnt:proj}, and (iii) the fields $\mathcal{B}$ and $\bar{\mathcal{B}}$ are chiral and satisfy the bilinear algebraic constraint \eqref{cons:BB}. 

In the holomorphic case, where $\bar\eta=0$, the irregular system admits a simple linear in $C$ solution, 
\begin{equation}
    \bar\Lambda^{\text{hol}}=0\,,\qquad \bar{\mathcal{B}}^{\text{hol}}(\bar z; Y|x)=C(Y|x)\,,
\end{equation}
which restores the {\it whole} holomorphic sector through a linear in $C$ connection $\Lambda$. It is also important to note that, unlike the system on the left of \eqref{conc:eq}, which has a single 0-form module $B(Z;Y)$, the irregular case features two chiral modules: 
\begin{equation}
    \mathcal{B}(z; Y)=C(Y|x)+\mathcal{B}_2(z; Y)[C,C]+\dots\,,\quad \bar{\mathcal{B}}(\bar z; Y)=C(Y|x)+\bar{\mathcal{B}}_2(\bar z; Y)[C,C]+\dots\,.
\end{equation}
This chirality is enforced by a very restrictive functional class associated with the (anti)holomor-\\phic interaction. It imposes a strict limitation on the consistent interactions. Due to this chiral splitting of the 0-form sector, and due to the fact that the linearized approximation of the irregular system \eqref{conc:eq} embraces the complete (anti)holomorphic sector, we refer to the perturbation theory for the proposed irregular system as {\it chiral}. Its expansion parameter, referred to as {\it degree}, represents the extent to which the two dual sectors interpenetrate. In addition, the proposed scheme guaranties that the (anti)holomorphic vertices emerge maximally local in all orders.

We have examined equations in \eqref{conc:eq} by considering perturbations for general values of $\eta$ and $\bar\eta$ at both leading and next-to-leading orders. Our analysis revealed the existence of a solution, which we have explicitly identified. This finding indicates that there are consistent interaction vertices of the order $O(\eta^N\bar\eta^{k=0,1,2})$ and their conjugates for all values of $N$ and the degree $k\leq 2$. Consequently, the necessary $b$-constraint \eqref{consis:CCC} turns out to be nontrivially satisfied. The results of our analysis include the following observations:
\begin{itemize}
    \item In the quadratic order, the consistency for the mixed interactions implied by the $b$-constraint \eqref{cons:BB} eliminates fermions from the spectrum.\footnote{Fermions can be included in the usual way by supersymmetrizing the HS algebra with the additional so-called extra Klein operators $k$ and $\bar k$ \cite{Vasiliev:1992av}. This makes the field spectrum doubly degenerate, containing each field in two copies. Other supersymmetric extensions can also be considered along the lines of \cite{Engquist:2002vr, Engquist:2002gy}. For clarity, we have restricted ourselves to the non-supersymmetric case.} This outcome was, of course, expected. However, it is of interest that this limitation arises from the quadratic condition for the Weyl module \eqref{constr:CC}. 
    
    \item In the cubic order, the consistency boils down to \eqref{consis:CCC}. It represents a cubic star-product relation on the Weyl module $C$ and its minimal-derivative $C^2$-correction encoded in the field $\mathcal{B}$. The validity of this relation is significant and highly nontrivial. Furthermore, it directly applies to the standard Vasiliev case. Unfortunately, we lack insight into why this condition is satisfied {\it a priori}, and we are uncertain whether the higher-order $b$-constraints will also hold. However, while solving for the connections $\Lambda$' in a few orders, we encountered not only gauge ambiguity but also a persistent ambiguity in a single function each time. This allowed us to maintain control over consistency in the subsequent order. Let us highlight that the data obtained at this point opens access to crucial cubic vertices $O(C^3)$ (quartic at the level of action).   

    \item The structure of chiral perturbation theory is unique in that its linear approximation captures all-order (anti)self-dual interactions. From a technical perspective, this results in a specific configuration for the module $\mathcal{B}$, as shown in Eq. \eqref{B0i:eq}. This configuration suggests a set of dual representations for HS vertices \eqref{structure:hol}. These relations significantly simplify the otherwise involved process of evaluating holomorphic vertices. More importantly, we believe they will help connect the holomorphic system proposed in \cite{Didenko:2022qga} with the standard framework of Vasiliev. Indeed, in our analysis of the perturbative equations, we found a solution for the module $\mathcal{B}$, whose components match the previously known (modulo irrelevant terms) holomorphic 0-form module from Vasiliev's theory.   
\end{itemize}
We anticipate that the proposed approach will not encounter any obstructions at higher orders. This expectation is grounded in the fact that the generating system \eqref{mixed:gen} consistently produces cubic and quartic vertices according to Lagrangian counting. Moreover, it generates consistent all-order vertices for degrees $k < 3$.

Assuming consistency, our formalism can be effectively applied to the issue of higher-spin locality. Our construction is benchmarked against the maximally local (anti)holomorphic sector of the theory, which imposes strict constraints on the mixed sector and limits the freedom in field redefinition for potentially problematic vertices. While we did not tackle this issue in our paper, we plan to address it in the future. A preliminary examination of the structure of nonlocal derivative contractions at degree $k = 2$  indicates that they are not arbitrary; rather, they are constrained by a moderate nonlocality inequality \eqref{mod-non}, which aligns with the proposal from \cite{Gelfond:2023fwe}.

As a different but related direction for future research, it would be interesting to see whether our approach could be applied to the Coxeter higher-spin models proposed in \cite{Vasiliev:2018zer}. In particular, it is of interest to see if there is an analog of the holomorphic sector in these models. While the linearized analysis of such models has been initiated in \cite{Tarusov:2025sre}, little is known about the interactions. We hope that the Coxeter higher-spin theories may provide a framework for the irregular formulation that effectively addresses the locality issue.

\section*{Acknowledgments}

I would like to thank A. Korybut, M. Povarnin,  A. Tarusov, and K. Ushakov for the discussion, comments on the draft, and valuable suggestions. 
This work is supported by the Russian Science Foundation Grant No. 26-12-00240, {\texttt https://rscf.ru/project/26-12-00240/}.

\newcounter{appendix}
\setcounter{appendix}{1}
\renewcommand{\theequation}{\Alph{appendix}.\arabic{equation}}
\addtocounter{section}{1} \setcounter{equation}{0}
 \renewcommand{\thesection}{\Alph{appendix}.}

\addcontentsline{toc}{section}{\,\,\,\,\,\,\,A. Convolution relations}

\section*{A. Convolution relations}\label{app:A}

There are two important identities involving total derivatives that are central to our approach: the 1-form and 0-form relations. These can be expressed as follows:
\begin{subequations}
    \begin{align}
       &\dr_z\left(\dr z^{\al}\int_{0}^{1}d\tau\phi(y)\moast \tau z_{\al}e^{i\tau\,z(y+q)}\right)=\phi(-q)e^{-iy q}*\gga\,, \label{App:pre-proj}\\
       &\dr_z\left(\int_{0}^{1}d\tau \phi^{\al}(y)\moast(1-\tau) z_{\al}e^{i\tau\,z(y+q)}\right)=i\,\dr z^{\al}\int_{0}^{1}d\tau\,(y+q)^{\gb}\phi_{\gb}(y)\moast\tau z_{\al}e^{i\tau\,z(y+q)}\,,\label{App:dzC0}
    \end{align}
\end{subequations}
where $\phi(y)$ and $\phi^{\al}(y)$ are analytic, 
while $q$ is a spinor parameter that is independent of both $y$ and $z$. The 2-form $\gga$ is defined in \eqref{gamma}. Similar identities hold in the antiholomorphic case, where $(y,z)\to(\bar y, \bar z)$ and $\moast\to\bar\moast$:
\begin{subequations}
    \begin{align}
       &\dr_{\bar z}\left(\dr \bar z^{\dal}\int_{0}^{1}d\tau'\phi(y)\bar\moast \tau' \bar z_{\dal}e^{i\tau'\,\bar z(\bar y+\bar q)}\right)=\phi(-\bar q)e^{-i\bar y \bar q}*\bar\gga\,, \label{App:pre-proj-cc}\\
       &\dr_{\bar z}\left(\int_{0}^{1}d\tau\phi^{\dal}(\bar y)\bar\moast(1-\tau') \bar z_{\dal}e^{i\tau'\,\bar z(\bar y+\bar q)}\right)=i\,\dr \bar z^{\dal}\int_{0}^{1}d\tau'\,(\bar y+\bar q)^{\dgb}\phi_{\dgb}(\bar y)\bar\moast\tau' \bar z_{\dal}e^{i\tau'\,\bar z(\bar y+\bar q)}\,.\label{App:dzC0-cc}
    \end{align}
\end{subequations}
Differentiating with respect to $q$, we obtain a generalization of Eq. \eqref{App:pre-proj} for any polynomial in $z$ prefactor, e.g.,
\begin{equation}\label{App:zz_ident}
    \dr_z\left(\dr z^{\al}\int_{0}^{1}d\tau\phi^{\gb}(y)\moast \tau^2 z_{\al}z_{\gb}e^{i\tau\,z(y+q)}\right)=-i\frac{\p}{\p q^{\gb}}\left(\phi^{\gb}(-q)e^{-iy q}\right)*\gga\,.
\end{equation}

It is important to note that the differential operator $\dr_z$ acts on the function from the space $\mathcal{C}^1$ in Eq. \eqref{App:pre-proj}, and on the function from the space $\mathbf{C}^0$ in Eq. \eqref{App:dzC0}. Similar identities apply in the conjugate sector, where the variables $(z, y)$ are replaced by $(\bar z, \bar y)$.

The relation \eqref{App:pre-proj} underlies the projector identities \eqref{idnt:proj}, while \eqref{App:dzC0} is useful for solving for the connections $\Lambda$ and $\bar\Lambda$ from \eqref{gen:L}. The proofs of both relations rely on the two-component Schouten identity and integration by parts. To illustrate this, let us prove \eqref{App:pre-proj}. We start with:
\begin{align}
    \frac{\p}{\p z^{\al}}\int_{0}^{1}d\tau\,\phi(y)\moast\tau z^{\al}e^{i\tau\,z(y+q)}=\int_{0}^{1}d\tau\,\phi(y)\moast (2\tau+i\tau^2\,z(y+q))e^{i\tau\,z(y+q)}=\nn\\
    =\int_{0}^{1}d\tau\,\phi(y)\moast (2\tau+\tau^2 \p_\tau)e^{i\tau\,z(y+q)}=\int_{0}^{1}d\tau\,\p_{\tau}\left(\tau^2\,\phi(y)\moast e^{i\tau\,z(y+q)}\right)=\label{App:dL-proof}\\
    =\phi(y)\moast e^{iz(y+q)}=\phi(-q)e^{iz(y+q)}=\phi(-q)e^{-iyq}*e^{izy}\,,\nn
\end{align}
where we made use of \eqref{convl:proj} in the last line. Now, the Schouten identity implies
\begin{equation}
    \dr_z(\dr z^{\al}A_{\al})=\frac{1}{2}\dr z_{\al}\wedge\dr z^{\al}\,\frac{\p}{\p z^{\gb}}A^{\gb}
\end{equation}
and we arrive at Eq. \eqref{App:pre-proj}. 

Another frequently used simple identity reads
\begin{equation}\label{App:frq-ident}
    \phi(y)\moast e^{i\tau\,z(y+q)}=e^{izy}\left(\phi(y_q)e^{-izy_q}\right)\,,\qquad y_q:=(1-\tau)y-\tau q\,.
\end{equation}

\renewcommand{\theequation}{\Alph{appendix}.\arabic{equation}}
\addtocounter{appendix}{1} \setcounter{equation}{0}
\addtocounter{section}{1}
\addcontentsline{toc}{section}{\,\,\,\,\, B. Projector identities}

\section*{B. Projector identities}\label{app:B}

In this appendix, we prove the identities from \eqref{idnt:proj}. They apply either to holomorphic or antiholomorphic functional classes. We will consider the holomorphic case generated by variables $z$ and $y$, leaving the unnecessary dependence on $\bar z$ and $\bar y$ implicit. With that being said, we need to demonstrate that for  
\begin{equation}
    \Lambda[\phi_1]=\dr z^{\al}\int_{0}^{1}d\tau\tau\int_{u,v}(z-v)_{\al}\phi_1(\tau(z-v), (1-\tau)y+u)e^{i\tau\, zy+iuv}\in\mathcal{C}^{1}\,,
\end{equation}
and 
\begin{equation}\label{App:f}
    f[\phi_0]=\int_{0}^{1}d\tau\frac{1-\tau}{\tau}\int_{u, v}\phi_0(\tau(z-v), (1-\tau)y+u)e^{i\tau\,zy+iuv}\in\mathbf{C}^{0}
\end{equation}
the identities in \eqref{idnt:proj} hold. Functions $\phi_{1,0}(z,y)$ are required to be analytic and $\phi_0(0,y)=0$ in order for the integration over $\tau$ in \eqref{App:f} to make sense. It is convenient to proceed using the source prescription and the convolution representation \eqref{C:circle} for classes $\mathbf{C}^{0}$ and $\mathbf{C}^{1}$. Specifically, 
\begin{align}
    &f(z,y)=\int_{0}^{1}d\tau\frac{1-\tau}{\tau}e^{iyA_1}\moast e^{i\tau\,z(y+B_1)}\phi_0(z', y')\Big|_{z'=y'=0}\,,\\
    &\Lambda(z,y)=\int_{0}^{1}d\tau e^{iyA_2}\moast \tau z_{\al}e^{i\tau\,z(y+B_2)}\phi_1(z'', y'')\Big|_{z''=y''=0}\,,
\end{align}
where 
\begin{equation}
    A_{1\al}=i\frac{\p}{\p y'^{\al}}\,,\quad
    B_{1\al}=i\frac{\p}{\p z'^{\al}}\,,\quad A_{2\al}=i\frac{\p}{\p y''^{\al}}\,,\quad
    B_{2\al}=i\frac{\p}{\p z''^{\al}}\,.
\end{equation}
Using \eqref{C0*C1}, we find that 
\begin{equation}
    f*\Lambda=\int_{[0,1]^2}d\tau d\gs \frac{1-\gs}{\gs}\left(e^{iyA_1}*e^{iyA_2}\right)\moast\tau z_{\al}e^{i\tau\,z(y+B_{1,2})}\phi_0(z', y')\phi_1(z'', y'')\Big|_{z'=z''=y'=y''=0}\,,
\end{equation}
where 
\begin{equation}\label{App:B12-coef}
    B_{1,2}=\gs(B_1+A_2)+(1-\gs)(B_2-A_1)\,.
\end{equation}
Now, from \eqref{App:pre-proj}, we have 
\begin{align}
    &\dr_z(f*\Lambda)=\int_{0}^{1}d\gs \frac{1-\gs}{\gs}\,e^{-iyB_{1,2}-iB_{1,2}(A_1+A_2)+iA_2A_1}\phi_0(z', y')\phi_1(z'', y'')\Big|_{z'=z''=y'=y''=0}*\gga=\\
    &=\int_{0}^{1}d\tau \frac{1-\tau}{\tau}\,e^{-iy(\tau(B_1+A_2)+(1-\tau)(B_2-A_1))-i(\tau B_1+(1-\tau)B_2)(A_1+A_2)}\phi_0(z', y')\phi_1(z'', y'')\Big|_{z'=z''=y'=y''=0}*\gga\,,\nn
\end{align}
where, in the last line, we renamed the integration variable $\gs\to\tau$ and substituted \eqref{App:B12-coef}. Now, we observe that the exponential above acquires the following representation:
\begin{align}\label{App:fLeq}
    e^{-iy(\tau(B_1+A_2)+(1-\tau)(B_2-A_1))-i(\tau B_1+(1-\tau)B_2)(A_1+A_2)}=\\
    =\left(e^{i\tau\tilde{z}(y+B_1)+i(1-\tau)yA_1+i\tau A_1B_1}*e^{-iB_2A_2-iyB_2}\right)\Big|_{\tilde{z}=-y+A_2}\nn\,,
\end{align}
where $\tilde{z}$ emerges as a spinor parameter in the star product, which is evaluated with respect to the variable $y$. Using the definition of the projector \eqref{proj}, the latter expression can be further massaged into 
\begin{align}
    =\left(e^{i\tau\tilde{z}(y+B_1)+i(1-\tau)yA_1+i\tau A_1B_1}*e^{-iB_2A_2-iyB_2}\right)\Big|_{\tilde{z}=-y+A_2}=\\
    \mathbf{\tilde{h}}_y\left[e^{-i\tau \tilde{z}(y+B_1)+i(1-\tau)yA_1+i\tau A_1B_1}*\mathbf{h}_y[e^{-izB_2+i(y+\tilde{y})A_2}]\right]\,.\nn
\end{align}
Eventually, using \eqref{C0:source} and that 
\begin{equation}
    e^{-izB_2+i(y+\tilde{y})A_2}\phi_1(z'', y'')\Big|_{z''=y''=0}=\phi_1(-z, y+\tilde{y})\,,
\end{equation}
gives us the  desired identity \eqref{proj:fL}
\begin{equation}
    \dr_z(f*\Lambda)=\mathbf{\tilde{h}}_y[f(-\tilde{z}, y)*\mathbf{h}_y\phi_1(-z, y+\tilde{y})]\,.
\end{equation}
The proof of \eqref{proj:Lf} is analogous.

\renewcommand{\theequation}{\Alph{appendix}.\arabic{equation}}
\addtocounter{appendix}{1} \setcounter{equation}{0}
\addtocounter{section}{1}
\addcontentsline{toc}{section}{\,\,\,\,\, C. Vasiliev's field $B$ in the second order}

\section*{C. Vasiliev's field $B$ in the second order}\label{app:C}

Here, we demonstrate that the second-order field $\mathcal{B}$ presented in \eqref{B2:final}, up to a normalization, coincides with the Vasiliev field $B_2$ from \cite{Vasiliev:2017cae} (see Eq. (5.23) in this reference), corresponding to the minimal-derivative holomorphic HS interactions. 

Ignoring the $\bar y$ dependence that emerges via the star-products of $\bar y$, the source representation for Vasiliev's $B_2$ in our normalization and convention is given by
\begin{equation}\label{B2:vas}
    B_2=\eta\int d_+^3\rho\,[-izy\,\gd(x)+\gd'(x)]e^{-iy(\rho_1\pp_1+\rho_2\pp_2+\rho_3z)+i\rho_3(z-\pp_1)(\pp_1-\pp_2)}\,,
\end{equation}
where the integration goes over $0\leq\rho_i\leq 1$, which, thanks to the presence of the delta-function $\gd(x)$ with $x=1-\rho_1-\rho_2-\rho_3$, implies integration over a simplex
\begin{equation}
    \Delta_{\rho}=\{\rho_1+\rho_2+\rho_3=1\,,\quad 0\leq\rho_i\leq 1\}
\end{equation}
and $\gd'(x)=\frac{\p}{\p x}\gd(x)$ is a derivative of the delta-function. Let us compare Eq. \eqref{B2:vas} with $\mathcal{B}_2$ from \eqref{B2:final}. Again, by ignoring the $\bar y$-part in Eq. \eqref{B2:final}, we have
\begin{equation}\label{App:B2}
    \mathcal{B}_2=\eta\, e^{izy}\int_{[0,1]^2}d\tau d\gs\,(1-\tau)\left(2+\mathrm{y}^{\al}_{q}\frac{\p}{\p \mathrm{y}^{\al}_{q}}\right)\Phi(\mathrm{y}_q)\,,
\end{equation}
where
\begin{equation}
        \Phi(y)=e^{iy(z-\gs\pp_2-(1-\gs)\pp_1)}\,,\qquad \mathrm{y}_q^{\al}=((1-\tau) y-\tau (\pp_1-\pp_2))^{\al}\,.
    \end{equation}
Substituting this into \eqref{App:B2} and introducing the new integration variables,
\begin{subequations}
    \begin{align}
        &\rho_1=(1-\tau)(1-\gs)\,,\\
        &\rho_2=\gs(1-\tau)\,,\qquad \rho_1+\rho_2+\rho_3=1\,,\qquad |J|=(1-\rho_3)^{-1}\,,\\
        &\rho_3=\tau\,,
    \end{align}
\end{subequations}
Eq. \eqref{App:B2} takes the following form:
\begin{equation}
    \mathcal{B}_2=\eta\int d_+^3\rho\,\gd(x)\left(2-izy+\sum_{i=1}^{3}\rho_i\frac{\p}{\p\rho_i}\right)e^{-iy(\rho_1\pp_1+\rho_2\pp_2+\rho_3z)+i\rho_3(z-\pp_1)(\pp_1-\pp_2)}\,.
\end{equation}
Using now that
\begin{equation}
    \gd'(x)=-\frac{\p}{\p\rho_1}\gd(x)=-\frac{\p}{\p\rho_2}\gd(x)=-\frac{\p}{\p\rho_1}\gd(x)\,,
\end{equation}
and, by integrating by parts, we arrive at 
\begin{equation}
    \mathcal{B}_2=B_2\,.
\end{equation}
Consequently, we conclude that our $\mathcal{B}_2$ identified in the analysis of the quadratic mixed sector coincides with Vasiliev's master field $B_2$ responsible for the minimal quadratic interaction within the holomorphic sector.

\renewcommand{\theequation}{\Alph{appendix}.\arabic{equation}}
\addtocounter{appendix}{1} \setcounter{equation}{0}
\addtocounter{section}{1}
\addcontentsline{toc}{section}{\,\,\,\,\, D. Cubic analysis}

\section*{D. Cubic analysis}\label{app:D}

In this appendix, we solve Eq. \eqref{L21:eq}
\begin{equation}\label{L21:App}
    \dr_{\bar z}\Lambda_{2,1}+\dr_{z}\bar\Lambda_{2,1}+\{\Lambda_{1,0}, \bar\Lambda_{1,1}\}_*=0\,,
\end{equation}
with (see Eqs. \eqref{L:source} and \eqref{L2})
\begin{subequations}
    \begin{align}
        &\Lambda_{1,0}=i\dr z^{\al} e^{-i\bar y \bar \pp}\Oast \int_{0}^{1}d\tau\tau z_{\al} e^{i\tau z(y+\pp)}\,,\label{App:L1}\\
        &\bar\Lambda_{1,1}=\dr \bar z^{\dal}y^{\al}\phi(Y)\Oast\int_{[0,1]^2}d\tau d\tau' \tau'(1-\tau)z_{\al}\bar z_{\dal}e^{i\tau z(y+\qq)+i\tau'\bar z(\bar y+\bar \qq)}\,,
    \end{align}
\end{subequations}
and
\begin{subequations}\label{App:fAq}
\begin{align}
    &\phi(y,\bar y)=\int_{0}^{1}d\gs e^{iyA+i\bar y\bar A'}\,,\\
    &\qq_{\al}=(\pp_1-\pp_2)_{\al}\,,\qquad \bar\qq_{\dal}=(\bar\pp_1+\bar\pp_2)_{\dal}\,,\\
    &A_{\al}=-(\gs \pp_2+(1-\gs)\pp_1)_{\al}\,,\quad \bar A'_{\dal}=(-\gs\bar\pp_1+i(1-\gs)\bar \pp_{2})_{\dal}\,.
\end{align}
\end{subequations}
Using \eqref{prod}, we find that 
\begin{equation}\label{App:eq1}
    \Lambda_{1,0}*\bar\Lambda_{1,1}=i\dr z^{\al}\wedge\dr\bar z^{\dal}\int_{0}^{1}d\rho\rho\,y^{\gb}\phi(y, \bar y+\bar\pp_1)e^{-i\bar y\bar\pp_1}\Oast\int_{[0,1]^2}d\tau d\tau'\,\tau'\tau^2\,z_{\al}z_{\gb}\bar z_{\dal}e^{i\tau z(y+\qq_{3})+i\tau'\bar z(\bar y+\bar{\qq}_{3}')}\,,
\end{equation}
where 
\begin{subequations}
    \begin{align}
        &\qq_3=\rho
\pp_1+(1-\rho-\rho(1-\gs))\pp_2-(\rho\gs+1-\rho)\pp_3\,,\\
&\bar\qq'_{3}=\bar \pp_1+\bar \pp_2+\bar \pp_3
    \end{align}
\end{subequations}
or in a manifest form,
\begin{align}\label{L1*barL2}
\Lambda_{1,0}*\bar\Lambda_{1,1}=i\dr z^{\al}\wedge\dr\bar z^{\dal}y^{\gb}\int_{[0,1]^2}d\rho d\gs\,
\rho e^{iy(-\gs \pp_3-(1-\gs)\pp_2)+i\bar y(-\bar \pp_1-\gs\bar
\pp_2+(1-\gs)\bar \pp_3)+i\bar \pp_1(-\gs\bar \pp_2+(1-\gs)\bar
\pp_3)}\circledast\nn\\
\circledast\int_{[0,1]^2}d\tau d\tau'\,\tau^2\tau' z_{\al}z_{\gb}\bar{z}_{\dal}e^{i\tau
z(y+\rho
\pp_1+(1-\rho-\rho(1-\gs))\pp_2-(\rho\gs+1-\rho)\pp_3)+i\tau'\bar z(\bar
y+\bar \pp_1+\bar \pp_2+\bar \pp_3)}\,.
\end{align}
Recall that when performing the above calculation, we need to consider how the resulting expression in \eqref{App:eq1} acts on the $CCC$. Specifically, the differential operators in $\Lambda_{1,0}$ will act on the first $C$ from the left, while those in $\bar\Lambda_{1,1}$ will act on the last two $C$s. As a result, we must make the following substitutions: set $\pp \to \pp_1$ in \eqref{App:L1} and update $\pp_1 \to \pp_2$ and $\pp_2 \to \pp_3$ in \eqref{App:fAq}.

Similarly, we find the opposite product:
\begin{align}\label{barL2*L1}
\bar\Lambda_{1,1}*\Lambda_{1,0}=&-i\dr z^{\al}\wedge\dr\bar z^{\dal}y^{\gb}\int_{[0,1]^2}d\rho d\gs\,\rho e^{iy(-\gs \pp_2-(1-\gs)\pp_1)+i\bar y(-\bar \pp_3-\gs\bar
\pp_1+(1-\gs)\bar \pp_2)-i\bar \pp_3(-\gs\bar \pp_1+(1-\gs)\bar
\pp_2)}\circledast\nn\\
&\circledast\int_{[0,1]^2}d\tau d\tau'\,\tau^2\tau' z_{\al}z_{\gb}\bar{z}_{\dal}e^{i\tau
z(y+(1-\gs\rho) \pp_1+(\rho\gs-1+\rho)\pp_2+\rho \pp_3)+i\tau'\bar
z(\bar y+\bar \pp_1+\bar \pp_2-\bar \pp_3)}\,.
\end{align}
Consider now \eqref{L1*barL2} and make the following change of
integration variables that introduces the simplex $\Delta_\rho$: 
\begin{align}
&\rho_1:=1-\rho\,,\\
&\rho_2:=\gs\rho\quad\Rightarrow\qquad \sum\rho_i=1\,,\quad
0\leq\rho_i\leq1\,,\quad \gs=\ff{\rho_2}{\rho_2+\rho_3}\,,\\
&\rho_3:=\rho(1-\gs)\,.
\end{align}
This results in the following form of \eqref{L1*barL2}:
\begin{align}\label{L1*barL2:new}
\Lambda_{1,0}*\bar\Lambda_{1,1}=i\dr z^{\al}\wedge\dr\bar z^{\dal}y^{\gb}\int_{\Delta_\rho}d^3\rho
e^{iy\mathcal{P}_y+i(\bar y+\bar
\pp_1)\bar{\mathcal{P}}_{\bar y}}\circledast \int_{[0,1]^2}d\tau d\tau'\,\tau^2\tau'
z_{\al}z_{\gb}\bar z_{\dal}e^{i\tau z(y+\qq)+i\tau' \bar z(\bar
y+\bar \qq')}\,,
\end{align}
where
\begin{align}
\mathcal{P}_y=-\frac{\rho_2}{\rho_2+\rho_3}\pp_3-\frac{\rho_3}{\rho_2+\rho_3}\pp_2\,,\qquad
\bar{\mathcal{P}}_{\bar y}=-\bar
\pp_1-\frac{\rho_2}{\rho_2+\rho_3}\bar\pp_2+\frac{\rho_3}{\rho_2+\rho_3}\bar \pp_3\,,
\end{align}
and
\be\label{q}
\qq=(\rho_2+\rho_3)\pp_1+(\rho_1-\rho_3)\pp_2-(\rho_1+\rho_2)\pp_3\,,\qquad
\bar \qq'=\bar \pp_1+\bar \pp_2+\bar \pp_3\,.
\ee
Now we proceed with Eq. \eqref{barL2*L1} in such a way as to make the right-hand sides of the convolutions in \eqref{L1*barL2} and \eqref{barL2*L1} coincide. This turns out to be feasible in two steps. First, we 
simultaneously replace\footnote{We can do this because of the bosonic condition $C(y,-\bar y)=C(-y,\bar y)$.} in \eqref{barL2*L1} 
\be
\pp_3\to-\pp_3\,,\qquad \bar \pp_3\to -\bar \pp_{3}
\ee
and then introduce the new simplex integration variables in \eqref{barL2*L1},
\begin{align}
&\rho_1:=\gs\rho\,,\\
&\rho_2:=(1-\gs)\rho\quad\Rightarrow\qquad \sum\rho_i=1\,,\quad
0\leq\rho_i\leq1\,,\quad \gs=\ff{\rho_1}{\rho_1+\rho_2}\,,\\
&\rho_3:=1-\rho\,.
\end{align}
This gives us
\begin{align}\label{barL2*L1:new}
\bar\Lambda_{1,1}*\Lambda_{1,0}=&-i\dr z^{\al}\wedge\dr\bar z^{\dal}y^{\gb}\int e^{iy\mathcal{P}'_y+i(\bar y+\bar
\pp_3)\bar{\mathcal{P}}'_{\bar y}}\circledast \int_{[0,1]^2}\tau^2\tau'
z_{\al}z_{\gb}\bar z_{\dal}e^{i\tau z(y+\qq)+i\tau' \bar z(\bar
y+\bar \qq')}\,,
\end{align}
with
\begin{align}
\mathcal{P}'_y=-\frac{\rho_1}{\rho_1+\rho_2}\pp_2-\frac{\rho_2}{\rho_1+\rho_2}\pp_1\,,\qquad
\bar{\mathcal{P}}'_{\bar y}=\bar
\pp_3-\frac{\rho_1}{\rho_1+\rho_2}\bar
\pp_1+\frac{\rho_2}{\rho_1+\rho_2}\bar \pp_2\,.
\end{align}
It is important to note that because the fields are bosonic and we are integrating over a simplex, there is a match between the right-hand side's $\Oast$-factors from \eqref{L1*barL2:new} and \eqref{barL2*L1:new}.
Combining now \eqref{L1*barL2:new} and \eqref{barL2*L1:new} via an additional integration over $\gs\in [0,1]$, we have
\begin{align}\label{comm:L1L2}
&\Lambda_{1,0}*\bar\Lambda_{1,1}+\bar\Lambda_{1,1}*\Lambda_{1,0}=
i\dr z^{\al}\wedge\dr\bar z^{\dal}\times\nn\\
&\times y^{\gb}\int_{0}^{1}d\gs\, \frac{\p}{\p\gs}e^{i\gs\big[y\mathcal{P}_y+(\bar
y+\bar \pp_1)\bar{\mathcal{P}}_{\bar
y}\big]+i(1-\gs)\big[y\mathcal{P}'_y+(\bar y+\bar
\pp_3)\bar{\mathcal{P}}'_{\bar y}\big]}\circledast \int_{[0,1]^2}\tau^2\tau'
z_{\al}z_{\gb}\bar z_{\dal}e^{i\tau z(y+\qq)+i\tau' \bar z(\bar
y+\bar \qq')}\,.
\end{align}
To proceed, the following identities are useful:
\begin{subequations}\label{App:idnt}
    \begin{align}
&\mathcal{P}_y-\mathcal{P}'_y=\frac{\rho_2}{(\rho_1+\rho_2)(\rho_2+\rho_3)}\qq\,,\\
&\bar{\mathcal{P}}_{\bar y}-\bar{\mathcal{P}}'_{\bar
y}=\frac{\rho_2}{(\rho_1+\rho_2)(\rho_2+\rho_3)}(\rho_1 \bar
\pp_1-\rho_2\bar \pp_2+\rho_3\bar \pp_3-\bar \qq')\,,\\
&\bar \pp_1\bar{\mathcal{P}}_{\bar y}-\bar
\pp_3\bar{\mathcal{P}}'_{\bar
y}=\frac{\rho_2}{(\rho_1+\rho_2)(\rho_2+\rho_3)}\bar \qq'(\rho_1
\bar \pp_1-\rho_2\bar \pp_2+\rho_3\bar \pp_3)\,.
\end{align}
\end{subequations}
Now, differentiation gives us a prefactor
\be
\frac{\p}{\p\gs}\to iy(\mathcal{P}_y-\mathcal{P}'_y)+i\bar
y(\bar{\mathcal{P}}_{\bar y}-\bar{\mathcal{P}}'_{\bar y})+i(\bar
\pp_1\bar{\mathcal{P}}_{\bar y}-\bar \pp_3\bar{\mathcal{P}}'_{\bar
y})\,.
\ee
Using the identities \eqref{App:idnt}, we have
\begin{align}\label{[L1,L2]}
&\Lambda_{1,0}*\bar\Lambda_{1,1}+\bar\Lambda_{1,1}*\Lambda_{1,0}=
\dr z^{\al}\wedge\dr\bar z^{\dal}\times\\
&\times y^{\gb}\int_{\gs, \Delta_\rho}
\frac{\rho_2}{(\rho_1+\rho_2)(\rho_2+\rho_3)}\big[
\overbrace{(y+\qq)y}^{I}+\overbrace{(\bar y+\bar \qq')(-\bar
y-\rho_1\bar p_1+\rho_2\bar p_2-\rho_3\bar p_3)}^{II}
\big]\nn\\
&e^{i\gs\big[y\mathcal{P}_y+i(\bar y+\bar
\pp_1)\bar{\mathcal{P}}_{\bar
y}\big]+i(1-\gs)\big[y\mathcal{P}'_y+i(\bar y+\bar
\pp_3)\bar{\mathcal{P}}'_{\bar y}\big]}\circledast\int_{[0,1]^2}\tau^2\tau'
z_{\al}z_{\gb}\bar z_{\dal}e^{i\tau z(y+\qq)+i\tau' \bar z(\bar
y+\bar \qq')}\,.\nn
\end{align}
The two pieces $I$ and $II$ compose total derivatives,
$\dr_z$ and $\dr_{\bar z}$, respectively. Indeed, using
\eqref{App:dzC0}, for any analytic function $\phi^{\al\gb}(y)$, one finds
\be\label{App:dz-quadr}
\dr_z\int_{0}^{1}d\tau\,
\phi^{\al\gb}(y)\circledast\tau(1-\tau)z_{\al}z_{\gb}e^{i\tau
z(y+q)}=i\dr z^{\al}\int_{0}^{1}d\tau\,
(y+q)^{\gga}\phi_{\gga}{}^{\gb}(y)\circledast
\tau^2z_{\al}z_{\gb}e^{i\tau z(y+q)}\,.
\ee
and analogously, for any $\phi^{\dal}(\bar y)$, we have 
\be\label{App:dz-lin}
\dr_{\bar z}\int_{0}^{1}d\tau'\, \phi^{\dal}(\bar y)\circledast (1-\tau')\bar
z_{\dal}e^{i\tau'\bar z(\bar y+\bar q)}=i\dr\bar z^{\dal}\int_{0}^{1}d\tau'\,
(\bar y+\bar q)^{\dgb}\phi_{\dgb}(\bar y)\circledast\tau'\bar
z_{\dal}e^{i\tau'\bar z(\bar y+\bar q)}\,.
\ee
Notice how from \eqref{App:dz-quadr} and \eqref{App:dz-lin} it follows that 
\begin{equation}
    \phi_{\al\gb}\sim\phi_{\al\gb}+(y+q)_{\al}(y+q)_{\gb}\psi(Y)\,,\qquad \phi_{\dal}\sim\phi_{\dal}+(\bar y+\bar q)_{\dal}\bar\psi(Y)\,,
\end{equation}
where $\psi$ and $\bar\psi$ are arbitrary functions. 
Having this, Eq. \eqref{[L1,L2]} reduces to
\be
\Lambda_{1,0}*\bar\Lambda_{1,1}+\bar\Lambda_{1,1}*\Lambda_{1,0}=-\dr_{z}\bar\Lambda_{2,1}-\dr_{\bar
z}\Lambda_{2,1}\,,
\ee
where
\begin{align}
&\bar\Lambda_{2,1}=-i\,\dr\bar z^{\dal}\int_{\Delta_{\rho, \gs}}\frac{\rho_2}{(\rho_1+\rho_2)(\rho_2+\rho_3)}
(y^{\al}y^{\gb}+(y+\qq)^{\al}(y+\qq)^\gb\psi(Y))\times\\
&\times e^{i\gs\big[y\mathcal{P}_y+(\bar y+\bar
\pp_1)\bar{\mathcal{P}}_{\bar
y}\big]+i(1-\gs)\big[y\mathcal{P}'_y+(\bar y+\bar
\pp_3)\bar{\mathcal{P}}'_{\bar y}\big]}\circledast\tau(1-\tau)\tau'
z_{\al}z_{\gb}\bar z_{\dal}e^{i\tau z(y+\qq)+i\tau' \bar z(\bar
y+\bar \qq')}+\dr_{\bar z}\gep\,.\nn
\end{align}
Here $\dr_{\bar z}\gep\in\mathcal{C}^{0,1}$, and an otherwise arbitrary function representing gauge ambiguity, while $\psi(Y)$ is an arbitrary function encoding field redefinition. While the choice of these functions might be important at higher orders, in this paper, we choose to  set 
\begin{equation}
\gep=\psi=0\label{App:gep-psi}\,.    
\end{equation}
The $II$-piece yields 
\begin{align}
&\Lambda^{(2,1)}=-i\dr z^{\al}\int_{\Delta_{\rho, \gs}}\frac{\rho_2}{(\rho_1+\rho_2)(\rho_2+\rho_3)}
\big[(\bar y+\rho_1 \bar
\pp_1-\rho_2\bar \pp_2+\rho_3\bar \pp_3)^{\dal}+(\bar y+\bar \qq')^{\dal}\bar\psi(Y)\big]y^{\gb}\times\\
&\times e^{i\gs\big[y\mathcal{P}_y+(\bar y+\bar
\pp_1)\bar{\mathcal{P}}_{\bar
y}\big]+i(1-\gs)\big[y\mathcal{P}'_y+(\bar y+\bar \pp_3)\bar{\mathcal{P}}'_{\bar y}\big]}\circledast\tau^2(1-\tau')
z_{\al}z_{\gb}\bar z_{\dal}e^{i\tau z(y+\qq)+i\tau' \bar z(\bar
y+\bar \qq')}+\dr_{z}\gep\,,\nn
\end{align}
with the same gauge freedom $\dr_z\gep\in\mathcal{C}^{1}$ and an ambiguity in one function $\bar\psi$, which we also set to zero:
\begin{equation}
    \bar\psi=0\label{App:gep-psi:conj}\,.    
\end{equation}

\paragraph{Zero-form module}

The chiral 0-forms \eqref{B20} and \eqref{B11} result from the obtained $\Lambda$'s and our choice of ambiguity freedom in Eqs. \eqref{App:gep-psi} and \eqref{App:gep-psi:conj}. The form \eqref{B20:res} for $\mathcal{B}_{2,0}$ straightforwardly emerges from \eqref{B20} using \eqref{App:frq-ident} and represents a spin-local contribution to HS vertices; see \cite{Didenko:2020bxd},
\begin{align}
    \mathcal{B}_{2,0}(z; Y)=&
    -y^{\al}y^{\gb}\int_{\Delta_{ \rho,\gs}}\frac{\rho_2}{(\rho_1+\rho_2)(\rho_2+\rho_3)} e^{i\gs  y\mathcal{P}_y+i(1-\gs)y\mathcal{P}'_y}\moast\int_{0}^{1}d\tau \tau(1-\tau)\,z_{\al}z_{\gb} e^{i\tau z(y+\qq)}\times\nn\\
    &\times C(y_1', \bar y)\bar *C(y_2', \bar y)\bar *C(y_3', \bar y)\Big|_{y_i'=0}\,.
\end{align}

Field $\bar{\mathcal{B}}_{1,1}$ in \eqref{B11} is new. Let us look closer at its structure. To proceed, we rewrite it in the following form:
\begin{equation}   
\bar{\mathcal{B}}_{1,1}(\bar z; Y)=-i\int_{\Delta_{ \rho,\gs}}(\bar y+\bar\xi)^{\dal}\left(2+\qq^{\al}\frac{\p}{\p\qq^{\al}}\right)\left(\phi(-\qq, \bar y)e^{-iy\qq}\right)\bar\moast\int_{0}^{1}d\tau'(1-\tau')\bar z_{\dal}e^{i\tau' \bar z(\bar y+\bar\qq')}\,,\label{App:B11}
\end{equation}
where $\bar\xi_{\dal}$, $\qq_{\al}$, $\bar\qq'_{\dal}$ and $\phi(y,\bar y)$ are defined in \eqref{def:q-xi} and \eqref{def:phi2}. Using this and also that 
\begin{subequations}
    \begin{align}
        &\sum_{i=1}^3\rho_i\frac{\p}{\p\rho_i}\mathcal{P}_y=\sum_{i=1}^3\rho_i\frac{\p}{\p\rho_i}\mathcal{P}'_y=0\,,\\
        &\sum_{i=1}^3\rho_i\frac{\p}{\p\rho_i}\bar{\mathcal{P}}_{\bar y}=\sum_{i=1}^3\rho_i\frac{\p}{\p\rho_i}\bar{\mathcal{P}}'_{\bar y}=0\,,
    \end{align}
\end{subequations}
we obtain 
\begin{align}
  \bar{\mathcal{B}}_{1,1}(\bar z; Y)=-i\int_{\Delta_{ \rho,\gs}}\frac{\rho_2}{(\rho_1+\rho_2)(\rho_2+\rho_3)}(\bar y+\bar\xi)^{\dal}\left(2+\sum_{i=1}^3\rho_i\frac{\p}{\p\rho_i}\right)\times\\
  \times e^{-iy\qq-i\qq\mathcal{P}_y+i\gs(\bar y+\bar\pp_1)\bar{\mathcal{P}}_{\bar y}+i(1-\gs)(\bar y+\bar\pp_3)\bar{\mathcal{P}}'_{\bar y}}\,\bar\moast\int_{0}^{1}d\tau'(1-\tau')\bar z_{\dal}e^{i\tau' \bar z(\bar y+\bar\qq')}\,,\nn
\end{align}
This expression can be further reduced by integrating by parts with respect to the variables $\rho_i$. To this end, we note that 
\begin{equation}
    \sum_{i=1}^3\rho_i\frac{\p}{\p\rho_i}\frac{\rho_2}{(\rho_1+\rho_2)(\rho_2+\rho_3)}=-\frac{\rho_2}{(\rho_1+\rho_2)(\rho_2+\rho_3)}\,,\qquad \sum_{i=1}^3\rho_i\frac{\p}{\p\rho_i}\xi_{\dal}=\xi_{\dal}\,.
\end{equation}
We also rewrite
\begin{equation}
    \int_{\Delta_\rho}d^3\rho:=\int d^3\rho\,\gd(x)\prod_{i=1}^3\theta(\rho_i)\theta(1-\rho_i)\,,\qquad x:=1-\rho_1-\rho_2-\rho_3\,, 
\end{equation}
where 
\begin{align}
        \theta(x)=\begin{cases}
        1 \,,\quad x\geq 0\,, \\
        0\,,\quad x<0\,,
    \end{cases}
    \end{align}
and the domain $\Delta_{\rho}$ is specified in \eqref{domain:rho-gs}. Using that $\frac{\p}{\p\rho_i}\gd(x)=-\gd'(x)$, after simple algebra, we arrive at the final result:
\begin{align}\label{App:B11-fin}
  \bar{\mathcal{B}}_{1,1}(\bar z; Y)=-i\int_{\Delta_{\rho, \gs}} d^3\rho d\gs\,\frac{\rho_2}{(\rho_1+\rho_2)(\rho_2+\rho_3)}[\gd(x)\bar y^{\dal}+\gd'(x)(\bar y+\bar\xi)^{\dal}]\times\\
  \times e^{-iy\qq-i\qq\mathcal{P}_y+i\gs(\bar y+\bar\pp_1)\bar{\mathcal{P}}_{\bar y}+i(1-\gs)(\bar y+\bar\pp_3)\bar{\mathcal{P}}'_{\bar y}}\,\bar\moast\int_{0}^{1}d\tau'(1-\tau')\bar z_{\dal}e^{i\tau' \bar z(\bar y+\bar\qq')}\,.\nn
\end{align}
Notice that there are no contributions from differentiating the boundary $\theta(\rho_i)$ and $\theta(1-\rho_i)$.

    % основной текст -- от абстракта до acknowledgments
    
%\input{parts/appendix.tex}
    % аппендикс


\addcontentsline{toc}{section}{Bibliography}

\begin{thebibliography}{20}

%\cite{Vasiliev:1999ba}
\bibitem{Vasiliev:1999ba}
M.~A.~Vasiliev,
{\it Higher spin gauge theories: Star product and AdS space,}
%doi:10.1142/9789812793850{\_}0030
[arXiv:hep-th/9910096 [hep-th]].

%\cite{Bekaert:2004qos}
\bibitem{Bekaert:2004qos}
X.~Bekaert, S.~Cnockaert, C.~Iazeolla and M.~A.~Vasiliev,
{\it Nonlinear higher spin theories in various dimensions,}
[arXiv:hep-th/0503128 [hep-th]].

%\cite{Didenko:2014dwa}
\bibitem{Didenko:2014dwa}
V.~E.~Didenko and E.~D.~Skvortsov,
{\it Elements of Vasiliev Theory,}
Lect. Notes Phys. \textbf{1028}, 269-456 (2024)
%doi:10.1007/978-3-031-59656-8{\_}3
[arXiv:1401.2975 [hep-th]].

%\cite{Vasiliev:1990en}
\bibitem{Vasiliev:1990en}
M.~A.~Vasiliev, {\it Consistent equation for interacting gauge fields of all spins in (3+1)-dimensions,}
Phys. Lett. B \textbf{243}, 378-382 (1990)
%doi:10.1016/0370-2693(90)91400-6

%\cite{Vasiliev:1992av}
\bibitem{Vasiliev:1992av}
M.~A.~Vasiliev,
{\it More on equations of motion for interacting massless fields of all spins in (3+1)-dimensions,}
Phys. Lett. B \textbf{285} (1992), 225-234
%doi:10.1016/0370-2693(92)91457-K

%\cite{Boulanger:2011dd}
\bibitem{Boulanger:2011dd}
N.~Boulanger and P.~Sundell,
{\it An action principle for Vasiliev's four-dimensional higher-spin gravity,}
J. Phys. A \textbf{44}, 495402 (2011)
%doi:10.1088/1751-8113/44/49/495402
[arXiv:1102.2219 [hep-th]].

%\cite{Flato:1978qz}
\bibitem{Flato:1978qz}
M.~Flato and C.~Fronsdal,
{\it One Massless Particle Equals Two Dirac Singletons: Elementary Particles in a Curved Space. 6.,}
Lett. Math. Phys. \textbf{2} (1978), 421-426
%doi:10.1007/BF00400170

%\cite{Sezgin:2002rt}
\bibitem{Sezgin:2002rt}
E.~Sezgin and P.~Sundell, {\it Massless higher spins and holography,}
Nucl. Phys. B \textbf{644} (2002), 303-370
[erratum: Nucl. Phys. B \textbf{660} (2003), 403-403]
%doi:10.1016/S0550-3213(02)00739-3
[arXiv:hep-th/0205131 [hep-th]].


%\cite{Klebanov:2002ja}
\bibitem{Klebanov:2002ja}
I.~R.~Klebanov and A.~M.~Polyakov,
{\it AdS dual of the critical O(N) vector model,}
Phys. Lett. B \textbf{550}, 213-219 (2002)
%doi:10.1016/S0370-2693(02)02980-5
[arXiv:hep-th/0210114 [hep-th]].

%\cite{Leigh:2003gk}
\bibitem{Leigh:2003gk}
R.~G.~Leigh and A.~C.~Petkou,
{\it Holography of the N=1 higher spin theory on AdS(4),}
JHEP \textbf{06}, 011 (2003)
%doi:10.1088/1126-6708/2003/06/011
[arXiv:hep-th/0304217 [hep-th]].

%\cite{Sezgin:2003pt}
\bibitem{Sezgin:2003pt}
E.~Sezgin and P.~Sundell,
{\it Holography in 4D (super) higher spin theories and a test via cubic scalar couplings,}
JHEP \textbf{07}, 044 (2005)
%doi:10.1088/1126-6708/2005/07/044
[arXiv:hep-th/0305040 [hep-th]].


%\cite{Giombi:2009wh}
\bibitem{Giombi:2009wh}
S.~Giombi and X.~Yin,
{\it Higher Spin Gauge Theory and Holography: The Three-Point Functions,}
JHEP \textbf{09} (2010), 115
%doi:10.1007/JHEP09(2010)115
[arXiv:0912.3462 [hep-th]].

%\cite{Giombi:2010vg}
\bibitem{Giombi:2010vg}
S.~Giombi and X.~Yin,
{\it Higher Spins in AdS and Twistorial Holography,}
JHEP \textbf{04}, 086 (2011)
%doi:10.1007/JHEP04(2011)086
[arXiv:1004.3736 [hep-th]].

%\cite{Didenko:2017lsn}
\bibitem{Didenko:2017lsn}
V.~E.~Didenko and M.~A.~Vasiliev,
{\it Test of the local form of higher-spin equations via AdS / CFT,}
Phys. Lett. B \textbf{775} (2017), 352-360
%doi:10.1016/j.physletb.2017.09.091
[arXiv:1705.03440 [hep-th]].

%\cite{Maldacena:2011jn}
\bibitem{Maldacena:2011jn}
J.~Maldacena and A.~Zhiboedov,
{\it Constraining Conformal Field Theories with A Higher Spin Symmetry,}
J. Phys. A \textbf{46}, 214011 (2013)
%doi:10.1088/1751-8113/46/21/214011
[arXiv:1112.1016 [hep-th]].

%\cite{Maldacena:2012sf}
\bibitem{Maldacena:2012sf}
J.~Maldacena and A.~Zhiboedov,
{\it Constraining conformal field theories with a slightly broken higher spin symmetry,}
Class. Quant. Grav. \textbf{30}, 104003 (2013)
%doi:10.1088/0264-9381/30/10/104003
[arXiv:1204.3882 [hep-th]].


%\cite{Colombo:2012jx}
\bibitem{Colombo:2012jx}
N.~Colombo and P.~Sundell, {\it Higher Spin Gravity Amplitudes From Zero-form Charges,}
[arXiv:1208.3880 [hep-th]].


%\cite{Didenko:2012tv}
\bibitem{Didenko:2012tv}
V.~E.~Didenko and E.~D.~Skvortsov,
{\it Exact higher-spin symmetry in CFT: all correlators in unbroken Vasiliev theory,}
JHEP \textbf{04} (2013), 158
%doi:10.1007/JHEP04(2013)158
[arXiv:1210.7963 [hep-th]].

%\cite{Gelfond:2013xt}
\bibitem{Gelfond:2013xt}
O.~A.~Gelfond and M.~A.~Vasiliev,
{\it Operator algebra of free conformal currents via twistors,}
Nucl. Phys. B \textbf{876}, 871-917 (2013)
%doi:10.1016/j.nuclphysb.2013.09.001
[arXiv:1301.3123 [hep-th]].

%\cite{Engquist:2005yt}
\bibitem{Engquist:2005yt}
J.~Engquist and P.~Sundell, {\it Brane partons and singleton strings,}
Nucl. Phys. B \textbf{752} (2006), 206-279
%doi:10.1016/j.nuclphysb.2006.06.040
[arXiv:hep-th/0508124 [hep-th]].

%\cite{Sezgin:2005pv}
\bibitem{Sezgin:2005pv}
E.~Sezgin and P.~Sundell, {\it An Exact solution of 4-D higher-spin gauge theory,}
Nucl. Phys. B \textbf{762} (2007), 1-37
%doi:10.1016/j.nuclphysb.2006.06.038
[arXiv:hep-th/0508158 [hep-th]].

%\cite{Colombo:2010fu}
\bibitem{Colombo:2010fu}
N.~Colombo and P.~Sundell, {\it Twistor space observables and quasi-amplitudes in 4D higher spin gravity,}
JHEP \textbf{11} (2011), 042
%doi:10.1007/JHEP11(2011)042
[arXiv:1012.0813 [hep-th]].

%\cite{Sezgin:2011hq}
\bibitem{Sezgin:2011hq}
E.~Sezgin and P.~Sundell, {\it Geometry and Observables in Vasiliev's Higher Spin Gravity,}
JHEP \textbf{07} (2012), 121
%doi:10.1007/JHEP07(2012)121
[arXiv:1103.2360 [hep-th]].


%\cite{Scalea:2023dpw}
\bibitem{Scalea:2023dpw}
A.~Scalea,
{\it On Correlation Functions as Higher-Spin Invariants,}
Symmetry \textbf{15}, no.4, 950 (2023)
%doi:10.3390/sym15040950
[arXiv:2303.11159 [hep-th]].

%\cite{Sleight:2016dba}
\bibitem{Sleight:2016dba}
C.~Sleight and M.~Taronna,
{\it Higher Spin Interactions from Conformal Field Theory: The Complete Cubic Couplings,}
Phys. Rev. Lett. \textbf{116}, no.18, 181602 (2016)
%doi:10.1103/PhysRevLett.116.181602
[arXiv:1603.00022 [hep-th]].

%\cite{Prokushkin:1998bq}
\bibitem{Prokushkin:1998bq}
S.~F.~Prokushkin and M.~A.~Vasiliev,
{\it Higher spin gauge interactions for massive matter fields in 3-D AdS space-time,}
Nucl. Phys. B \textbf{545} (1999), 385
%doi:10.1016/S0550-3213(98)00839-6
[arXiv:hep-th/9806236 [hep-th]].

%\cite{Boulanger:2015ova}
\bibitem{Boulanger:2015ova}
N.~Boulanger, P.~Kessel, E.~D.~Skvortsov and M.~Taronna,
{\it Higher spin interactions in four-dimensions: Vasiliev versus Fronsdal,}
J. Phys. A \textbf{49} (2016) no.9, 095402
%doi:10.1088/1751-8113/49/9/095402
[arXiv:1508.04139 [hep-th]].

%\cite{Bekaert:2015tva}
\bibitem{Bekaert:2015tva}
X.~Bekaert, J.~Erdmenger, D.~Ponomarev and C.~Sleight,
{\it Quartic AdS Interactions in Higher-Spin Gravity from Conformal Field Theory,}
JHEP \textbf{11} (2015), 149
%doi:10.1007/JHEP11(2015)149
[arXiv:1508.04292 [hep-th]].

%\cite{Sleight:2017pcz}
\bibitem{Sleight:2017pcz}
C.~Sleight and M.~Taronna,
{\it Higher-Spin Gauge Theories and Bulk Locality,}
Phys. Rev. Lett. \textbf{121} (2018) no.17, 171604
%doi:10.1103/PhysRevLett.121.171604
[arXiv:1704.07859 [hep-th]].

%\cite{Ponomarev:2017qab}
\bibitem{Ponomarev:2017qab}
D.~Ponomarev,
{\it A Note on (Non)-Locality in Holographic Higher Spin Theories,}
Universe \textbf{4}, no.1, 2 (2018)
%doi:10.3390/universe4010002
[arXiv:1710.00403 [hep-th]].

%\cite{Neiman:2023orj}
\bibitem{Neiman:2023orj}
Y.~Neiman,
{\it Quartic locality of higher-spin gravity in de Sitter and Euclidean anti-de Sitter space,}
Phys. Lett. B \textbf{843}, 138048 (2023)
%doi:10.1016/j.physletb.2023.138048
[arXiv:2302.00852 [hep-th]].

%\cite{Vasiliev:2012vf}
\bibitem{Vasiliev:2012vf}
M.~A.~Vasiliev,
{\it Holography, Unfolding and Higher-Spin Theory,}
J. Phys. A \textbf{46} (2013), 214013
%doi:10.1088/1751-8113/46/21/214013
[arXiv:1203.5554 [hep-th]].

%\cite{Diaz:2024kpr}
\bibitem{Diaz:2024kpr}
F.~Diaz, C.~Iazeolla and P.~Sundell,
{\it Fractional spins, unfolding, and holography. Part I. Parent field equations for dual higher-spin gravity reductions,}
JHEP \textbf{09} (2024), 109
%doi:10.1007/JHEP09(2024)109
[arXiv:2403.02283 [hep-th]].

%\cite{Diaz:2024iuz}
\bibitem{Diaz:2024iuz}
F.~Diaz, C.~Iazeolla and P.~Sundell,
{\it Fractional spins, unfolding, and holography. Part II. 4D higher spin gravity and 3D conformal dual,}
JHEP \textbf{10} (2024), 066
%doi:10.1007/JHEP10(2024)066
[arXiv:2403.02301 [hep-th]].



%\cite{Lysov:2022zlw}
\bibitem{Lysov:2022zlw}
V.~Lysov and Y.~Neiman,
{\it Higher-spin gravity{\textquoteright}s {\textquotedblleft}string{\textquotedblright}: new gauge and proof of holographic duality for the linearized Didenko-Vasiliev solution,}
JHEP \textbf{10}, 054 (2022)
%doi:10.1007/JHEP10(2022)054
[arXiv:2207.07507 [hep-th]].

%\cite{Neiman:2022enh}
\bibitem{Neiman:2022enh}
Y.~Neiman,
{\it New Diagrammatic Framework for Higher-Spin Gravity,}
Phys. Rev. Lett. \textbf{130}, no.17, 171601 (2023)
%doi:10.1103/PhysRevLett.130.171601
[arXiv:2209.02185 [hep-th]].

%\cite{Vasiliev:2016xui}
\bibitem{Vasiliev:2016xui}
M.~A.~Vasiliev,
{\it Current Interactions and Holography from the 0-Form Sector of Nonlinear Higher-Spin Equations,}
JHEP \textbf{10} (2017), 111
%doi:10.1007/JHEP10(2017)111
[arXiv:1605.02662 [hep-th]].

%\cite{Vasiliev:2017cae}
\bibitem{Vasiliev:2017cae}
M.~A.~Vasiliev,
{\it On the Local Frame in Nonlinear Higher-Spin Equations,}
JHEP \textbf{01}, 062 (2018)
%doi:10.1007/JHEP01(2018)062
[arXiv:1707.03735 [hep-th]].

%\cite{Gelfond:2018vmi}
\bibitem{Gelfond:2018vmi}
O.~A.~Gelfond and M.~A.~Vasiliev,
{\it Homotopy Operators and Locality Theorems in Higher-Spin Equations,}
Phys. Lett. B \textbf{786} (2018), 180-188
%doi:10.1016/j.physletb.2018.09.038
[arXiv:1805.11941 [hep-th]].

%\cite{Didenko:2018fgx}
\bibitem{Didenko:2018fgx}
V.~E.~Didenko, O.~A.~Gelfond, A.~V.~Korybut and M.~A.~Vasiliev,
{\it Homotopy Properties and Lower-Order Vertices in Higher-Spin Equations,}
J. Phys. A \textbf{51} (2018) no.46, 465202
%doi:10.1088/1751-8121/aae5e1
[arXiv:1807.00001 [hep-th]].

%\cite{Didenko:2019xzz}
\bibitem{Didenko:2019xzz}
V.~E.~Didenko, O.~A.~Gelfond, A.~V.~Korybut and M.~A.~Vasiliev,
{\it Limiting Shifted Homotopy in Higher-Spin Theory and Spin-Locality,}
JHEP \textbf{12} (2019), 086
%doi:10.1007/JHEP12(2019)086
[arXiv:1909.04876 [hep-th]].

%\cite{Gelfond:2019tac}
\bibitem{Gelfond:2019tac}
O.~A.~Gelfond and M.~A.~Vasiliev,
{\it Spin-Locality of Higher-Spin Theories and Star-Product Functional Classes,}
JHEP \textbf{03} (2020), 002
%doi:10.1007/JHEP03(2020)002
[arXiv:1910.00487 [hep-th]].

%\cite{Didenko:2020bxd}
\bibitem{Didenko:2020bxd}
V.~E.~Didenko, O.~A.~Gelfond, A.~V.~Korybut and M.~A.~Vasiliev,
{Spin-locality of $\eta^{2}$ and $ {\overline{\eta}}^2 $ quartic higher-spin vertices,}
JHEP \textbf{12} (2020), 184
%doi:10.1007/JHEP12(2020)184
[arXiv:2009.02811 [hep-th]].

%\cite{Gelfond:2021two}
\bibitem{Gelfond:2021two}
O.~A.~Gelfond and A.~V.~Korybut,
{\it Manifest form of the spin-local higher-spin vertex $\varUpsilon ^{\eta \eta }_{\omega CCC}$,}
Eur. Phys. J. C \textbf{81} (2021) no.7, 605
%doi:10.1140/epjc/s10052-021-09401-4
[arXiv:2101.01683 [hep-th]].


%\cite{Vasiliev:2022med}
\bibitem{Vasiliev:2022med}
M.~A.~Vasiliev,
{\it Projectively-compact spinor vertices and space-time spin-locality in higher-spin theory,}
Phys. Lett. B \textbf{834} (2022), 137401
%doi:10.1016/j.physletb.2022.137401
[arXiv:2208.02004 [hep-th]].

%\cite{Didenko:2022eso}
\bibitem{Didenko:2022eso}
V.~E.~Didenko and A.~V.~Korybut,
{\it On z-dominance, shift symmetry and spin locality in higher-spin theory,}
JHEP \textbf{05} (2023), 133
%doi:10.1007/JHEP05(2023)133
[arXiv:2212.05006 [hep-th]].


%\cite{Vasiliev:2023yzx}
\bibitem{Vasiliev:2023yzx}
M.~A.~Vasiliev,
{\it Differential contracting homotopy in higher-spin theory,}
JHEP \textbf{11} (2023), 048
%doi:10.1007/JHEP11(2023)048
[arXiv:2307.09331 [hep-th]].

%\cite{Kirakosiants:2025gpd}
\bibitem{Kirakosiants:2025gpd}
P.~T.~Kirakosiants, D.~A.~Valerev and M.~A.~Vasiliev,
{\it Quadratic Corrections to the Higher-Spin Equations by the Differential Homotopy Approach,} Nucl. Phys. B \textbf{1023} (2026), 117290
%doi:10.1016/j.nuclphysb.2025.117290
[arXiv:2506.16634 [hep-th]].

%\cite{Didenko:2022qga}
\bibitem{Didenko:2022qga}
V.~E.~Didenko,
{\it On holomorphic sector of higher-spin theory,}
JHEP \textbf{10} (2022), 191
%doi:10.1007/JHEP10(2022)191
[arXiv:2209.01966 [hep-th]].

%\cite{Korybut:2025vdn}
\bibitem{Korybut:2025vdn}
A.~V.~Korybut,
{\it On consistency of the interacting (anti)holomorphic higher-spin sector,}
Eur. Phys. J. C \textbf{85}, no.8, 885 (2025)
%doi:10.1140/epjc/s10052-025-14617-9
[arXiv:2505.13125 [hep-th]].

%\cite{Didenko:2024zpd}
\bibitem{Didenko:2024zpd}
V.~E.~Didenko and M.~A.~Povarnin,
{\it All vertices for unconstrained symmetric gauge fields,}
Phys. Rev. D \textbf{110} (2024) no.12, 126012
[erratum: Phys. Rev. D \textbf{111} (2025) no.10, 109901]
%doi:10.1103/PhysRevD.110.126012
[arXiv:2409.00808 [hep-th]].

%\cite{Vasiliev:1988sa}
\bibitem{Vasiliev:1988sa}
M.~A.~Vasiliev,
{\it Consistent Equations for Interacting Massless Fields of All Spins in the First Order in Curvatures,}
Annals Phys. \textbf{190} (1989), 59-106
%doi:10.1016/0003-4916(89)90261-3

%\cite{Misuna:2022cma}
\bibitem{Misuna:2022cma}
N.~Misuna,
{\it Unfolded dynamics approach and quantum field theory,}
JHEP \textbf{12} (2023), 119
%doi:10.1007/JHEP12(2023)119
[arXiv:2208.04306 [hep-th]].

%\cite{Misuna:2024ccj}
\bibitem{Misuna:2024ccj}
N.~Misuna,
{\it Scalar electrodynamics and Higgs mechanism in the unfolded dynamics approach,}
JHEP \textbf{12} (2024), 090
%doi:10.1007/JHEP12(2024)090
[arXiv:2402.14164 [hep-th]].

%\cite{Misuna:2024dlx}
\bibitem{Misuna:2024dlx}
N.~Misuna,
{\it Unfolded formulation of 4d Yang-Mills theory,}
Phys. Lett. B \textbf{870}, 139882 (2025)
%doi:10.1016/j.physletb.2025.139882
[arXiv:2408.13212 [hep-th]].

%\cite{Iazeolla:2025btr}
\bibitem{Iazeolla:2025btr}
C.~Iazeolla, P.~Sundell and B.~C.~Vallilo,
{\it Unfolding the six-dimensional tensor multiplet,}
J. Phys. A \textbf{58}, no.36, 365402 (2025)
%doi:10.1088/1751-8121/adfe46
[arXiv:2503.14673 [hep-th]].


%\cite{DeFilippi:2021xon}
\bibitem{DeFilippi:2021xon}
D.~De Filippi, C.~Iazeolla and P.~Sundell,
{\it Metaplectic representation and ordering (in)dependence in Vasiliev{\textquoteright}s higher spin gravity,}
JHEP \textbf{07}, 003 (2022)
%doi:10.1007/JHEP07(2022)003
[arXiv:2111.09288 [hep-th]].

%\cite{Didenko:2023vna}
\bibitem{Didenko:2023vna}
V.~E.~Didenko and A.~V.~Korybut,
{\it Interaction of symmetric higher-spin gauge fields,}
Phys. Rev. D \textbf{108} (2023) no.8, 086031
[erratum: Phys. Rev. D \textbf{109} (2024) no.6, 069901]
%doi:10.1103/PhysRevD.108.086031
[arXiv:2304.08850 [hep-th]].

%\cite{Didenko:2021vdb}
\bibitem{Didenko:2021vdb}
V.~E.~Didenko and A.~V.~Korybut,
{\it Planar solutions of higher-spin theory. Nonlinear corrections,}
JHEP \textbf{01} (2022), 125
%doi:10.1007/JHEP01(2022)125
[arXiv:2110.02256 [hep-th]].

%\cite{Didenko:2025xca}
\bibitem{Didenko:2025xca}
V.~E.~Didenko and I.~S.~Faliakhov,
{\it Symmetry breaking in the self-dual higher-spin theory,}
Phys. Rev. D \textbf{112}, no.10, 106010 (2025)
%doi:10.1103/nmxv-c8kx
[arXiv:2509.01477 [hep-th]].

%\cite{Vasiliev:2015wma}
\bibitem{Vasiliev:2015wma}
M.~A.~Vasiliev,
{\it Star-Product Functions in Higher-Spin Theory and Locality,}
JHEP \textbf{06}, 031 (2015)
%doi:10.1007/JHEP06(2015)031
[arXiv:1502.02271 [hep-th]].

%\cite{Metsaev:1991mt}
\bibitem{Metsaev:1991mt}
R.~R.~Metsaev,
{\it Poincare invariant dynamics of massless higher spins: Fourth order analysis on mass shell,}
Mod. Phys. Lett. A \textbf{06} (1991), 359-367
%doi:10.1142/S0217732391000348

\bibitem{Metsaev:PhD}
R.~R.~Metsaev, {\it Effective Action in String Theory.} PhD thesis, Lebedev Physical Institute, 1991

%\cite{Metsaev:2018xip}
\bibitem{Metsaev:2018xip}
R.~R.~Metsaev,
{\it Light-cone gauge cubic interaction vertices for massless fields in AdS(4),}
Nucl. Phys. B \textbf{936}, 320-351 (2018)
%doi:10.1016/j.nuclphysb.2018.09.021
[arXiv:1807.07542 [hep-th]].

%\cite{Ponomarev:2017nrr}
\bibitem{Ponomarev:2017nrr}
D.~Ponomarev,
{\it Chiral Higher Spin Theories and Self-Duality,}
JHEP \textbf{12} (2017), 141
%doi:10.1007/JHEP12(2017)141
[arXiv:1710.00270 [hep-th]].

%\cite{Serrani:2025owx}
\bibitem{Serrani:2025owx}
M.~Serrani,
{\it On classification of (self-dual) higher-spin gravities in flat space}
JHEP \textbf{08}, 032 (2025)
%doi:10.1007/JHEP08(2025)032
[arXiv:2505.12839 [hep-th]].

%\cite{Basile:2024raj}
\bibitem{Basile:2024raj}
T.~Basile,
{\it Massless chiral fields in six dimensions,}
SciPost Phys. \textbf{19}, no.3, 079 (2025)
%doi:10.21468/SciPostPhys.19.3.079
[arXiv:2409.12800 [hep-th]].

%\cite{Sharapov:2022nps}
\bibitem{Sharapov:2022nps}
A.~Sharapov, E.~Skvortsov, A.~Sukhanov and R.~Van Dongen,
{\it More on Chiral Higher Spin Gravity and convex geometry,}
Nucl. Phys. B \textbf{990} (2023), 116152
%doi:10.1016/j.nuclphysb.2023.116152
[arXiv:2209.15441 [hep-th]].

%\cite{Sharapov:2023erv}
\bibitem{Sharapov:2023erv}
A.~Sharapov, E.~Skvortsov and R.~Van Dongen, {\it Strong homotopy algebras for chiral higher spin gravity via Stokes theorem,}
JHEP \textbf{06}, 186 (2024)
%doi:10.1007/JHEP06(2024)186
[arXiv:2312.16573 [hep-th]].

%\cite{Engquist:2002vr}
\bibitem{Engquist:2002vr}
J.~Engquist, E.~Sezgin and P.~Sundell,
{\it On N=1, N=2, N=4 higher spin gauge theories in four-dimensions,}
Class. Quant. Grav. \textbf{19}, 6175-6196 (2002)
%doi:10.1088/0264-9381/19/23/316
[arXiv:hep-th/0207101 [hep-th]].

%\cite{Engquist:2002gy}
\bibitem{Engquist:2002gy}
J.~Engquist, E.~Sezgin and P.~Sundell,
{\it Superspace formulation of 4-D higher spin gauge theory,}
Nucl. Phys. B \textbf{664}, 439-456 (2003)
%doi:10.1016/S0550-3213(03)00411-5
[arXiv:hep-th/0211113 [hep-th]].

%\cite{Gelfond:2023fwe}
\bibitem{Gelfond:2023fwe}
O.~A.~Gelfond,
{\it Moderately non-local $\eta {\bar{\eta }}$ vertices in the $AdS_4$ higher-spin gauge theory,} Eur. Phys. J. C \textbf{83} (2023) no.12, 1154
%doi:10.1140/epjc/s10052-023-12308-x
[arXiv:2308.16281 [hep-th]].

%\cite{Vasiliev:2018zer}
\bibitem{Vasiliev:2018zer}
M.~A.~Vasiliev,
{\it From Coxeter Higher-Spin Theories to Strings and Tensor Models,}
JHEP \textbf{08}, 051 (2018)
%doi:10.1007/JHEP08(2018)051
[arXiv:1804.06520 [hep-th]].

%\cite{Tarusov:2025sre}
\bibitem{Tarusov:2025sre}
A.~A.~Tarusov, K.~A.~Ushakov and M.~A.~Vasiliev,
{\it Linearized Coxeter higher-spin theories,}
JHEP \textbf{08}, 052 (2025)
%doi:10.1007/JHEP08(2025)052
[arXiv:2503.05948 [hep-th]].

\end{thebibliography}
\end{document}